\documentclass[aps,prb,twocolumn,showpacs,floatfix,superscriptaddress,nofootinbib]{revtex4-2}
\usepackage{graphicx}%
\usepackage{dcolumn}%
\usepackage{bm}%
\usepackage{physics}
\usepackage{hyperref}%
\usepackage{color}
\usepackage{amsmath}
\usepackage{amssymb}
\usepackage{ulem} %
\usepackage{multirow}
\usepackage{booktabs}
\usepackage{blkarray}
\usepackage{tabularray}
\usepackage[ruled]{algorithm2e}

\definecolor{gr4} {RGB}{34,118,34}

\newcommand{\gtwo}{g^{(2)}(i)}

\SetCommentSty{mycommfont}

\begin{document}
\title{Catching thermal avalanches in the disordered XXZ model}

\author{Tomasz Szo\l{}dra} 
\affiliation{Doctoral School of Exact and Natural Sciences, Jagiellonian University, \L{}ojasiewicza 11, PL-30-348 Krak\'ow, Poland}
\affiliation{Instytut Fizyki Teoretycznej, 
Uniwersytet Jagiello\'nski,  \L{}ojasiewicza 11, PL-30-348 Krak\'ow, Poland}
\author{Piotr Sierant} 
\affiliation{ICFO-Institut de Ci\`encies Fot\`oniques, The Barcelona Institute of Science and Technology, Av. Carl Friedrich
Gauss 3, 08860 Castelldefels (Barcelona), Spain}
\author{Maciej Lewenstein} 
\affiliation{ICFO-Institut de Ci\`encies Fot\`oniques, The Barcelona Institute of Science and Technology, Av. Carl Friedrich
Gauss 3, 08860 Castelldefels (Barcelona), Spain}
\affiliation{ICREA, Passeig Lluis Companys 23, 08010 Barcelona, Spain}
\author{Jakub Zakrzewski} 
\affiliation{Instytut Fizyki Teoretycznej, 
Uniwersytet Jagiello\'nski,  \L{}ojasiewicza 11, PL-30-348 Krak\'ow, Poland}
\affiliation{Mark Kac Complex Systems Research Center, Uniwersytet Jagiello{\'n}ski, PL-30-348 Krak{\'o}w, Poland}

\date{\today}

\begin{abstract}

We study the XXZ model with a random magnetic field in contact with a weakly disordered spin chain, acting as a finite thermal bath. We revise Fermi's golden rule description of the interaction between the thermal bath and the XXZ spin chain, contrasting it with a nonperturbative quantum avalanche scenario for the thermalization of the system. We employ two-point correlation functions to define the extent $\xi_d$ of the thermalized region next to the bath. Unbounded growth of $\xi_d$ proportional to the logarithm of time or faster is a signature of an avalanche. Such behavior signifies the thermalization of the system, as we confirm numerically for a generic initial state in the ergodic and critical regimes of the XXZ spin chain. In the many-body localized regime, a clear termination of avalanches is observed for specifically prepared initial states and, surprisingly, is not visible for generic initial product states.
Additionally, we extract the localization length of the local integrals of motion and show that a bath made out of a weakly disordered XXZ chain has a similar effect on the system as a bath modeled by a Hamiltonian from a Gaussian orthogonal ensemble of random matrices. We also comment on the result of the earlier study (Phys. Rev. B 108, L020201 (2023)), arguing that the observed thermalization is due to external driving of the system and does not occur in the autonomous model. Our work reveals experimentally accessible signatures of quantum avalanches and identifies conditions under which termination of the avalanches may be observed. 

\end{abstract}

\maketitle

\section{Introduction}

The ergodic hypothesis of Boltzmann~\cite{Gallavotti95} stipulates that all microstates of a many-body system are equiprobable over a sufficiently long period of time. Such an equilibrium state can be described within the framework of statistical mechanics with several macroscopic variables. The eigenstate thermalization hypothesis (\textbf{ETH})~\cite{Deutsch91,Srednicki94,Rigol08,Alessio16} describes the equilibrium of isolated quantum many-body systems, warranting that, after the thermalization process is complete, local observables, including their higher order correlation functions, are described by appropriate ensembles of statistical mechanics~\cite{Foini19,Pappalardi22,Pappalardi23,pappalardi2023microcanonical,fava2023designs}.

\begin{figure}[hbt!]
    \centering
    \includegraphics[width=.85\columnwidth]{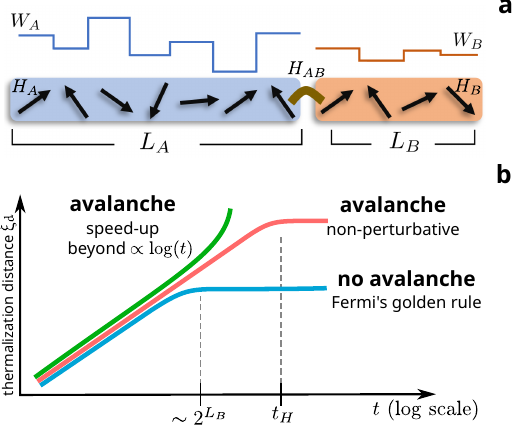}
    \caption{\textbf{a} Scheme of our setup: two spin-1/2 XXZ chains of sizes $L_A$, $L_B$, governed by Hamiltonians $H_A$, $H_B$, with different amplitudes of random disorder, connected at the interface by $H_{AB}$. \textbf{b} Sketch of the signatures of quantum avalanches obtained in Sec.~\ref{sec:ergVSmbl} for the model from panel \textbf{a}. First order perturbative scheme predicts the growth of thermalized region as $\propto \ln(t)$ up to $t\sim 2^{L_B}$ when the system starts to feel the discreteness of the bath spectrum (blue curve). In the avalanche scenario, the bath continually increases its size and the logarithmic growth continues until the Heisenberg time $t_H$ (red curve) or speeds up (green curve).}
    \label{fig:setup}
\end{figure}

While the applicability of the ETH has been confirmed for a wide range of quantum many-body systems, see e.g.~\cite{Rigol09, Santos10,  Steinigeweg13, Khatami13, Beugeling14, Schonle21}, a phenomenon of many-body localization (\textbf{MBL})~\cite{Gornyi05, Basko06,Oganesyan07,Znidaric08,Pal10} has been proposed as a general mechanism that prevents many-body systems from reaching the thermal equilibrium in the presence of a strong disorder. Numerical results for spin-1/2 XXZ chains \cite{DeLuca13,Luitz15}, bosonic models \cite{Sierant17,Sierant18, Orell19, Hopjan19}, or systems of spinful fermions \cite{Mondaini15, Prelovsek16, Zakrzewski18, Kozarzewski18} indicate a rapid slow-down of dynamics with increasing disorder strength~\cite{Luitz16, Bera17,Weiner19,Sels23dilute}, as confirmed experimentally in setups of ultracold atoms \cite{Schreiber15,Choi16,Luschen17,Luschen18, Lukin19,Rispoli19}, ions \cite{Smith16} or superconducting qubits \cite{Roushan17,Guo20,Guo21,Morong21}. Early studies interpreted this behavior as a stable MBL \textit{phase}~\cite{Nandkishore15,Alet18, Abanin19}. In such a dynamical phase of matter, the information about details of the initial state of the system is kept indefinitely in the values of local integrals of motion (\textbf{LIOMs})~\cite{Huse14,Ros15,Serbyn13b, Imbrie16, Wahl17, Mierzejewski18, Thomson18}. At the same time, the transport is suppressed \cite{Znidaric16}, and the entanglement spreads slowly \cite{Bardarson12,serbyn2013universal,iemini2016signatures}. However, it has recently become apparent that the numerical evidence about the dramatic slow down of the dynamics with increasing disorder strength~\cite{Doggen18, Chanda19, Doggen19, Chanda20m}, and the related properties of highly excited eigenstates~\cite{Panda20,Sierant20thouless, Sierant20polfed,Kiefer20, Kiefer21, Abanin21,Sels20,Krajewski22}, do not necessarily imply an existence of a stable MBL phase and could be rather interpreted in terms of an MBL \textit{regime}~\cite{Suntajs20e, Sierant22challenges, Morningstar22}, in which the thermalization occurs only at enormous time scales and only for sufficiently large system size~\cite{Sierant21constrained, Deroeck23}.

The question of the existence of a stable MBL phase is actively pursued, and the answer may be model-dependent as suggested by recent numerical studies~\cite{Sierant21constrained, Sierant22f, Krajewski23}. One approach to tackle this question, fundamental for our understanding of non-equilibrium many-body physics, is to investigate a mechanism that underlies the slow thermalization of a strongly disordered many-body system. Spatially uncorrelated disorder, assumed in the standard models of MBL, necessarily leads to the emergence of Griffiths regions, where the disorder is locally weaker. The Griffiths regions play an essential role in the physics of quantum phase transitions~\cite{Vojta10} and affect the slow dynamics of disordered many-body systems~\cite{Gopalakrishnan16,Agarwal15, Agarwal17, Pancotti18}. The significance of Griffiths regions was further highlighted when an \textit{avalanche} mechanism for MBL transition was proposed~\cite{DeRoeck17}. The avalanche mechanism assumes that the Griffiths regions of anomalously small disorder become \textit{ergodic bubbles}, and describes, under certain simplifying assumptions, how the ergodic bubbles grow by thermalizing their surroundings. The simplified model of avalanche spreading is itself an interesting model of ergodicity breaking~\cite{Luitz17, Suntajs22sun, Suntajs23, Deroeck23, pawlik2023manybody}, and constitutes the basis of various phenomenological approaches to MBL~\cite{Thiery18,Goremykina19,Dumitrescu19,Morningstar19}. Nevertheless, understanding the relevance of the avalanche mechanism for the dynamics of disordered many-body systems~\cite{Potirniche19, Herviou19, Szoldra21, Hemery22, Foo23, Colmenarez24} and how to model the spread of avalanches~\cite{Morningstar22, Sels22bath, Ha23, Tu23avalanche} remain the major challenges in the field of MBL.

In this work, to study the fate of such ergodic bubbles in a controlled setting, we follow the approach of~\cite{Leonard23, Peacock23, Tu23} and divide the lattice into subsystems $A$ and $B$ with different disorder strengths $W_A$ and $W_B<W_A$. The subsystem $A$ plays the role of an MBL system, while the subsystem $B$ is the ergodic bubble, see Fig.~\ref{fig:setup}a. By monitoring the growth of correlations between the bath $B$ and the MBL system $A$, we investigate how the bath influences the dynamics of the subsystem $A$, and whether it induces its thermalization. Quantitative analysis of the correlations between the subsystems allows us to pin-point the onset of the propagation of avalanches that destabilize the MBL system. We distinguish three different types of behavior of the system, presented in Fig.~\ref{fig:setup}b. If the evolution starts from a generic initial state, the system thermalizes in the ergodic and critical regime and the quantum avalanches spread faster than $\propto \ln(t)$. For the same initial state, but in the MBL regime, we still observe a non-trivial increase of the thermalization distance $\xi_d(t)\propto \ln (t)$. The latter behavior is an artifact of the initial state. Considering a product state of a highly excited eigenstate of subsystem $A$ and arbitrary state in subsystem $B$ as the initial state, we demonstrate that in the MBL regime the growth of $\xi_d(t)$ terminates at a time set by the bath size signifying the halt of spreading of avalanches.

The rest of this paper is structured as follows. In Sec.~\ref{sec:ergVSmbl}, we outline a general picture of the possible scenarios for interaction of an ergodic bath with an MBL system. Subsequently, in Sec.~\ref{sec:ham}, we present the disordered spin-1/2 XXZ model and describe the correlation functions that allow us to investigate the propagation of avalanches, followed by numerically extracted signatures of avalanche propagation in our setup. In Sec.~\ref{sec:num}, we discuss a situation when the Hamiltonian of the subsystem $B$ is replaced by a GOE random matrix, study the sensitivity to the disorder strength in the bath, and discuss the influence of the subsystem $A$ on the dynamics of the bath $B$. Finally, in Sec.~\ref{sec:comp}, we show that periodic turning on and off of the coupling between subsystems $A$ and $B$ leads to much stronger tendency towards delocalization in comparison to the case of the time independent coupling.

\section{Ergodic bubbles in MBL system}
\label{sec:ergVSmbl}

Typically, MBL is studied in systems with uncorrelated on-site disorder, with values drawn from a uniform distribution on an interval $[-W,W]$. The probability that a stretch of $\ell$ consecutive sites, denoted as subsystem $B$, experiences a weaker disorder $W_B<W$ is exponentially small in $\ell$ and equal to $p_\ell = (W_B/W)^{\ell}$. However, the probability of finding an ergodic bubble, i.e., a thermal subsystem with weak disorder strength $W_B$ of any finite size $\ell$, tends to unity with increasing system size $L$. Such a subsystem $B$ plays a role of a \textit{thermal bath} for the neighboring spins. The fate of this bath, and its influence on the dynamics of the surrounding spins are of vital importance for the stability of MBL in the system. 

We consider a 1D disordered spin chain described by spin-1/2 operators $\vec{S}_i=(\hat X_i, \hat Y_i, \hat Z_i)$ on sites $i=1,\ldots,L$, and model the situation of interest by the following Hamiltonian
\begin{equation}
    H = H_A + H_B + H_{AB}
\end{equation}
where $H_A$ describes a generic MBL subsystem $A$, $H_B$ is the Hamiltonian of the ergodic bubble, identified here as the subsystem $B$ and $H_{AB}$ is the coupling between the subsystems $A$ and $B$.

The density of states of the thermal subsystem is, at the leading order, given by $\rho_B \sim 2^{L_B}$. Since the subsystem $B$ is ergodic, the matrix elements of a local operator $O_i$ localized around site $i$ in $B$, in the eigenbasis of $H_B$, $H_B\ket{\psi_n} = E_n\ket{\psi_n}$, are given by the 
ETH ansatz \cite{Srednicki94, Deutsch91, Srednicki99, Alessio16}
\begin{equation}
\label{eq:ETH}
    \bra{\psi_n}O_i\ket{\psi_m} = O(\bar{E})\delta_{nm} + 
    \rho_B^{-\frac{1}{2} } f_O(\bar{E}, \omega) R_{mn}, 
\end{equation}
where $\bar{E} = (E_m + E_n) / 2$, $\omega = E_n - E_m$ is the energy difference, 
$R_{mn}$ is a random variable of zero mean and unit variance, and the factor $\rho_B^{-\frac{1}{2} }$ encodes the thermodynamic entropy for the eigenstates in the middle of the spectrum of $H_B$~\cite{Burke23}. Importantly, $O(\bar{E})$ and $f_O(\bar{E}, \omega)$ are both smooth functions of their arguments.

The MBL Hamiltonian $H_A$ can be expressed in terms of the set of LIOM operators, $\lbrace \tau^z_i\rbrace$, $i=1,\dots L_A$ \cite{Abanin19}. Each of the LIOMs can be expanded in the microscopic spin operators $(\hat X_j, \hat Y_j, \hat Z_j)$ as
\begin{equation}
 \label{Eq:OPE}
\hat \tau^z_j=\zeta \,\hat Z_j+ \sum_{n=1}^\infty a^{(n)}_{j}{\hat A^{(n)}}_j,
\end{equation}
where $\zeta$ is the overlap of $\hat \tau^z_j$ with $Z_j$, and $\hat A^{(n)}_j$ is an operator acting non-trivially only at sites $ j-n,\ldots,j,\ldots j+n$. The coefficients of longer-range operators decay as ${a^{(n)}_j\sim e^{-n/\xi}}$~\cite{Ros15}. This decay defines the characteristic lengthscale $\xi$ and is a crucial property that determines the phenomenology of the MBL systems by ensuring the quasi-locality of LIOMs.

The bath-MBL coupling $H_{AB}$ involves microscopic spin operators supported at the interface between the subsystems $A$ and $B$. Due to the exponential decay of LIOMs~\eqref{Eq:OPE}, this coupling can rewritten as $H_{AB} = \sum_{j=1}^{L_A} H_{AB}^{(j)}$ with
\begin{equation}
    H_{AB}^{(j)} =  \left( V e^{-r/\xi }O_i \tau^{-}_j + \text{h.c.} \right),
    \label{eq:int}
\end{equation}
where $O_i$ acts on the degrees of freedom of the ergodic subsystem $B$, and $r=|i-j|$ is the distance between the LIOM and the bath. To write~\eqref{eq:int}, we have neglected all sub-leading terms involving higher order products of LIOM operators and we have kept only the local operator $O_i$. These assumptions are sufficient for the qualitative analysis of the leading terms dictating the influence of the bath on the MBL system. 

Let us assume that at time $t=0$ we prepare the system in the state 

\begin{equation}
    \ket{\phi_{\mathrm{in}}} = \ket{\tau_1 \dots  \tau_j \dots \tau_{L_A}} \otimes \ket{\psi_n}
    \label{phi0}
\end{equation} 
with a fixed configuration of LIOMs, $\tau_j=\uparrow,\downarrow$, and with the ergodic bath initialized in the eigenstate $\ket{\psi_n}$ with energy $E_n$\footnote{The eigenstate of the bath $\ket{\psi_n}$ is chosen as the initial state for simplicity. Due to the ergodicity of the bath, product states in real space become superpositions of eigenstates $\ket{\psi_n}$ with structureless coefficients, leading to qualitatively the same behavior. In contrast, the assumption of an initial state being an eigenstate in the subsystem $A$ is of crucial importance, see Sec.~\ref{signatures}.}.

The initial state of the MBL subsystem $A$ may slowly thermalize due to the coupling $H_{AB}$ to the ergodic bath $B$. In the following, we analyze two extreme scenarios for time evolution of the system. First, we assume that the ergodic bath remains unaltered during the time evolution, and employ the first order of perturbation theory to estimate the decay rate of the LIOM configuration~\cite{Tannoudji02quantum}. We contrast this perturbative scenario with the non-perturbative avalanche mechanism~\cite{DeRoeck17} which assumes that the thermalized spins become members of the ergodic bubble, increasing the potential of the bath to thermalize the subsequent degrees of freedom.

\subsubsection{Perturbative coupling to a static bath}

Let us estimate the time scale for thermalization of the LIOM $\hat{\tau}_j$ at site $j$. To that end, from the bath-MBL interaction we keep only the coupling term $H_{AB}^{(j)}$, and note that, in  the first order perturbation theory, the initial state $\ket{\phi_{\mathrm{in} }}$~\eqref{phi0} is coupled to 
\begin{equation}
    \ket{\phi_\mathrm{fin} } = \ket{\tau_1 \dots  \overline{\tau}_j \dots \tau_{L_A}} \otimes \ket{\psi_m},
    \label{phi1}
\end{equation} 
where $\overline{\tau}_j$ is the opposite spin to $\tau_j$ and $\ket{\psi_m}$ is an eigenstate of the bath with $m \neq n$. The probability that the LIOM $\hat{\tau}_j$ remains in its initial eigenstate $\tau_j$ falls down exponentially in time with a decay rate given by the Fermi golden rule~\cite{Tannoudji02quantum},
\begin{equation}
     \Gamma_j \sim |\mathcal{T}_j|^2 \rho_{B},
     \label{eq:FGR1}
\end{equation} 
where the matrix element $\mathcal{T}_j = \bra{\phi_\mathrm{fin}} H_{AB}^{(j)} \ket{ \phi_{\mathrm{in}} }$ can be rewritten with the help of the ETH ansatz \eqref{eq:ETH}, giving rise to
\begin{equation}
     \Gamma_j \sim V^2 e^{-2 r/\xi},
\label{rate}
\end{equation}
where we have neglected the energy dependence of the spectral function $f_O$ and of the density of states. We note that the  rate $\Gamma_j$ is independent of the bath size $L_B$ since the density of states $\rho_{B}$ in \eqref{eq:FGR1} is canceled by the square of $\rho_B^{-\frac{1}{2} }$ from \eqref{eq:ETH}. Moreover, \eqref{rate} is valid only on timescales shorter than $\rho_B^{-1}$. At larger times, the discreteness of the spectrum of $H_B$ is resolved, and transitions from the state $\ket{\phi_{\mathrm{in}}}$ occur only in the rare cases of matching between the energies of $\ket{\phi_\mathrm{in} }$, $\ket{\phi_\mathrm{fin} }$.
Therefore, LIOM at site $j$ is thermalized at timescale $t_j \sim 1/\Gamma_j \propto e^{2 r/\xi}$ increasing exponentially with the distance $r$ from the bath, provided that $t_j < \rho_B^{-1}$. In other words, the distance $\xi_d$ of penetration of the ergodic bath $B$ into the MBL subsystem $A$ will grow in time as $\xi_d \propto \ln t$, and then will saturate at a time proportional to $2^{L_B}$. 
Therefore, the MBL in the subsystem $A$ remains stable in this scenario. Indeed, the difference of energies of the states $\ket{\phi_\mathrm{in} }$, $\ket{\phi_\mathrm{fin} }$ is, at minimum (in a typical case), proportional to $\rho_B^{-1}$. Hence, the first order perturbation theory correction to the state $\ket{\phi_\mathrm{in} }$ is of the order of
\begin{equation}
   h^{\mathrm{stat}}_r \sim \frac{ \mathcal{T}_{L_A - r} }{ \rho_{B}^{-1}  } \sim e^{(- 1/\xi)r},
    \label{eq:PER}
\end{equation}
i.e., it is vanishing exponentially with the  increasing distance $r$ from the bath.

\subsubsection{Avalanche mechanism}
\label{sec:avalanche_mechanism}
The above description assumes that the events of thermalization of subsequent LIOMs are taking place independently, and that the thermalized LIOMs do not influence the dynamics of the system at the later stages of the process. In the following, we analyze the consequences of an opposite assumption.

Let us now consider thermalization of the LIOM $\hat \tau_{L_A}$ closest to the bath. Equation~\eqref{rate} implies that the probability remaining in the eigenstate  $\tau_{L_A}$ of the LIOM is significantly below unity after time $t_{L_A} \propto  e^{2/\xi}$. According to the avalanche mechanism~\cite{DeRoeck17}, once the state of the LIOM $\hat \tau_{L_A}$ has relaxed, the spin at site $j=L_A$ becomes a member of the ergodic bath, and, consequently the density of states the new bath $B'$ is increased by a factor of $2$, $\rho_{B'} = 2\rho_{B}$.
After $r$ iterations of such a process, the density of the states of the enlarged bath is $\rho_{B'} = 2^r\rho_{B}$.
We consider a state $\ket{\tau_1 \dots \tau_{L_A-r}} \otimes \ket{\phi_n}$, where $\ket{\phi_n} $ is an eigenstate of the enlarged bath comprised  of $L_B+r$ spins, and calculate the first order perturbation theory correction to this state due to the coupling $H_{AB}$, which leads us to
\begin{equation}
    h^{\mathrm{aval}}_r \sim \frac{ \mathcal{T}_{L_A - r} }{ (2^r\rho_{B})^{-1} } \sim e^{((\ln 2)/2 - 1/\xi)r}.
    \label{eq:AVA}
\end{equation}
In contrast to the perturbative expression for the static bath $B$, Eq.~\eqref{eq:PER}, the hybridization ratio $h^{\mathrm{aval}}_r$ is exponentially enhanced by a factor $e^{(r\ln 2)/2}$, associated with the growth of $\rho_{B'}$ upon incorporation of the neighboring spins. As a consequence, thermalization of the MBL subsystem may be a self-sustaining process if the growth of the bath density of states overcomes the decrease of the interaction strength with the distance $r$, i.e., when $(\ln 2)/2 - 1/\xi > 0$. When this condition is met, an avalanche propagates throughout the system, and the system thermalizes. According to the Fermi golden rule rate \eqref{rate}, the avalanche propagation may still be exponentially slow, i.e. the distance $\xi_d$ of penetration of the MBL system may grow in time $\xi_d \propto \ln t$. However, the density of states of the enlarged bath increases after each act of thermalizing and incorporation of the neighboring spin, hence, \eqref{rate} is valid up to the time scale $\rho_{B'}^{-1}$, which continually increases when the avalanche propagates. The discussed scenario predicts an unbounded logarithmic growth of the thermalization length $\xi_d \propto \ln t$. Higher order perturbative corrections, as well as non-perturbative mechanisms associated with the growth of the thermal inclusion, beyond the simplified processes discussed above, may speed up the spreading of avalanche when $(\ln 2)/2 - 1/\xi > 0$. Thus, we \textit{define} that avalanches spread in the investigated disordered many-body system whenever the presence of a thermal inclusion induces logarithmic, or any faster, growth of $\xi_d$ in time. This definition does not agree with that of Ref.~\cite{Leonard23} in which $\xi_d \propto \ln t$ is classified as a lack of an avalanche. Disagreement comes from the fact that the expression for the decay rate $\Gamma_j$, Eq.~\eqref{rate} is valid only at times $t < \rho_B^{-1}$, which, in the absence of the avalanche spreading, is bounded by the constant $2^{L_B}$.

\section{Disordered XXZ spin chain with engineered ergodic bubble}
\label{sec:ham}

The simple scenarios for the time evolution of an MBL system in contact with an ergodic bath described in Sec.~\ref{sec:ergVSmbl} rely on a number of assumptions which are not straightforwardly testable. For that reason, the direct relevance of the avalanche mechanism for the dynamics of disordered systems remains not clear. The first assumption regards the form of LIOM operators, see Eq.~\eqref{Eq:OPE}. While there are numerical approaches that allow to construct LIOMs for strongly disordered systems~\cite{Chandran15, Mierzejewski18,Rademaker16, Ortuno19, Pekker17, Kelly20, Thomson21flow, Thomson23,Brien16, Peng19, Johns19, Adami22}, probing of the properties of LIOMs at large time scales, and especially in the vicinity of the crossover to ETH phase remains a challenge, and one has to resort to perturbative schemes which are not rigorously controlled~\cite{Ros15}. 
Moreover, the avalanche mechanism assumes that each thermalized spin joins the ergodic bath which still can be described by the ETH ansatz~\eqref{eq:ETH}. A strict validity of this assumption is unclear~\cite{Potirniche19}. Finally, ergodic bubbles in strongly disordered systems form due to rare disorder configurations, and hence, their dynamics can be probed only indirectly, with specific tools~\cite{Szoldra22}. 

The above reasons motivate us to analyze numerically the process of avalanche propagation in a system in which we assure the presence of the ergodic bubble by dividing the lattice into two subsystems as shown in Fig.~\ref{fig:setup}a.

\subsection{The system}

We consider the XXZ-spin model with random magnetic field, widely studied in the context of MBL~\cite{Berkelbach10, Luitz15, Agarwal15, Bera15, Enss17, Bera17, Herviou19,Serbyn16,Bertrand16,Santos04a,Santos04, Colmenarez19, Sierant19b, Sierant20model, Schiulaz20, TorresHerrera20,Szoldra23,Mace18,Laflorencie20,Suntajs20, Chanda19,Doggen18, Gray18,Sierant20polfed,Suntajs20e}, coupled to a bath modelled also by XXZ-spin system. The Hamiltonian reads:
\begin{eqnarray}
    H &=& H_A + H_B + H_{AB}, \\\nonumber
    H_A &=& J \sum_{i=1}^{L_A-1} X_i X_{i+1} + Y_i Y_{i+1}  + \Delta_A Z_i Z_{i+1}\\\nonumber 
    &&+ \sum_{i=1}^{L_A} W_i Z_i,\\\nonumber
    H_B &=& J \sum_{i=L_A+1}^{L-1} X_i X_{i+1} + Y_i Y_{i+1} + \Delta_B Z_i Z_{i+1} \\\nonumber 
    &&+ \sum_{i=L_A+1}^{L} W_i Z_i,\\\nonumber
    H_{AB} &=& J \left(X_i X_{i+1} + Y_i Y_{i+1} + \Delta_{AB} Z_i Z_{i+1}\right),\nonumber
    \label{eq:H}
\end{eqnarray}
where $\vec{S}_i=(X_i, Y_i, Z_i)$ are spin-1/2 operators at lattice site $i$, $J=1$ fixes the energy unit (and $J^{-1}$ is the characteristic time scale which we call a tunneling time having in mind the equivalent spinless fermions expression of the model  which arises upon the Jordan-Wigner transformation). Open boundary conditions are assumed. Unless stated otherwise, we set the interaction strengths $\Delta_A=\Delta_B=\Delta_{AB}=1$. The chain consists of two subsystems of sizes $L_A$ and $L_B$, with the total system size $L=L_A+L_B$. The on-site magnetic fields, $W_i$, are independent random variables drawn from the uniform distribution in the interval that differs between the two subsystems:
\begin{equation}
    W_i \in \begin{cases}
  [-W_A, W_A]  & \text{ if }1\leq i \leq L_A, \\
  [-W_B, W_B] & \text{ otherwise},
\end{cases}
\end{equation}
see Fig.~\ref{fig:setup}a. The subsystem $B$ serves as a thermal bath that seeds the avalanche and initializes the process of thermalization of the subsystem $A$, and, consequently, of the whole chain. For that reason, we fix the disorder in the subsystem $B$ to be weak. If not stated otherwise, we put $W_B=0.5$. At this disorder strength the Heisenberg spin chain thermalizes quickly, since $W_B=0.5$ is well below the various estimates for critical disorder strength for MBL transition \cite{Luitz15, Devakul15, Gray18, Mace19, Sierant20polfed}. We work in the ${Z=0}$ total magnetization sector. Time evolution is computed exactly using the Chebyshev polynomial expansion of the time evolution operator \cite{TalEzer84, Fehske08}.

Observables describing the system (except for imbalance from Fig.~\ref{fig:kicked_vs_notkicked}) are calculated as a median over $1000$ random disorder realizations unless specified explicitly. Using a robust estimator, such as median, instead of a usual arithmetic average, ensures convergence of the measured quantities in the number of disorder realizations. Median captures a typical behavior of the system and is not sensitive to few accidental disorder realizations, particularly affecting the mean value at the largest disorder strengths, deep in the MBL regime. For a comparison between results obtained using the mean and the median, see Appendix~\ref{app:median}.

\subsection{The avalanche spreading}

In order to describe the process of the avalanche spreading and of the ensuing thermalization of the subsystem $A$ quantitatively, we fix the initial state as the N\'{e}el state and calculate the two-point connected correlation function $\expval{Z_i Z_j}_c=\expval{Z_i Z_j} - \expval{Z_i}\expval{Z_j}$ for sites $i$ and $j$. 

\begin{figure*}
    \centering
    \includegraphics[width=.9\textwidth]{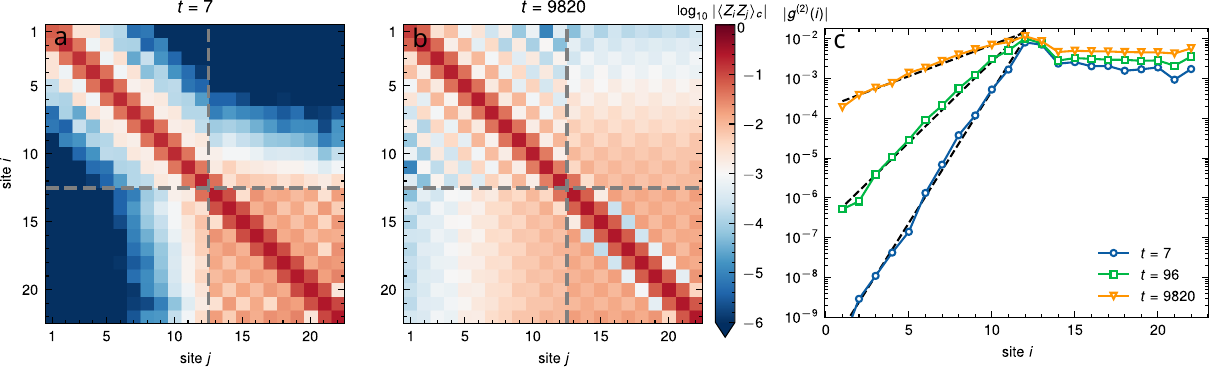}
    \caption{Model parameters are $L_A=12$, $L_B=10$, $W_A=4$, and $W_B=0.5$. \textbf{a} Initially, correlations within subsystem $A$ (first 12 sites) are short-ranged, while correlations within the bath (sites 13-22) spread over the whole bath. \textbf{b} After a sufficiently long time, correlations between the bath and the MBL subsystem are developed and the system thermalizes. Checkerboard pattern appears due to a specific initial Neel state. \textbf{c} Average (over bath) of two-body correlations for each lattice site. Correlation decay length $\xi_d$ is obtained by fitting an exponential decay with distance, Eq.~\eqref{eq:xi_d_definition} (dashed lines).}
    \label{fig:corr_mat}
\end{figure*}

Typical behavior of the $\expval{Z_i Z_j}_c$ correlation function for $W_A=4$ and $W_B=0.5$ is presented in Fig.~\ref{fig:corr_mat}. Initially, at time scale of one tunneling time only the short-range correlations are present throughout the system among the neighboring sites $|i-j| \approx 1$ (plot not shown). After several tunneling times, $t \approx 7$, the sites within the bath $B$ develop mutual correlations, while the correlations in the subsytem $A$ remain short-ranged, see Fig.\ref{fig:corr_mat}a. Subsystems $A$ and $B$ begin to develop non-trivial correlations due to their connection via a bond between the sites $L_A$, $L_A+1$. During the course of time evolution of the system, the correlations gradually increase, and, at a large time $t=9820$, we observe that the whole subsystem $A$ develops significant correlations with the bath $B$. The correlation between a given site $i_A$ from $A$ is, to a good approximation, the same with all of the sites $i_B$ from the bath $B$. This occurs because the spins in the bath are strongly correlated with each other and as such can no longer be treated individually. In contrast, the strength of the correlation between the spins from subsystem $A$ with spins of the bath $B$ decreases with the distance to the edge of the bath.

Our aim is to find out if the correlations between the bath $B$ and the subsystem $A$ are a non-perturbative effect of the propagation of a quantum avalanche, or are only a perturbative result of the bath proximity, as described in Sec.~\ref{sec:ergVSmbl}. To quantify how the correlations with the bath penetrate through the subsystem $A$, following Ref.~\cite{Leonard23}, we define a bath-averaged correlation function for each site, 
\begin{equation}
    \gtwo = \overline{\expval{Z_i Z_j}_c}|_{j\in L_B}.
\end{equation}
An example of the $\gtwo$ function is shown in Fig.~\ref{fig:corr_mat}c. Clearly, the correlations with the bath in the subsystem $A$ fall off exponentially with the distance from the interface between $A$ and $B$. This leads us to the definition of the correlation decay length $\xi_d$ which is obtained by fitting 
\begin{equation}
    |\gtwo| = c \exp \left( -\frac{L_A-i}{\xi_d} \right)
\label{eq:xi_d_definition}
\end{equation} 
at a given instant of time. We note that the employed definition allows us to accurately determine the decay length $\xi_d$ without introducing unwanted artifacts of the approach employed in Ref.~\cite{Leonard23}. In particular, this method is less sensitive to finite-size effects and the discreteness of the lattice, see Appendix \ref{app:xi} for more details. In the following, we will probe the thermalization of the subsystem $A$ mainly focusing on the time dependence of the decay length $\xi_d$.

\subsection{Signatures of avalanches due to the bath}
\label{signatures}
Exact diagonalization studies of the disordered XXZ model \cite{Sierant20polfed} suggest the presence of three distinct regimes of the system characteristics, depending on the disorder strength $W$. For the lowest disorder values $W<W^T(L)$ the system follows ETH, its energy level statistics follow GOE predictions and the system thermalizes in a generic way. In contrast, for disorder larger than a certain value $W>W^*(L)$, the spectrum shows signs of emergent integrability (MBL). If 
an MBL transition occurs
in the system, then $\lim_{L\to\infty}W^T(L)=\lim_{L\to\infty}W^*(L)=W_c$, where $W_c$ is the critical disorder strength. 
In that case, for finite systems, there is a finite transient critical regime $W^T(L) < W < W^*(L)$ with multifractal entanglement structure in the eigenstates \cite{Herviou19}. This might suggest that, in these three regimes, the system will react differently to the attached finite thermal bath. Our expectation is that in the ergodic and critical regimes, any bath will lead to a growth of $\xi_d(t)$ faster than $\propto \ln(t)$, consistent with an avalanche scenario. In the MBL case, the growth $\propto \ln(t)$ should saturate around time $t\sim 2^{L_B}$ when the subsystem $A$ starts to resolve the discreteness of the bath's energy spectrum.

\begin{figure*}
    \includegraphics[width=\textwidth]{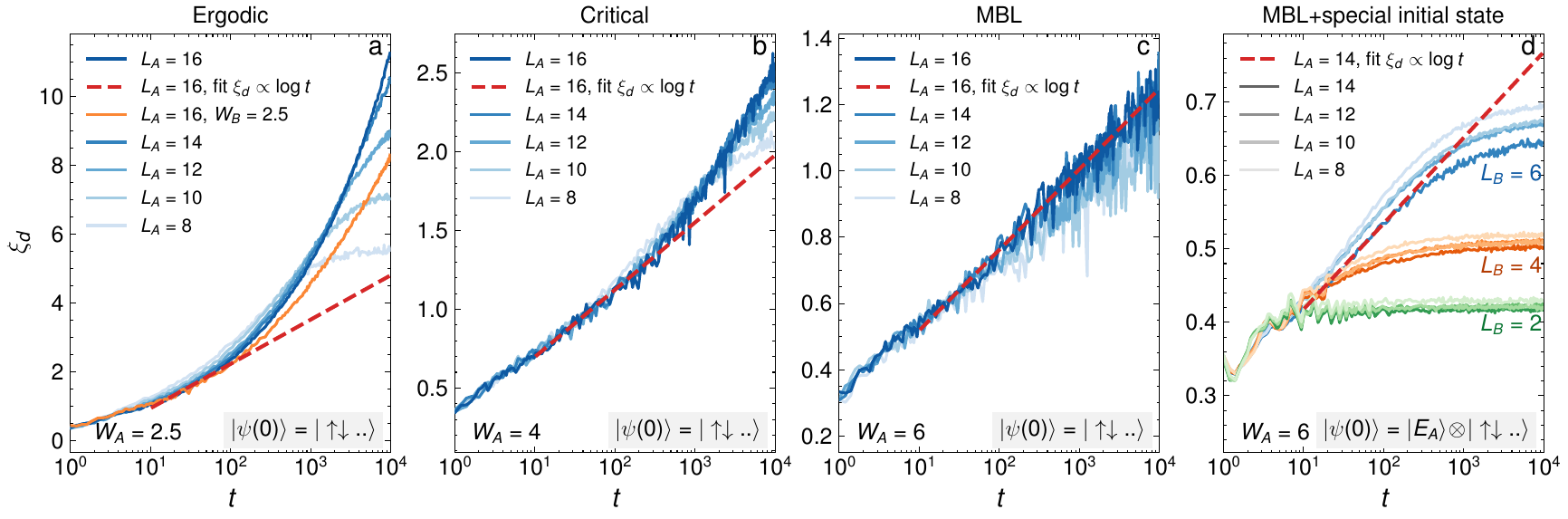}
    \caption{Growth of the correlation decay length $\xi_d(t)$ in the three regimes of disorder $W_A$, different subsystem $A$ sizes $L_A$, bath size $L_B=6$ and initial N\'{e}el state (panels abc) or $L_B=2,4,6$ and a special initial state (panel d). Thermal avalanches are observed in the ergodic and critical regimes (panels ab) as a growth faster than $\xi_d\propto\ln(t)$. Even though there are no quantum avalanches in the MBL regime (panel c), the growth of $\xi_d$ continues as $\propto \ln(t)$ much beyond the timescale $t\sim 2^{L_B}$. To see termination of avalanches in the MBL regime, one can start with a fine-tuned initial state, a product state of the subsystem $A$ eigenstate energy $E_A$ (we choose $E_A$  {closest to $0$}), 
    and a N\'{e}el state in subsystem $B$ (panel d). For reference, linear fits $\xi_d \propto \ln(t)$ are performed between times $t=10-100$ (abd) or $t=10-10000$ (c) and denoted by dashed red lines. In panel d, $5\cdot10^4$ disorder realizations were used. For panel d we employ a special fitting method described in Appendix \ref{app:xi}.}
    
    \label{fig:xi_d_growth_4panels}
\end{figure*}

The results of our calculations of the correlation decay length $\xi_d(t)$ in all three regimes with a bath of size $L_B=6$ and the N\'{e}el initial state are presented in Fig.~\ref{fig:xi_d_growth_4panels}abc. Figure \ref{fig:xi_d_growth_4panels}a shows the system behavior in the ergodic regime with $W_A=2.5$. The avalanche is clearly spreading through the system, as suggested by a positive curvature of the decay length $\xi_d(t)$, cf. linear fit between times $t=10-100$, implying faster-than-logarithmic growth with time. It suggests that some assumptions of the perturbation theory introduced in Sec.~\ref{sec:ergVSmbl} are not valid.
In this case, the subsystem $A$ cannot described by a set of LIOMs. For the smallest system sizes, the positive curvature is also visible, but the growth of $\xi_d(t)$ saturates before reaching $t=10^4$. It is a finite-size effect occuring when $\xi_d$ becomes of the order of $L_A$. Increasing $L_A$ systematically increases accessible $\xi_d$ before saturation. 
We also notice that, in fact, no thermal bath is needed to see features compatible with avalanches - the curve for $W_B=W_A=2.5$ and $L_A=16$ behaves qualitatively the same as the case with a bath $W_B=0.5$. This means that the system can serve as a bath for itself, in agreement with the usual understanding of the thermalization in quantum many-body systems.

In Fig.~\ref{fig:xi_d_growth_4panels}b we switch to the critical regime $W_A=4$. The growth of $\xi_d(t)$ is slower than in the ergodic case, yet it is still faster than $\xi_d\sim\ln(t)$, suggesting the spread of avalanche. With increasing subsystem size $L_A$, the correlation decay length $\xi_d$ slightly increases, ie., thermalization is enhanced. This is compatible with exact diagonalization studies \cite{Sierant20polfed} which show that increase of the system size increases the disorder strength at which the ETH-MBL crossover is observed.

Figure~\ref{fig:xi_d_growth_4panels}c shows the apparent instability of the strongly disordered MBL system ($W_A=6$) in contact with a thermalizing bath when starting from an initial N\'{e}el state. We observe that $\xi_d$ grows as $\propto \ln(t)$ for the whole simulation time $t=10^4$. Recall that according to the perturbative arguments presented in Sec.~\ref{sec:ergVSmbl}, if the logarithmic growth of $\xi_d$ continues beyond $t\approx 2^{L_B}$, one deals with a continually increasing timescale of thermalization by the bath (see Fermi's golden rule in Eq.~\eqref{rate}). This suggests spreading of an avalanche which thermalizes the system and may be surprising as other signatures of MBL, such as energy level statistics, entanglement entropy, Thouless time scaling \cite{Sierant20polfed} suggest a deeply localized regime for this value of disorder in the absence of the bath. On the other hand, a non-trivial long-time dynamics is also observed for the bipartite von Neumann entanglement entropy which grows as $\propto \ln t$ \cite{Sierant22challenges} even in MBL regime. Other works \cite{Morningstar22} predict a much larger value of disorder $W>18$ needed for the MBL phase. Looking for the explanation of this counterintuitive behavior of the MBL system, we have checked that increasing the disorder strength to $W_A=20$ does not remove the logarithmic growth until the latest accessible times $t=10^4$. Removing the interactions in subsystem $A$ and going to Anderson localized phase by setting $\Delta_A=0$ in subsystem $A$ also leads to a similar long-time logarithmic growth, see details in Appendix \ref{app:AL}. 

It turns out that, in order to observe the termination of a thermal avalanche, as predicted by the perturbation theory for sufficiently small LIOM localization lengths, one has to necessarily fulfill one of its assumption: the initial state must be an eigenstate of the subsystem $A$. Similar setup was considered in Ref.~\cite{Tu23} with the purpose of analyzing energy-resolved features of the spectrum and in Ref.~\cite{Pulikkottil23} for perturbatively coupled chaotic systems. One can still start with the N\'{e}el state in subsystem $B$ as its eigenstates are featureless, and so is their superposition. Results for this initial condition are presented in Fig.~\ref{fig:xi_d_growth_4panels}d. Clearly, the penetration of the thermal bath into the MBL subsystem is logarithmic with time and then stops around time on the order of $2^{L_B}$, in full agreement with the perturbative argument. Behavior of $\xi_d(t)$ is qualitatively the same for many subsystem sizes $L_A$ and fixed bath $L_B$. 

Summarizing, the long-time growth of $\xi_d(t)$ for $W_A=6$ in Fig.~\ref{fig:xi_d_growth_4panels}c seems to be a feature of the MBL regime rather than a signature of an avalanche spreading and thermalization of the system. When the initial state is an eigenstate of the subsystem $A$, we do not observe any signatures of avalanche spreading for the same system parameters, see Fig.~\ref{fig:xi_d_growth_4panels}d. The influence of the bath is fully captured with the first order perturbation theory and the highly non-perturbative effects of the avalanche mechanism are not observed at $W_A=6$. This dependence of the $\xi_d$ behavior on the initial condition is not occurring at smaller disorder strengths, when the system thermalization genuinely begins, for instance at $W_A=4$ (data not shown, see Appendix \ref{app:AL}). Therefore our analysis indicates that the access to the eigenstates of $H_A$ is required for observation of the stop of logarithmic growth of $\xi_d$ in time. Preparation of eigenstate of $H_A$ is experimentally a nontrivial task, increasing the difficulty of an experiment that could observe the halt of quantum avalanche spreading. Finally, we emphasize that we are able to observe the stopping of avalanches only at the considered (experimentally relevant) time scales and system sizes. We cannot exclude that the avalanche starts to spread and the system thermalizes at $W_A=6$ for larger $L_A$ or longer times.

\subsection{Extraction of the LIOM localization length from time evolution}
\label{sec:loclength}
Linear fits from Fig.~\ref{fig:xi_d_growth_4panels} allow us to extract the localization length of the LIOMs $\xi$, see Eq.~\eqref{Eq:OPE}, and compare it with a predicted threshold for avalanche, Eq.~\eqref{eq:AVA}, $\xi_{\text{aval}} = 2/\ln (2) \approx 2.9$. In both perturbative and nonperturbative pictures, the LIOM at distance $r$ decays around time $t_j \propto \exp(2r/\xi)$. In numerical data from Fig.~\ref{fig:xi_d_growth_4panels}, distance $r$ is measured by the correlation decay rate $\xi_d$. Thus, one can perform a linear fit
\begin{equation}
    \xi_d(t) = \frac{\xi}{2} \ln t + \text{const}
\end{equation}
and extract the LIOM localization length $\xi$. For ${W_A=2.5}$ we get $\xi=1.1 < \xi_{\text{aval}}$ between times $t=10-100$ which grows to $xi=5.4 > \xi_{\text{aval}}$ between times $t=1000-10000$ and is compatible with the presence of an avalanche in Fig.~\ref{fig:xi_d_growth_4panels}a. In the critical regime $W_A=4$, the obtained value $\xi=0.4$ is close to the threshold $\xi_{\text{aval}}$, as expected. In the localized regime $W_A=6$, the LIOM localization length extracted from the initial N\'{e}el state scenario is $\xi=0.21$ and is also well below the threshold for avalanche. A lower $\xi=0.10$ is obtained for the special initial state, signaling some additional effects happening in the N\'{e}el state case which enhance the non-triviality of the dynamics in that case.

\section{Bath features}
\label{sec:num}

This section is devoted to the features of the bath that might be relevant for the thermalization process. In Sec.~\ref{subsec:goe} we show that the bath can be approximated well by all-to-all interacting system modeled by a GOE matrix. In Sec.~\ref{subsec:Wb} we find that the system dynamics does not change qualitatively upon changing disorder amplitude $W_B$ in the bath (provided $W_B$ is small enough so the bath is ergodic by itself). Finally, in Sec.~ \ref{subsec:backreaction}, we study the evolution of correlations within the bath when connected to subsystem $A$ in its three different regimes, capturing the moment when the bath starts to grow. Additional analyses, such as change of the decay length for a fixed bath and changing subsystem $A$ size $L_A$, or changing disorder strength $W_A$, are shown in Appendix \ref{app:further}.

\subsection{Modeling the bath as a GOE matrix}
\label{subsec:goe}
Many approaches to thermalization in MBL through avalanche scenario assumed that the bath can be modeled by a random matrix from the Gaussian orthogonal ensemble (GOE) which should behave identically to a region thermalized due to accidentally low disorder values/differences \cite{DeRoeck17, Luitz17, Crowley20, Crowley22}.
The full Hamiltonian for such a setting reads
\begin{equation}
   H' = J \sum_{i=1}^{L_A} \vec{S}_i \cdot \vec{S}_{i+1} + \sum_{i=1}^{L_A} W_i Z_i + \alpha H_{B, \text{GOE}},
\label{eq:H_GOE}
\end{equation}
where $H_{\text{B, GOE}}$ is a GOE matrix defined for $L_B$ sites in subsystem B, and $\alpha$ is an energy rescaling factor we will specify later. Having access to the measures directly probing the avalanches, we verify this assumption numerically by comparing the two scenarios. Algorithm for numerical construction of the $H_{B, \text{GOE}}$ in the total $S_z$ magnetization conserving basis is given in Appendix~\ref{app:goe}. Before, we proceed to present the results, we explain a caveat that needs to be taken into account when comparing the two models.

\begin{figure}
    \centering
    \includegraphics[width=.9\columnwidth]{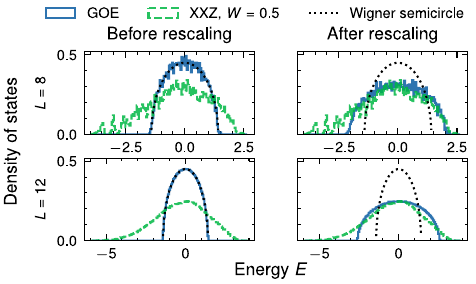}
\caption{Comparison of the density of states for the XXZ model with uniform $W=0.5$ disorder (green dashed line) and the GOE density of states (blue solid line) for the same system size $L$. Before rescaling, the GOE density of states reproduces the Wigner semicircle law (black dotted line). After rescaling the GOE density by a constant multiplicative factor, as described in the text, the two densities start overlapping in the major part of the spectrum. Histograms are averaged over $100$ disorder and GOE realizations.}
    \label{fig:dos_xxz_vs_goe_rescaling}
\end{figure}

The density of states of $H_{\text{B, GOE}}$ is given by the Wigner semicircle law \cite{Wigner55, Wigner58},
\begin{equation}
    \rho(E) = \frac{1}{\pi} \sqrt{2 - E^2}.
\end{equation}
Its spectrum spans an interval of energies ${-\sqrt{2}<E<\sqrt{2}}$. In order to reproduce bandwidth of the bath in the XXZ model, the GOE Hamiltonian is multiplied by a factor $\alpha=(p_{95} - p_5) / (2 \sqrt{2})$ where $p_{95}$ and $p_5$ are the $95$th and $5$th percentile of the disorder-averaged XXZ density of states at $W=0.5$. The result of rescaling energies is presented in Fig.~\ref{fig:dos_xxz_vs_goe_rescaling}. The rescaling gives a real symmetric random matrix whose middle spectrum energy gaps reproduce the middle spectrum gaps of the XXZ spin chain with disorder $W=0.5$ in the zero magnetization sector. Numerical values of the rescaling factors for different lengths of the chain $L=2-12$ and other "replicated" disorder strengths $W=0,0.5,1$ are listed in the Appendix \ref{app:goe}. 

\begin{figure}
    \centering
    \includegraphics[width=.8\columnwidth]{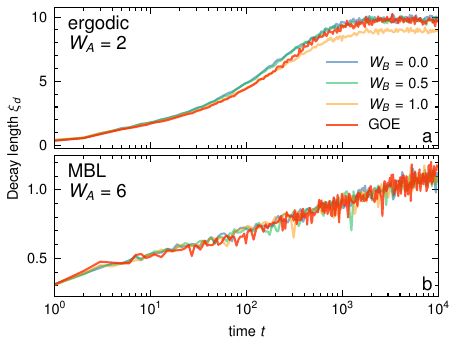}
\caption{
Decay length growth with time. ${L_A=8}$, ${L_B=6}$. Precise value of the bath disorder strength $W_B$ does not qualitatively change the behavior, irrespective if it is in the ergodic (panel \textbf{a}) or the MBL side of the crossover (panel \textbf{b}). Thus, our analysis is restricted to $W_B=0.5$ for the rest of the paper. We compare it with the bath modelled as a GOE matrix coupled to a single spin at the end of the chain. The GOE matrix is rescaled to replicate the density of states of the bath with $W_B=0.5$, see text.}
    \label{fig:bath_disorder_strength}
\end{figure}

Naively, one could expect that replacing the nearest-neighbor interacting low-disorder spin bath with an all-to-all interacting GOE bath will speed up the thermalization in the system. The GOE Hamiltonian scrambles quantum information very fast, on the timescale $\mathcal{O}(1)$ \cite{Sekino08, Cotler17}. In contrast, the XXZ model has Thouless time that scales as $\mathcal{O}(L^2)$ at best \cite{Sierant20thouless, Suntajs20e}. In spite of this, we do not observe any significant differences between these two cases when inspecting the correlation decay length $\xi_d(t)$ for subsystem $A$ in both ergodic ($W_A=2$) and MBL regime ($W_A=6$), see Fig.~\ref{fig:bath_disorder_strength}. To our understanding, it is not a surprise that the long-time $t \geq \mathcal{O}(L_B^2)$ behaviors are the same for the GOE bath and low-disorder XXZ model bath, as in both cases the baths have already thermalized and are in a similar featureless quantum state with random probability amplitudes dependent only on the initial energy. However, an unprecedented agreement between these scenarios also at initial stages of time evolution remains a puzzle. One possible explanation is that if the subsystem $A$ is connected with the bath only at its end by a local term in the Hamiltonian, thermalization of the distant site from the interface is mostly governed by the rate of propagation of correlations through subsystem $A$ up to the interface, which does not depend on the bath properties.

\subsection{Disorder strength $W_B$}
\label{subsec:Wb}
As mentioned in Sec.~\ref{subsec:goe}, the low-disorder bath can be well modelled by the GOE Hamiltonian. What is more, $\xi_d(t)$ calculated in Fig.~\ref{fig:bath_disorder_strength} does not change with the disorder strength of the bath $W_B$ as long as $W_B$ stays far on the ergodic side of the crossover, $W_B=0, 0.5, 1$. This is particularly important as we are trying to use the "engineered" bath to model "accidental" baths naturally present in the system due to accidentally small disorder values. Interestingly, even for the case of $W_B=0$ the thermalizing behavior is consistently the same. Integrability of the bath is broken in this case by the interaction with subsystem A.

\subsection{Correlations within the bath}
\label{subsec:backreaction}
We now concentrate on the reaction of the bath when it is interacting with the disordered XXZ chain. As it was suggested in earlier studies \cite{Potirniche19,Nandkishore15proximity}, the ergodic bath may be destroyed when in contact with an MBL system by the so-called "MBL proximity" effect. In an opposite scenario, a quantum avalanche may extend the bath into the MBL subsystem. Here we study the off-diagonal correlations of spins within the bath to characterize the bath during the time evolution. 

\begin{figure}
    \centering
    \includegraphics[width=.85\columnwidth]{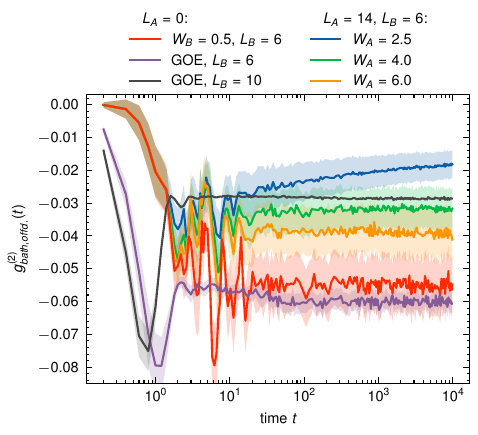}
    \caption{Mean offdiagonal correlations within the bath as function of time. Shades denote standard deviations of the median. Results for the disconnected bath ($L_A=0$) with weak disorder $W_B=0.5$ (red line) and the GOE bath of different sizes (purple and gray line) are compared with the $L_A=14$, $L_B=6$ case. Quantum avalanche is visible in the $W_A=2.5$ case since growth of offdiagonal correlations indicates an increase of the effective size of the bath. 
    }
    \label{fig:bath_stationary}
\end{figure}

Figure~\ref{fig:bath_stationary} shows the average (over the bath) of the offdiagonal correlations within the bath,
\begin{equation}
    g^{(2)}_{\text{bath, offd.}} = \frac{1}{L_B(L_B-1)}\sum_{\substack{i,j=L_A+1\\ i\neq j}}^L \langle Z_i Z_j \rangle_c.
\end{equation}
As a limiting case, we check a bath that is modeled by the GOE Hamiltonian from Sec.~\ref{subsec:goe} with $L_A=0$ (no subsystem A) and $L_B=6, 10$. Clearly, the correlations saturate fast at a negative value and remain stationary - all pairs of spins within the bath develop correlations between each other until no more correlations can be built further. The larger the system, the smaller the absolute value of the bath averaged correlations, since there are more possibilities to transition from a single site. A similar thermalizing behavior is observed for a weakly disordered spin bath ($W_B=0.5$), also for $L_A=0$, but reaching the equilibrium takes longer compared to the GOE case for the same size $L_B=6$, see discussion in Sec.~\ref{subsec:goe}. 

Let us check what happens when the subsystem $A$ is attached in Fig.~\ref{fig:bath_stationary}. We examine different disorder strengths $W_A=2.5,4,6$ for fixed $L_A=14$ and $L_B=6$. At $W_A=6$, we see clear signatures of saturation of bath correlations around time $t\approx10$. Similar behavior is observed in the critical regime $W_A=4$ but the saturation occurs at a larger value. Since the correlations saturate at a steady value, we may conclude that the bath does not change its size at least to the accuracy offered by this probe. Comparison to the baseline values for the free and GOE baths 
suggests an increase of the effective bath size for $W_A=4$ in comparison to the $W_A=6$ case. 
Comparing with GOE bath for $L_B=10$ leads us to the conclusion that the size of the effective bath for $W_A=4$ case cannot be larger than 
10 sites. Interestingly, a continuous dynamic drift in the offdiagonal correlations is observed for $W_A=2.5$. This is another sign of a quantum avalanche spreading, in addition to the correlation decay length presented in Fig.~\ref{fig:xi_d_growth_4panels}. Additional sites from subsystem $A$ join the bath by developing correlations with its constituents. The increase in the effective size of the bath weakens the correlations therein. 

All in all, the dynamics of the correlation functions within the bath allows to probe the thermalization of the full system from a different perspective, providing indirect information about the increase of the effective bath size.

\begin{figure*}
    \centering
    \includegraphics[width=2.\columnwidth]{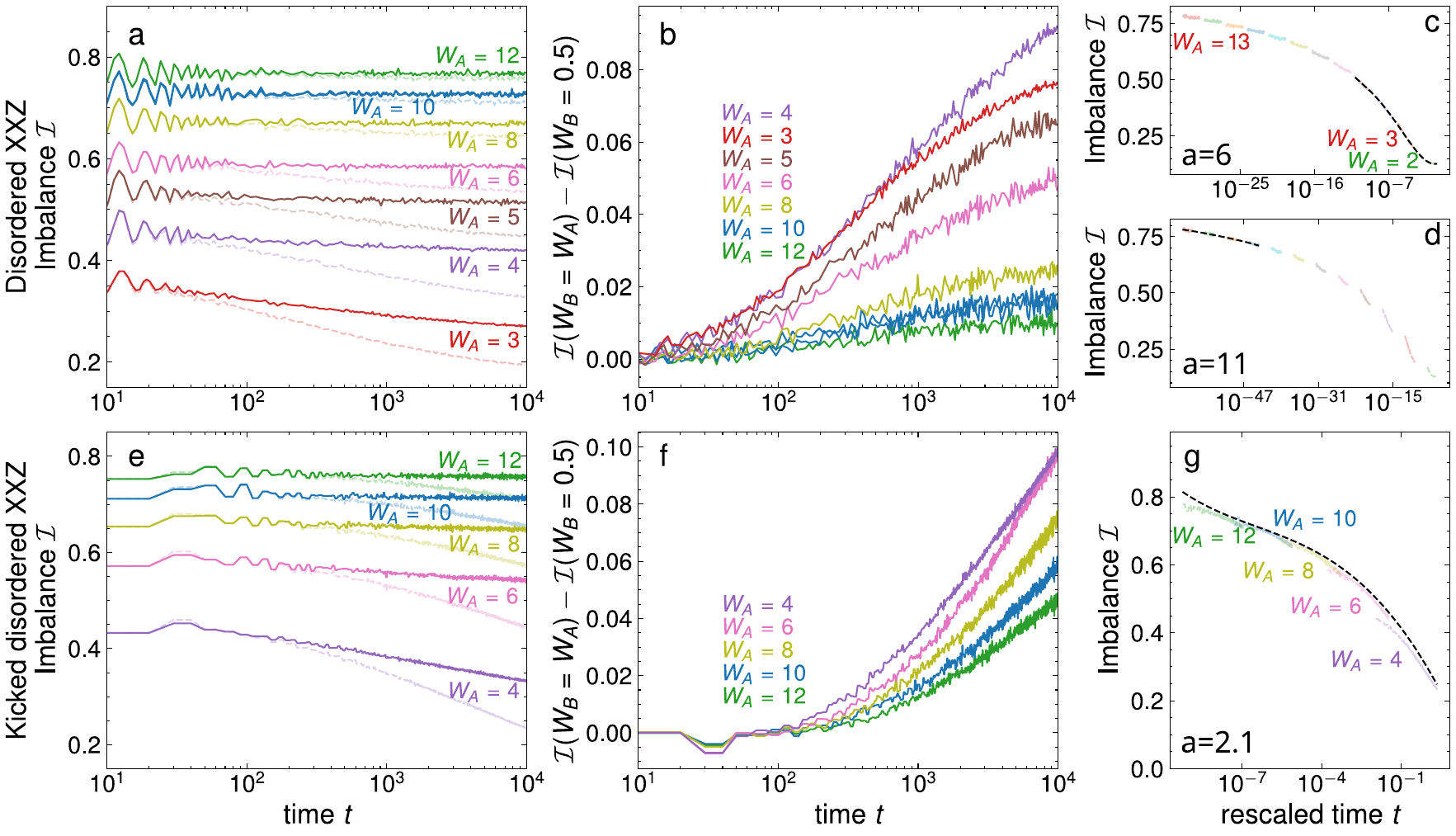}
    \caption{Comparison of imbalance on sites $3-10$ between kicked XXZ model by Peacock and Sels and the energy-conserving model for $L_A=12$, $L_B=10$. Solid lines: $W_A=W_B$, bright lines: $W_B=0.5$ (for panels acdeg). The speedup of imbalance decay happens due to the lack of energy conservation (e) because it is not observed in the time-independent case for the same parameters (a). Panels b, f: differences of imbalance with $W_B=W_A$ and $W_B=0.5$. Speedup of imbalance decay is visible for the kicked model, while saturation is observed for the time-independent model. Right column: Imbalance with rescaled time axis $t\rightarrow t e^{-aW_A}$ for the case $W_B=0.5$. Optimal constant $a$ is obtained by minimizing the error of a fourth degree polynomial fit (dashed black line) and a chosen range of disorder. While in the kicked case (panel g), there is a single universal constant $a\approx 2.1$ describing the curve for all $W_A\in [2,12]$, in the disordered XXZ model case the optimal constant $a$ drfits from $a\approx6$ for $W_A\in [2,5]$ to $a\in[8,15]$ for $W_A\in[10,13]$. See text and Appendix \ref{app:scaling} for more details on the fitting procedure.}
    \label{fig:kicked_vs_notkicked}
\end{figure*}

\section{Kicked XXZ model}
\label{sec:comp}
In their recent work, Peacock and Sels \cite{Peacock23} study a model similar to ours: a strongly disordered Heisenberg spin chain $A$ with a bath $B$ consisting of weakly disordered spins ($W_B=0.5$). 
However, in their model, the Hamiltonian is periodic in time.
The unitary evolution operator over one period of time reads
\begin{equation}
    U = e^{-i H_{AB} \tau} e^{-i (H_A + H_B)\tau},
    \label{eq:kick}
\end{equation}
where $H_{A, B, AB}$ are defined in Eq.~\eqref{eq:H}. The subsystems $A$ and $B$ evolve separately for half of the period $T=2\tau=20$ and for the second half only the $A-B$ coupling yields the dynamics with $H_A$ and $H_B$ being put to zero. We dub this system  ``kicked'' as the form of the unitary, \eqref{eq:kick}, allows one to interpret the dynamics  as evolving with $H_A + H_B$ for period $\tau$ and then undergoing a $\delta$-kick
of strength $H_{AB} \tau$ resembling thus famous kicked rotator or kicked top models \cite{Haake}.    
The time dynamics is studied by looking at the imbalance
\begin{equation}
    \mathcal{I}(t) = \frac{4}{L_A-4} \sum_{i=3}^{L_A-2} \langle Z_i(0) Z_i(t) \rangle,
\end{equation} 
for system $A$ of length $L_A$ and $B$ of length $L_B$ -- note that two spins on each side of the system $A$ are not included to minimalize edge effects.
 
We compare the imbalance obtained for the kicked model \eqref{eq:kick} with that obtained for autonomous system governed by Eq.~\eqref{eq:H} for the same parameter values. Reproducing the results of \cite{Peacock23} we note that in the presence of the thermal inclusion the system $A$ shows a much faster decay than in the case $W_B=W_A$ without a thermal inclusion. This difference is noticeable even for strongly localized chains at $W_A=12$.  At sufficiently long times, the imbalance decays, suggesting thermalization of the system in the  wide range of disorder values $W_A=4-12$. The onset of the decay of the imbalance scales with disorder as $t \rightarrow t e^{-a W_A}$, with constant $a$ found to be $a=2.1$ for $L_A=12$, $L_B=10$, $W_B=0.5$. The latter behavior bears some resemblance to the Thouless time scaling $t_{Th} = L^2 e^{W/\Omega}$ (where $\Omega$ is a constant) proposed for XXZ spin chain in Ref.~\cite{Suntajs20e}.

 Intuitively, one could expect that since the bath $B$ is connected with the subsystem $A$ at all times in our case,
contrary to kicked type of dynamics generated by \eqref{eq:kick}, the autonomous system should thermalize faster. On the other hand, the energy is absorbed from the driving in the kicked case, favoring the thermalization of the kicked system, unless the driving is very fast \cite{Abanin17}.

The comparison of both models is presented in Fig.~\ref{fig:kicked_vs_notkicked}.
The first column of Fig.~\ref{fig:kicked_vs_notkicked} (panels "a" and "e") shows the imbalance as a function of time in both models. It is clear that although the signs of thermalization in the kicked model start to appear at longer times $(t\approx 200-300)$ than in the time-independent model, the acceleration of the imbalance decay at largest disorder values is much more evident in the kicked case. This is further demonstrated in the middle column of Fig.~\ref{fig:kicked_vs_notkicked} (panels "b" and "f") which shows how the difference of imbalance in the case of no bath ($W_A=W_B$) and with the bath ($W_B=0.5$) accelerates with time for the kicked model even for $W_A=12$. In contrast, the growth of the imbalance difference does not accelerate for the time-independent model. On the contrary, it rather saturates, at sufficiently large $W_A$, to a constant (which is possibly dependent on $L_A$). This result is consistent with the MBL regime and is in a stark difference with the kicked case.

Finally, in the rightmost column of Fig.~\ref{fig:kicked_vs_notkicked} (panels "c", "d", and "g") we check whether the rescaling of time $t \rightarrow t e^{-a W_A}$, which captures the dynamics of the kicked model in a broad regime of $W_A$, applies also to the energy-conserving model. We find the optimal $a$ describing the data by fitting a fourth degree polynomial to the imbalance and minimizing the fit error by tuning $a$. To check if $a$ is universal and does not depend on disorder $W_A$, our fits are performed separately for small disorders ($W_A\in[2,5]$) on the ergodic side of the crossover (Fig.~\ref{fig:kicked_vs_notkicked}c) and for large disorders ($W_A\in[10,13]$) on the localized side. Fitting errors are measured only in the corresponding fitting range. The resulting optimal $a$ are $a=6$ and $a=11$, respectively. In the latter case, the data can be in fact almost equally well described by any $a\in[8,15]$, as the fit error does not change in this interval by more than a few percent. Notably, due to the presence of the exponent, the difference between $a=6$ and $a=11$ cases is huge, the rescaled times differ by $15$(!) orders of magnitude in the middle of the interval of disorder strengths considered. The lack of agreement with $a$ found for low disorder values suggests that the time-independent system does not have a universal timescale dictated by the disorder value. On the other hand, for the kicked model, the same procedure gives $a$ ranging from $a=1.6$ for the largest disorders $W_A\in[10,12]$ to $a=2.3$ for the smallest disorders $W_A\in[4,6]$, and the optimal $a$ is well-defined in both cases, staying in agreement with $a=2.1$ from Ref.~\cite{Peacock23}. We note that the dependence of $a$ on the interval of disorder strengths used to determine its value, is opposite in the two models. This highlights further the difference in the behavior of the two models at large $W_A$. Details on how we determine values of the constant $a$ are described in Appendix \ref{app:scaling}.

Summarizing, the results of Ref.~\cite{Peacock23} are reproduced for the kicked case (note that we use numerically exact Chebyshev time propagation rather than the tensor network time-evolved block decimation method (TEBD) that is approximate). The results for the kicked and time independent case differ qualitatively for the considered system size. The tendency towards the thermalization of the setup of \cite{Peacock23}, even at the strong disorder $W_A=12$, does not translate to the energy-conserving setup. While the rescaling $t \rightarrow t e^{-a W_A}$ suggests uniformity of the thermalizing behavior in the kicked case across the broad interval of disorder strengths, the same rescaling does not allow to collapse the results for our autonomous case with a reasonable accuracy. Importantly, it is the autonomous, time independent case which is relevant for our understanding of an avalanche spreading in isolated quantum spin chains. Indeed, if disorder fluctuations produce a region of anomalously weak disorder, its interaction with the surrounding spins is captured by a time independent Hamiltonian.

\section{Conclusions and discussion}
We have analyzed the effects of rare thermal bubbles, which arise in disordered spin chains due to disorder fluctuations on thermalization of the system. While the behavior of the ergodic bubbles determines the stability of MBL phase, their direct investigation is not straightforward since the probability of appearance of an ergodic bubble is exponentially small in its size. For that reason, we consider a scenario of a finite thermal bath (XXZ spin chain with low disorder) attached to an XXZ spin chain with uniformly distributed random disorder which plays the role of an MBL subsystem $A$. Based on the time evolution of the system, we study the decay length $\xi_d(t)$ of connected two-body correlations 
$\langle Z_i Z_j \rangle_c$ of the MBL subsystem with the bath. 
Our perturbative analysis of the system-bath coupling identifies the conditions under which the MBL system is not thermalized by the bath. In this case, $\xi_d(t) \propto \ln(t)$ for a time scaling exponentially in the bath size, followed by a saturation. In contrast, the quantum avalanche mechanism includes non-perturbative effects, leading to thermalization, manifested as a growth $\xi_d(t) \propto \ln(t)$ or faster until time scale which scales exponentially with the size of the total system.

We perform an extensive numerical study, obtaining a superlogarithmic growth of $\xi_d(t)$ in the critical and ergodic regimes. Such behavior is consistent with the quantum avalanche mechanism, although we cannot exclude that a different microscopic mechanism plays a role in the observed thermalization process. Surprisingly, the $\xi_d(t)\propto \ln(t)$ growth until the Heisenberg time is also observed in the deep MBL regime if the initial state is set to be a simple product state. However, the above result is not 
a sign of thermalization but rather a long-timescale feature of the dynamics, not captured by the theory. The fact that the system does not thermalize becomes apparent only if the time evolution starts from an initial state being a product state of an energy eigenstate in the MBL subsystem $A$ and N\'{e}el state in the bath. In this case, $\xi_d(t)$ saturates at a constant after a time fixed by the bath size. The onset of thermalization is visible in the same setting in the ergodic and critical regimes. Thus, we conclude that the observation of $\xi_d(t)\propto \ln(t)$ long-time growth from an initial product state does not allow us to unambiguously determine whether the avalanches propagate or not.

As we show in Sec.~\ref{sec:loclength} the rate of growth of $\xi_d(t)$ allows us to extract the LIOM localization length $\xi$. Extracted LIOM localization lengths in all three regimes (ergodic, critical or MBL) of subsystem $A$ are consistent with the predicted ETH-MBL crossover at $\xi=(\ln 2)/2$. We further demonstrate that the thermalization process does not depend on the details of the bath, as long as the bath is an ergodic system on its own. In particular, changing the bath disorder strength between $W_B=0-1$ does not qualitatively change the decay length growth with time. Moreover, instead of the XXZ spin chain with low disorder $W_B$ serving as a bath, one can attach a bath whose Hamiltonian is a properly rescaled GOE matrix and the thermalization characteristics manifested by the behavior of $\xi_d(t)$, remain the same to a good approximation.

We also inspect the offdiagonal correlations within the bath during the time evolution. The correlations in the bath disconnected from the MBL subsystem $A$ quickly saturate to time independent values. Similarly, when the bath is connected to XXZ spin chain in the MBL regime, the correlations saturate at a similar time scale, reflecting the lack of non-trivial dynamics between the MBL system and the bath. In contrast, when the XXZ spin chain in subsystem $A$ is in an ergodic regime, we observe a steady relaxation of the off-diagonal correlations within the bath, which is a direct consequence of a quantum avalanche increasing the effective bath size.

Finally, we contrast our calculations of the imbalance with the results from the kicked XXZ model \cite{Peacock23}. Unlike in the 
kicked model, the time rescaling $t \rightarrow t e^{-a W_A}$ does not allow us to collapse the results for time evolution of the imbalance for a wide range of disorder strength, $W_A$, in our energy conserving setup. Also, in our model, the change in the behavior of imbalance with time, with and without a bath, is not as dramatic as for the kicked case. This suggests that thermalization in the kicked model is enhanced by the driving and one should not generically expect the same behavior for the energy-conserving Hamiltonian model, relevant for the question of MBL in autonomous systems.

Our results shed new light on the avalanche mechanism present in the disordered XXZ model. We have provided numerical evidence consistent with the avalanche spreading from an engineered thermal inclusion and have shown that their fate depends on the disorder strength $W_A$. At the same time, we have demonstrated that an observation of the termination of spreading of quantum avalanches is more challenging as the system has to be prepared in an eigenstate of the MBL subsystem $A$. Notably, preparation of area-law entanglement initial states of the MBL subsystem $A$, is experimentally feasible~\cite{Shuo23}.

Our work demonstrates that the disorder strengths in the XXZ spin chain at which the avalanches propagate and are terminated correspond to regimes of disorder strengths identified respectively as ergodic and MBL in the exact diagonalization studies of the model, for example see Ref.~\cite{Sierant20polfed}. The parametric shift of the disorder strengths required for termination of avalanches observed in \cite{Morningstar22, Sels22bath} with respect to the exact diagonalization studies may, therefore, be attributed to the assumptions about the nature of the ergodic bubble made in Refs.~\cite{Morningstar22, Sels22bath}. The results presented in our work are valid for ergodic bubbles placed on several lattice sites and do not support those assumptions. Nevertheless, we cannot exclude the possibility that ergodic bubbles of much larger sizes, which can be found in thermodynamically large systems, seed avalanches that propagate in presence of much stronger disorder. The length and time scales associated with such rare events are, most likely, not relevant for numerical or experimental investigations of disordered many-body systems.

{\it Note} The numerical data presented in this work are freely available from  \url{https://chaos.if.uj.edu.pl/ZOA/opendata/} or from the authors upon a reasonable request.

\acknowledgements  
P.S. acknowledges interesting discussions with M. Hopjan, R. Świ\c{e}tek, and L. Vidmar.
This research has been supported by the National Science Centre (Poland) under project 2019/35/B/ST2/00034 (T.S.) and under the OPUS call within the WEAVE program 2021/43/I/ST3/01142 (J.Z.). 
The research was also supported by a grant from the Priority Research Area (Visibility and Mobility) and  Research Area DigiWorld under the Strategic Programme Excellence Initiative at Jagiellonian University (T.S.).  We gratefully acknowledge Polish high-performance computing infrastructure PLGrid (HPC Centers: ACK Cyfronet AGH) for providing computer facilities and support within computational grant no. PLG/2023/016370. 

P.S. and M.L. acknowledge support from: European Research Council AdG NOQIA; MCIN/AEI (PGC2018-0910.13039/501100011033,  CEX2019-000910-S/10.13039/501100011033, Plan National FIDEUA PID2019-106901GB-I00, Plan National STAMEENA PID2022-139099NB, I00, project funded by MCIN/AEI/10.13039/501100011033 and by the “European Union NextGenerationEU/PRTR" (PRTR-C17.I1), FPI); QUANTERA MAQS PCI2019-111828-2);  QUANTERA DYNAMITE PCI2022-132919, QuantERA II Programme co-funded by European Union’s Horizon 2020 program under Grant Agreement No 101017733); Ministry for Digital Transformation and of Civil Service of the Spanish Government through the QUANTUM ENIA project call - Quantum Spain project, and by the European Union through the Recovery, Transformation and Resilience Plan - NextGenerationEU within the framework of the Digital Spain 2026 Agenda; Fundació Cellex; Fundació Mir-Puig; Generalitat de Catalunya (European Social Fund FEDER and CERCA program, AGAUR Grant No. 2021 SGR 01452, QuantumCAT \ U16-011424, co-funded by ERDF Operational Program of Catalonia 2014-2020); Barcelona Supercomputing Center MareNostrum (FI-2023-1-0013); Funded by the European Union. Views and opinions expressed are however those of the author(s) only and do not necessarily reflect those of the European Union, European Commission, European Climate, Infrastructure and Environment Executive Agency (CINEA), or any other granting authority.  Neither the European Union nor any granting authority can be held responsible for them (EU Quantum Flagship PASQuanS2.1, 101113690, EU Horizon 2020 FET-OPEN OPTOlogic, Grant No 899794),  EU Horizon Europe Program (This project has received funding from the European Union’s Horizon Europe research and innovation program under grant agreement No 101080086 NeQSTGrant Agreement 101080086 — NeQST); ICFO Internal “QuantumGaudi” project; European Union’s Horizon 2020 program under the Marie Sklodowska-Curie grant agreement No 847648;  “La Caixa” Junior Leaders fellowships, La Caixa” Foundation (ID 100010434): CF/BQ/PR23/11980043.


\begin{thebibliography}{152}%
	\makeatletter
	\providecommand \@ifxundefined [1]{%
		\@ifx{#1\undefined}
	}%
	\providecommand \@ifnum [1]{%
		\ifnum #1\expandafter \@firstoftwo
		\else \expandafter \@secondoftwo
		\fi
	}%
	\providecommand \@ifx [1]{%
		\ifx #1\expandafter \@firstoftwo
		\else \expandafter \@secondoftwo
		\fi
	}%
	\providecommand \natexlab [1]{#1}%
	\providecommand \enquote  [1]{``#1''}%
	\providecommand \bibnamefont  [1]{#1}%
	\providecommand \bibfnamefont [1]{#1}%
	\providecommand \citenamefont [1]{#1}%
	\providecommand \href@noop [0]{\@secondoftwo}%
	\providecommand \href [0]{\begingroup \@sanitize@url \@href}%
	\providecommand \@href[1]{\@@startlink{#1}\@@href}%
	\providecommand \@@href[1]{\endgroup#1\@@endlink}%
	\providecommand \@sanitize@url [0]{\catcode `\\12\catcode `\$12\catcode
		`\&12\catcode `\#12\catcode `\^12\catcode `\_12\catcode `\%12\relax}%
	\providecommand \@@startlink[1]{}%
	\providecommand \@@endlink[0]{}%
	\providecommand \url  [0]{\begingroup\@sanitize@url \@url }%
	\providecommand \@url [1]{\endgroup\@href {#1}{\urlprefix }}%
	\providecommand \urlprefix  [0]{URL }%
	\providecommand \Eprint [0]{\href }%
	\providecommand \doibase [0]{https://doi.org/}%
	\providecommand \selectlanguage [0]{\@gobble}%
	\providecommand \bibinfo  [0]{\@secondoftwo}%
	\providecommand \bibfield  [0]{\@secondoftwo}%
	\providecommand \translation [1]{[#1]}%
	\providecommand \BibitemOpen [0]{}%
	\providecommand \bibitemStop [0]{}%
	\providecommand \bibitemNoStop [0]{.\EOS\space}%
	\providecommand \EOS [0]{\spacefactor3000\relax}%
	\providecommand \BibitemShut  [1]{\csname bibitem#1\endcsname}%
	\let\auto@bib@innerbib\@empty
	\bibitem [{\citenamefont {Gallavotti}(1995)}]{Gallavotti95}%
	\BibitemOpen
	\bibfield  {author} {\bibinfo {author} {\bibfnamefont {G.}~\bibnamefont
			{Gallavotti}},\ }\bibfield  {title} {\bibinfo {title} {Ergodicity, ensembles,
			irreversibility in boltzmann and beyond},\ }\href
	{https://doi.org/10.1007/BF02180143} {\bibfield  {journal} {\bibinfo
			{journal} {Journal of Statistical Physics}\ }\textbf {\bibinfo {volume}
			{78}},\ \bibinfo {pages} {1571} (\bibinfo {year} {1995})}\BibitemShut
	{NoStop}%
	\bibitem [{\citenamefont {Deutsch}(1991)}]{Deutsch91}%
	\BibitemOpen
	\bibfield  {author} {\bibinfo {author} {\bibfnamefont {J.~M.}\ \bibnamefont
			{Deutsch}},\ }\bibfield  {title} {\bibinfo {title} {Quantum statistical
			mechanics in a closed system},\ }\href
	{https://doi.org/10.1103/PhysRevA.43.2046} {\bibfield  {journal} {\bibinfo
			{journal} {Phys. Rev. A}\ }\textbf {\bibinfo {volume} {43}},\ \bibinfo
		{pages} {2046} (\bibinfo {year} {1991})}\BibitemShut {NoStop}%
	\bibitem [{\citenamefont {Srednicki}(1994)}]{Srednicki94}%
	\BibitemOpen
	\bibfield  {author} {\bibinfo {author} {\bibfnamefont {M.}~\bibnamefont
			{Srednicki}},\ }\bibfield  {title} {\bibinfo {title} {Chaos and quantum
			thermalization},\ }\href {https://doi.org/10.1103/PhysRevE.50.888} {\bibfield
		{journal} {\bibinfo  {journal} {Phys. Rev. E}\ }\textbf {\bibinfo {volume}
			{50}},\ \bibinfo {pages} {888} (\bibinfo {year} {1994})}\BibitemShut
	{NoStop}%
	\bibitem [{\citenamefont {Rigol}\ \emph {et~al.}(2008)\citenamefont {Rigol},
		\citenamefont {Dunjko},\ and\ \citenamefont {Olshanii}}]{Rigol08}%
	\BibitemOpen
	\bibfield  {author} {\bibinfo {author} {\bibfnamefont {M.}~\bibnamefont
			{Rigol}}, \bibinfo {author} {\bibfnamefont {V.}~\bibnamefont {Dunjko}},\ and\
		\bibinfo {author} {\bibfnamefont {M.}~\bibnamefont {Olshanii}},\ }\bibfield
	{title} {\bibinfo {title} {Thermalization and its mechanism for generic
			isolated quantum systems},\ }\href {https://doi.org/10.1038/nature06838}
	{\bibfield  {journal} {\bibinfo  {journal} {Nature}\ }\textbf {\bibinfo
			{volume} {452}},\ \bibinfo {pages} {854 EP } (\bibinfo {year}
		{2008})}\BibitemShut {NoStop}%
	\bibitem [{\citenamefont {D'Alessio}\ \emph {et~al.}(2016)\citenamefont
		{D'Alessio}, \citenamefont {Kafri}, \citenamefont {Polkovnikov},\ and\
		\citenamefont {Rigol}}]{Alessio16}%
	\BibitemOpen
	\bibfield  {author} {\bibinfo {author} {\bibfnamefont {L.}~\bibnamefont
			{D'Alessio}}, \bibinfo {author} {\bibfnamefont {Y.}~\bibnamefont {Kafri}},
		\bibinfo {author} {\bibfnamefont {A.}~\bibnamefont {Polkovnikov}},\ and\
		\bibinfo {author} {\bibfnamefont {M.}~\bibnamefont {Rigol}},\ }\bibfield
	{title} {\bibinfo {title} {From quantum chaos and eigenstate thermalization
			to statistical mechanics and thermodynamics},\ }\href
	{https://doi.org/10.1080/00018732.2016.1198134} {\bibfield  {journal}
		{\bibinfo  {journal} {Advances in Physics}\ }\textbf {\bibinfo {volume}
			{65}},\ \bibinfo {pages} {239} (\bibinfo {year} {2016})},\ \Eprint
	{https://arxiv.org/abs/https://doi.org/10.1080/00018732.2016.1198134}
	{https://doi.org/10.1080/00018732.2016.1198134} \BibitemShut {NoStop}%
	\bibitem [{\citenamefont {Foini}\ and\ \citenamefont
		{Kurchan}(2019)}]{Foini19}%
	\BibitemOpen
	\bibfield  {author} {\bibinfo {author} {\bibfnamefont {L.}~\bibnamefont
			{Foini}}\ and\ \bibinfo {author} {\bibfnamefont {J.}~\bibnamefont
			{Kurchan}},\ }\bibfield  {title} {\bibinfo {title} {Eigenstate thermalization
			hypothesis and out of time order correlators},\ }\href
	{https://doi.org/10.1103/PhysRevE.99.042139} {\bibfield  {journal} {\bibinfo
			{journal} {Phys. Rev. E}\ }\textbf {\bibinfo {volume} {99}},\ \bibinfo
		{pages} {042139} (\bibinfo {year} {2019})}\BibitemShut {NoStop}%
	\bibitem [{\citenamefont {Pappalardi}\ \emph {et~al.}(2022)\citenamefont
		{Pappalardi}, \citenamefont {Foini},\ and\ \citenamefont
		{Kurchan}}]{Pappalardi22}%
	\BibitemOpen
	\bibfield  {author} {\bibinfo {author} {\bibfnamefont {S.}~\bibnamefont
			{Pappalardi}}, \bibinfo {author} {\bibfnamefont {L.}~\bibnamefont {Foini}},\
		and\ \bibinfo {author} {\bibfnamefont {J.}~\bibnamefont {Kurchan}},\
	}\bibfield  {title} {\bibinfo {title} {Eigenstate thermalization hypothesis
			and free probability},\ }\href
	{https://doi.org/10.1103/PhysRevLett.129.170603} {\bibfield  {journal}
		{\bibinfo  {journal} {Phys. Rev. Lett.}\ }\textbf {\bibinfo {volume} {129}},\
		\bibinfo {pages} {170603} (\bibinfo {year} {2022})}\BibitemShut {NoStop}%
	\bibitem [{\citenamefont {Pappalardi}\ \emph {et~al.}(2023)\citenamefont
		{Pappalardi}, \citenamefont {Fritzsch},\ and\ \citenamefont
		{Prosen}}]{Pappalardi23}%
	\BibitemOpen
	\bibfield  {author} {\bibinfo {author} {\bibfnamefont {S.}~\bibnamefont
			{Pappalardi}}, \bibinfo {author} {\bibfnamefont {F.}~\bibnamefont
			{Fritzsch}},\ and\ \bibinfo {author} {\bibfnamefont {T.}~\bibnamefont
			{Prosen}},\ }\href@noop {} {\bibinfo {title} {General eigenstate
			thermalization via free cumulants in quantum lattice systems}} (\bibinfo
	{year} {2023}),\ \Eprint {https://arxiv.org/abs/2303.00713} {arXiv:2303.00713
		[cond-mat.stat-mech]} \BibitemShut {NoStop}%
	\bibitem [{\citenamefont {Pappalardi}\ \emph {et~al.}(2024)\citenamefont
		{Pappalardi}, \citenamefont {Foini},\ and\ \citenamefont
		{Kurchan}}]{pappalardi2023microcanonical}%
	\BibitemOpen
	\bibfield  {author} {\bibinfo {author} {\bibfnamefont {S.}~\bibnamefont
			{Pappalardi}}, \bibinfo {author} {\bibfnamefont {L.}~\bibnamefont {Foini}},\
		and\ \bibinfo {author} {\bibfnamefont {J.}~\bibnamefont {Kurchan}},\
	}\bibfield  {title} {\bibinfo {title} {Microcanonical windows on quantum
			operators},\ }\href {https://doi.org/10.22331/q-2024-01-11-1227} {\bibfield
		{journal} {\bibinfo  {journal} {Quantum}\ }\textbf {\bibinfo {volume} {8}},\
		\bibinfo {pages} {1227} (\bibinfo {year} {2024})}\BibitemShut {NoStop}%
	\bibitem [{\citenamefont {Fava}\ \emph {et~al.}(2023)\citenamefont {Fava},
		\citenamefont {Kurchan},\ and\ \citenamefont {Pappalardi}}]{fava2023designs}%
	\BibitemOpen
	\bibfield  {author} {\bibinfo {author} {\bibfnamefont {M.}~\bibnamefont
			{Fava}}, \bibinfo {author} {\bibfnamefont {J.}~\bibnamefont {Kurchan}},\ and\
		\bibinfo {author} {\bibfnamefont {S.}~\bibnamefont {Pappalardi}},\
	}\href@noop {} {\bibinfo {title} {Designs via free probability}} (\bibinfo
	{year} {2023}),\ \Eprint {https://arxiv.org/abs/2308.06200} {arXiv:2308.06200
		[quant-ph]} \BibitemShut {NoStop}%
	\bibitem [{\citenamefont {Rigol}(2009)}]{Rigol09}%
	\BibitemOpen
	\bibfield  {author} {\bibinfo {author} {\bibfnamefont {M.}~\bibnamefont
			{Rigol}},\ }\bibfield  {title} {\bibinfo {title} {Breakdown of thermalization
			in finite one-dimensional systems},\ }\href
	{https://doi.org/10.1103/PhysRevLett.103.100403} {\bibfield  {journal}
		{\bibinfo  {journal} {Phys. Rev. Lett.}\ }\textbf {\bibinfo {volume} {103}},\
		\bibinfo {pages} {100403} (\bibinfo {year} {2009})}\BibitemShut {NoStop}%
	\bibitem [{\citenamefont {Santos}\ and\ \citenamefont
		{Rigol}(2010)}]{Santos10}%
	\BibitemOpen
	\bibfield  {author} {\bibinfo {author} {\bibfnamefont {L.~F.}\ \bibnamefont
			{Santos}}\ and\ \bibinfo {author} {\bibfnamefont {M.}~\bibnamefont {Rigol}},\
	}\bibfield  {title} {\bibinfo {title} {Onset of quantum chaos in
			one-dimensional bosonic and fermionic systems and its relation to
			thermalization},\ }\href {https://doi.org/10.1103/PhysRevE.81.036206}
	{\bibfield  {journal} {\bibinfo  {journal} {Phys. Rev. E}\ }\textbf {\bibinfo
			{volume} {81}},\ \bibinfo {pages} {036206} (\bibinfo {year}
		{2010})}\BibitemShut {NoStop}%
	\bibitem [{\citenamefont {Steinigeweg}\ \emph {et~al.}(2013)\citenamefont
		{Steinigeweg}, \citenamefont {Herbrych},\ and\ \citenamefont
		{Prelov\ifmmode~\check{s}\else \v{s}\fi{}ek}}]{Steinigeweg13}%
	\BibitemOpen
	\bibfield  {author} {\bibinfo {author} {\bibfnamefont {R.}~\bibnamefont
			{Steinigeweg}}, \bibinfo {author} {\bibfnamefont {J.}~\bibnamefont
			{Herbrych}},\ and\ \bibinfo {author} {\bibfnamefont {P.}~\bibnamefont
			{Prelov\ifmmode~\check{s}\else \v{s}\fi{}ek}},\ }\bibfield  {title} {\bibinfo
		{title} {Eigenstate thermalization within isolated spin-chain systems},\
	}\href {https://doi.org/10.1103/PhysRevE.87.012118} {\bibfield  {journal}
		{\bibinfo  {journal} {Phys. Rev. E}\ }\textbf {\bibinfo {volume} {87}},\
		\bibinfo {pages} {012118} (\bibinfo {year} {2013})}\BibitemShut {NoStop}%
	\bibitem [{\citenamefont {Khatami}\ \emph {et~al.}(2013)\citenamefont
		{Khatami}, \citenamefont {Pupillo}, \citenamefont {Srednicki},\ and\
		\citenamefont {Rigol}}]{Khatami13}%
	\BibitemOpen
	\bibfield  {author} {\bibinfo {author} {\bibfnamefont {E.}~\bibnamefont
			{Khatami}}, \bibinfo {author} {\bibfnamefont {G.}~\bibnamefont {Pupillo}},
		\bibinfo {author} {\bibfnamefont {M.}~\bibnamefont {Srednicki}},\ and\
		\bibinfo {author} {\bibfnamefont {M.}~\bibnamefont {Rigol}},\ }\bibfield
	{title} {\bibinfo {title} {Fluctuation-dissipation theorem in an isolated
			system of quantum dipolar bosons after a quench},\ }\href
	{https://doi.org/10.1103/PhysRevLett.111.050403} {\bibfield  {journal}
		{\bibinfo  {journal} {Phys. Rev. Lett.}\ }\textbf {\bibinfo {volume} {111}},\
		\bibinfo {pages} {050403} (\bibinfo {year} {2013})}\BibitemShut {NoStop}%
	\bibitem [{\citenamefont {Beugeling}\ \emph {et~al.}(2014)\citenamefont
		{Beugeling}, \citenamefont {Moessner},\ and\ \citenamefont
		{Haque}}]{Beugeling14}%
	\BibitemOpen
	\bibfield  {author} {\bibinfo {author} {\bibfnamefont {W.}~\bibnamefont
			{Beugeling}}, \bibinfo {author} {\bibfnamefont {R.}~\bibnamefont
			{Moessner}},\ and\ \bibinfo {author} {\bibfnamefont {M.}~\bibnamefont
			{Haque}},\ }\bibfield  {title} {\bibinfo {title} {Finite-size scaling of
			eigenstate thermalization},\ }\href
	{https://doi.org/10.1103/PhysRevE.89.042112} {\bibfield  {journal} {\bibinfo
			{journal} {Phys. Rev. E}\ }\textbf {\bibinfo {volume} {89}},\ \bibinfo
		{pages} {042112} (\bibinfo {year} {2014})}\BibitemShut {NoStop}%
	\bibitem [{\citenamefont {Sch\"onle}\ \emph {et~al.}(2021)\citenamefont
		{Sch\"onle}, \citenamefont {Jansen}, \citenamefont {Heidrich-Meisner},\ and\
		\citenamefont {Vidmar}}]{Schonle21}%
	\BibitemOpen
	\bibfield  {author} {\bibinfo {author} {\bibfnamefont {C.}~\bibnamefont
			{Sch\"onle}}, \bibinfo {author} {\bibfnamefont {D.}~\bibnamefont {Jansen}},
		\bibinfo {author} {\bibfnamefont {F.}~\bibnamefont {Heidrich-Meisner}},\ and\
		\bibinfo {author} {\bibfnamefont {L.}~\bibnamefont {Vidmar}},\ }\bibfield
	{title} {\bibinfo {title} {Eigenstate thermalization hypothesis through the
			lens of autocorrelation functions},\ }\href
	{https://doi.org/10.1103/PhysRevB.103.235137} {\bibfield  {journal} {\bibinfo
			{journal} {Phys. Rev. B}\ }\textbf {\bibinfo {volume} {103}},\ \bibinfo
		{pages} {235137} (\bibinfo {year} {2021})}\BibitemShut {NoStop}%
	\bibitem [{\citenamefont {Gornyi}\ \emph {et~al.}(2005)\citenamefont {Gornyi},
		\citenamefont {Mirlin},\ and\ \citenamefont {Polyakov}}]{Gornyi05}%
	\BibitemOpen
	\bibfield  {author} {\bibinfo {author} {\bibfnamefont {I.~V.}\ \bibnamefont
			{Gornyi}}, \bibinfo {author} {\bibfnamefont {A.~D.}\ \bibnamefont {Mirlin}},\
		and\ \bibinfo {author} {\bibfnamefont {D.~G.}\ \bibnamefont {Polyakov}},\
	}\bibfield  {title} {\bibinfo {title} {Interacting electrons in disordered
			wires: {A}nderson localization and low-$t$ transport},\ }\href
	{https://doi.org/10.1103/PhysRevLett.95.206603} {\bibfield  {journal}
		{\bibinfo  {journal} {Phys. Rev. Lett.}\ }\textbf {\bibinfo {volume} {95}},\
		\bibinfo {pages} {206603} (\bibinfo {year} {2005})}\BibitemShut {NoStop}%
	\bibitem [{\citenamefont {Basko}\ \emph {et~al.}(2006)\citenamefont {Basko},
		\citenamefont {Aleiner},\ and\ \citenamefont {Altschuler}}]{Basko06}%
	\BibitemOpen
	\bibfield  {author} {\bibinfo {author} {\bibfnamefont {D.}~\bibnamefont
			{Basko}}, \bibinfo {author} {\bibfnamefont {I.}~\bibnamefont {Aleiner}},\
		and\ \bibinfo {author} {\bibfnamefont {B.}~\bibnamefont {Altschuler}},\
	}\bibfield  {title} {\bibinfo {title} {Metal-insulator transition in a weakly
			interacting many-electron system with localized single-particle states},\
	}\href@noop {} {\bibfield  {journal} {\bibinfo  {journal} {Ann. Phys. (NY)}\
		}\textbf {\bibinfo {volume} {321}},\ \bibinfo {pages} {1126} (\bibinfo {year}
		{2006})}\BibitemShut {NoStop}%
	\bibitem [{\citenamefont {Oganesyan}\ and\ \citenamefont
		{Huse}(2007)}]{Oganesyan07}%
	\BibitemOpen
	\bibfield  {author} {\bibinfo {author} {\bibfnamefont {V.}~\bibnamefont
			{Oganesyan}}\ and\ \bibinfo {author} {\bibfnamefont {D.~A.}\ \bibnamefont
			{Huse}},\ }\bibfield  {title} {\bibinfo {title} {Localization of interacting
			fermions at high temperature},\ }\href
	{https://doi.org/10.1103/PhysRevB.75.155111} {\bibfield  {journal} {\bibinfo
			{journal} {Phys. Rev. B}\ }\textbf {\bibinfo {volume} {75}},\ \bibinfo
		{pages} {155111} (\bibinfo {year} {2007})}\BibitemShut {NoStop}%
	\bibitem [{\citenamefont {\v{Z}nidari\v{c}}\ \emph {et~al.}(2008)\citenamefont
		{\v{Z}nidari\v{c}}, \citenamefont {Prosen},\ and\ \citenamefont
		{Prelov\v{s}ek}}]{Znidaric08}%
	\BibitemOpen
	\bibfield  {author} {\bibinfo {author} {\bibfnamefont {M.}~\bibnamefont
			{\v{Z}nidari\v{c}}}, \bibinfo {author} {\bibfnamefont {T.}~\bibnamefont
			{Prosen}},\ and\ \bibinfo {author} {\bibfnamefont {P.}~\bibnamefont
			{Prelov\v{s}ek}},\ }\bibfield  {title} {\bibinfo {title} {Many-body
			localization in the {H}eisenberg {X}{X}{Z} magnet in a random field},\ }\href
	{https://doi.org/10.1103/PhysRevB.77.064426} {\bibfield  {journal} {\bibinfo
			{journal} {Phys. Rev. B}\ }\textbf {\bibinfo {volume} {77}},\ \bibinfo
		{pages} {064426} (\bibinfo {year} {2008})}\BibitemShut {NoStop}%
	\bibitem [{\citenamefont {Pal}\ and\ \citenamefont {Huse}(2010)}]{Pal10}%
	\BibitemOpen
	\bibfield  {author} {\bibinfo {author} {\bibfnamefont {A.}~\bibnamefont
			{Pal}}\ and\ \bibinfo {author} {\bibfnamefont {D.~A.}\ \bibnamefont {Huse}},\
	}\bibfield  {title} {\bibinfo {title} {Many-body localization phase
			transition},\ }\href {https://doi.org/10.1103/PhysRevB.82.174411} {\bibfield
		{journal} {\bibinfo  {journal} {Phys. Rev. B}\ }\textbf {\bibinfo {volume}
			{82}},\ \bibinfo {pages} {174411} (\bibinfo {year} {2010})}\BibitemShut
	{NoStop}%
	\bibitem [{\citenamefont {De~Luca}\ and\ \citenamefont
		{Scardicchio}(2013)}]{DeLuca13}%
	\BibitemOpen
	\bibfield  {author} {\bibinfo {author} {\bibfnamefont {A.}~\bibnamefont
			{De~Luca}}\ and\ \bibinfo {author} {\bibfnamefont {A.}~\bibnamefont
			{Scardicchio}},\ }\bibfield  {title} {\bibinfo {title} {Ergodicity breaking
			in a model showing many-body localization},\ }\href
	{https://doi.org/10.1209/0295-5075/101/37003} {\bibfield  {journal} {\bibinfo
			{journal} {{EPL} (Europhysics Letters)}\ }\textbf {\bibinfo {volume}
			{101}},\ \bibinfo {pages} {37003} (\bibinfo {year} {2013})}\BibitemShut
	{NoStop}%
	\bibitem [{\citenamefont {Luitz}\ \emph {et~al.}(2015)\citenamefont {Luitz},
		\citenamefont {Laflorencie},\ and\ \citenamefont {Alet}}]{Luitz15}%
	\BibitemOpen
	\bibfield  {author} {\bibinfo {author} {\bibfnamefont {D.~J.}\ \bibnamefont
			{Luitz}}, \bibinfo {author} {\bibfnamefont {N.}~\bibnamefont {Laflorencie}},\
		and\ \bibinfo {author} {\bibfnamefont {F.}~\bibnamefont {Alet}},\ }\bibfield
	{title} {\bibinfo {title} {Many-body localization edge in the random-field
			{H}eisenberg chain},\ }\href {https://doi.org/10.1103/PhysRevB.91.081103}
	{\bibfield  {journal} {\bibinfo  {journal} {Phys. Rev. B}\ }\textbf {\bibinfo
			{volume} {91}},\ \bibinfo {pages} {081103} (\bibinfo {year}
		{2015})}\BibitemShut {NoStop}%
	\bibitem [{\citenamefont {Sierant}\ \emph {et~al.}(2017)\citenamefont
		{Sierant}, \citenamefont {Delande},\ and\ \citenamefont
		{Zakrzewski}}]{Sierant17}%
	\BibitemOpen
	\bibfield  {author} {\bibinfo {author} {\bibfnamefont {P.}~\bibnamefont
			{Sierant}}, \bibinfo {author} {\bibfnamefont {D.}~\bibnamefont {Delande}},\
		and\ \bibinfo {author} {\bibfnamefont {J.}~\bibnamefont {Zakrzewski}},\
	}\bibfield  {title} {\bibinfo {title} {Many-body localization due to random
			interactions},\ }\href {https://doi.org/10.1103/PhysRevA.95.021601}
	{\bibfield  {journal} {\bibinfo  {journal} {Phys. Rev. A}\ }\textbf {\bibinfo
			{volume} {95}},\ \bibinfo {pages} {021601} (\bibinfo {year}
		{2017})}\BibitemShut {NoStop}%
	\bibitem [{\citenamefont {Sierant}\ and\ \citenamefont
		{Zakrzewski}(2018)}]{Sierant18}%
	\BibitemOpen
	\bibfield  {author} {\bibinfo {author} {\bibfnamefont {P.}~\bibnamefont
			{Sierant}}\ and\ \bibinfo {author} {\bibfnamefont {J.}~\bibnamefont
			{Zakrzewski}},\ }\bibfield  {title} {\bibinfo {title} {Many-body localization
			of bosons in optical lattices},\ }\href
	{https://doi.org/10.1088/1367-2630/aabb17} {\bibfield  {journal} {\bibinfo
			{journal} {New Journal of Physics}\ }\textbf {\bibinfo {volume} {20}},\
		\bibinfo {pages} {043032} (\bibinfo {year} {2018})}\BibitemShut {NoStop}%
	\bibitem [{\citenamefont {Orell}\ \emph {et~al.}(2019)\citenamefont {Orell},
		\citenamefont {Michailidis}, \citenamefont {Serbyn},\ and\ \citenamefont
		{Silveri}}]{Orell19}%
	\BibitemOpen
	\bibfield  {author} {\bibinfo {author} {\bibfnamefont {T.}~\bibnamefont
			{Orell}}, \bibinfo {author} {\bibfnamefont {A.~A.}\ \bibnamefont
			{Michailidis}}, \bibinfo {author} {\bibfnamefont {M.}~\bibnamefont
			{Serbyn}},\ and\ \bibinfo {author} {\bibfnamefont {M.}~\bibnamefont
			{Silveri}},\ }\bibfield  {title} {\bibinfo {title} {Probing the many-body
			localization phase transition with superconducting circuits},\ }\href
	{https://doi.org/10.1103/PhysRevB.100.134504} {\bibfield  {journal} {\bibinfo
			{journal} {Phys. Rev. B}\ }\textbf {\bibinfo {volume} {100}},\ \bibinfo
		{pages} {134504} (\bibinfo {year} {2019})}\BibitemShut {NoStop}%
	\bibitem [{\citenamefont {Hopjan}\ and\ \citenamefont
		{Heidrich-Meisner}(2020)}]{Hopjan19}%
	\BibitemOpen
	\bibfield  {author} {\bibinfo {author} {\bibfnamefont {M.}~\bibnamefont
			{Hopjan}}\ and\ \bibinfo {author} {\bibfnamefont {F.}~\bibnamefont
			{Heidrich-Meisner}},\ }\bibfield  {title} {\bibinfo {title} {Many-body
			localization from a one-particle perspective in the disordered
			one-dimensional {Bose-Hubbard} model},\ }\href
	{https://doi.org/10.1103/PhysRevA.101.063617} {\bibfield  {journal} {\bibinfo
			{journal} {Phys. Rev. A}\ }\textbf {\bibinfo {volume} {101}},\ \bibinfo
		{pages} {063617} (\bibinfo {year} {2020})}\BibitemShut {NoStop}%
	\bibitem [{\citenamefont {Mondaini}\ and\ \citenamefont
		{Rigol}(2015)}]{Mondaini15}%
	\BibitemOpen
	\bibfield  {author} {\bibinfo {author} {\bibfnamefont {R.}~\bibnamefont
			{Mondaini}}\ and\ \bibinfo {author} {\bibfnamefont {M.}~\bibnamefont
			{Rigol}},\ }\bibfield  {title} {\bibinfo {title} {Many-body localization and
			thermalization in disordered {H}ubbard chains},\ }\href
	{https://doi.org/10.1103/PhysRevA.92.041601} {\bibfield  {journal} {\bibinfo
			{journal} {Phys. Rev. A}\ }\textbf {\bibinfo {volume} {92}},\ \bibinfo
		{pages} {041601} (\bibinfo {year} {2015})}\BibitemShut {NoStop}%
	\bibitem [{\citenamefont {Prelov\v{s}ek}\ \emph {et~al.}(2016)\citenamefont
		{Prelov\v{s}ek}, \citenamefont {Bari\v{s}i\'{c}},\ and\ \citenamefont
		{\v{Z}nidari\v{c}}}]{Prelovsek16}%
	\BibitemOpen
	\bibfield  {author} {\bibinfo {author} {\bibfnamefont {P.}~\bibnamefont
			{Prelov\v{s}ek}}, \bibinfo {author} {\bibfnamefont {O.~S.}\ \bibnamefont
			{Bari\v{s}i\'{c}}},\ and\ \bibinfo {author} {\bibfnamefont {M.}~\bibnamefont
			{\v{Z}nidari\v{c}}},\ }\bibfield  {title} {\bibinfo {title} {Absence of full
			many-body localization in the disordered {H}ubbard chain},\ }\href
	{https://doi.org/10.1103/PhysRevB.94.241104} {\bibfield  {journal} {\bibinfo
			{journal} {Phys. Rev. B}\ }\textbf {\bibinfo {volume} {94}},\ \bibinfo
		{pages} {241104} (\bibinfo {year} {2016})}\BibitemShut {NoStop}%
	\bibitem [{\citenamefont {Zakrzewski}\ and\ \citenamefont
		{Delande}(2018)}]{Zakrzewski18}%
	\BibitemOpen
	\bibfield  {author} {\bibinfo {author} {\bibfnamefont {J.}~\bibnamefont
			{Zakrzewski}}\ and\ \bibinfo {author} {\bibfnamefont {D.}~\bibnamefont
			{Delande}},\ }\bibfield  {title} {\bibinfo {title} {Spin-charge separation
			and many-body localization},\ }\href
	{https://doi.org/10.1103/PhysRevB.98.014203} {\bibfield  {journal} {\bibinfo
			{journal} {Phys. Rev. B}\ }\textbf {\bibinfo {volume} {98}},\ \bibinfo
		{pages} {014203} (\bibinfo {year} {2018})}\BibitemShut {NoStop}%
	\bibitem [{\citenamefont {Kozarzewski}\ \emph {et~al.}(2018)\citenamefont
		{Kozarzewski}, \citenamefont {Prelov\v{s}ek},\ and\ \citenamefont
		{Mierzejewski}}]{Kozarzewski18}%
	\BibitemOpen
	\bibfield  {author} {\bibinfo {author} {\bibfnamefont {M.}~\bibnamefont
			{Kozarzewski}}, \bibinfo {author} {\bibfnamefont {P.}~\bibnamefont
			{Prelov\v{s}ek}},\ and\ \bibinfo {author} {\bibfnamefont {M.}~\bibnamefont
			{Mierzejewski}},\ }\bibfield  {title} {\bibinfo {title} {Spin subdiffusion in
			the disordered {H}ubbard chain},\ }\href
	{https://doi.org/10.1103/PhysRevLett.120.246602} {\bibfield  {journal}
		{\bibinfo  {journal} {Phys. Rev. Lett.}\ }\textbf {\bibinfo {volume} {120}},\
		\bibinfo {pages} {246602} (\bibinfo {year} {2018})}\BibitemShut {NoStop}%
	\bibitem [{\citenamefont {Luitz}\ \emph {et~al.}(2016)\citenamefont {Luitz},
		\citenamefont {Laflorencie},\ and\ \citenamefont {Alet}}]{Luitz16}%
	\BibitemOpen
	\bibfield  {author} {\bibinfo {author} {\bibfnamefont {D.~J.}\ \bibnamefont
			{Luitz}}, \bibinfo {author} {\bibfnamefont {N.}~\bibnamefont {Laflorencie}},\
		and\ \bibinfo {author} {\bibfnamefont {F.}~\bibnamefont {Alet}},\ }\bibfield
	{title} {\bibinfo {title} {Extended slow dynamical regime close to the
			many-body localization transition},\ }\href
	{https://doi.org/10.1103/PhysRevB.93.060201} {\bibfield  {journal} {\bibinfo
			{journal} {Phys. Rev. B}\ }\textbf {\bibinfo {volume} {93}},\ \bibinfo
		{pages} {060201} (\bibinfo {year} {2016})}\BibitemShut {NoStop}%
	\bibitem [{\citenamefont {Bera}\ \emph {et~al.}(2017)\citenamefont {Bera},
		\citenamefont {De~Tomasi}, \citenamefont {Weiner},\ and\ \citenamefont
		{Evers}}]{Bera17}%
	\BibitemOpen
	\bibfield  {author} {\bibinfo {author} {\bibfnamefont {S.}~\bibnamefont
			{Bera}}, \bibinfo {author} {\bibfnamefont {G.}~\bibnamefont {De~Tomasi}},
		\bibinfo {author} {\bibfnamefont {F.}~\bibnamefont {Weiner}},\ and\ \bibinfo
		{author} {\bibfnamefont {F.}~\bibnamefont {Evers}},\ }\bibfield  {title}
	{\bibinfo {title} {Density propagator for many-body localization: Finite-size
			effects, transient subdiffusion, and exponential decay},\ }\href
	{https://doi.org/10.1103/PhysRevLett.118.196801} {\bibfield  {journal}
		{\bibinfo  {journal} {Phys. Rev. Lett.}\ }\textbf {\bibinfo {volume} {118}},\
		\bibinfo {pages} {196801} (\bibinfo {year} {2017})}\BibitemShut {NoStop}%
	\bibitem [{\citenamefont {Weiner}\ \emph {et~al.}(2019)\citenamefont {Weiner},
		\citenamefont {Evers},\ and\ \citenamefont {Bera}}]{Weiner19}%
	\BibitemOpen
	\bibfield  {author} {\bibinfo {author} {\bibfnamefont {F.}~\bibnamefont
			{Weiner}}, \bibinfo {author} {\bibfnamefont {F.}~\bibnamefont {Evers}},\ and\
		\bibinfo {author} {\bibfnamefont {S.}~\bibnamefont {Bera}},\ }\bibfield
	{title} {\bibinfo {title} {Slow dynamics and strong finite-size effects in
			many-body localization with random and quasiperiodic potentials},\ }\href
	{https://doi.org/10.1103/PhysRevB.100.104204} {\bibfield  {journal} {\bibinfo
			{journal} {Phys. Rev. B}\ }\textbf {\bibinfo {volume} {100}},\ \bibinfo
		{pages} {104204} (\bibinfo {year} {2019})}\BibitemShut {NoStop}%
	\bibitem [{\citenamefont {Sels}\ and\ \citenamefont
		{Polkovnikov}(2023)}]{Sels23dilute}%
	\BibitemOpen
	\bibfield  {author} {\bibinfo {author} {\bibfnamefont {D.}~\bibnamefont
			{Sels}}\ and\ \bibinfo {author} {\bibfnamefont {A.}~\bibnamefont
			{Polkovnikov}},\ }\bibfield  {title} {\bibinfo {title} {Thermalization of
			dilute impurities in one-dimensional spin chains},\ }\href
	{https://doi.org/10.1103/PhysRevX.13.011041} {\bibfield  {journal} {\bibinfo
			{journal} {Phys. Rev. X}\ }\textbf {\bibinfo {volume} {13}},\ \bibinfo
		{pages} {011041} (\bibinfo {year} {2023})}\BibitemShut {NoStop}%
	\bibitem [{\citenamefont {Schreiber}\ \emph {et~al.}(2015)\citenamefont
		{Schreiber}, \citenamefont {Hodgman}, \citenamefont {Bordia}, \citenamefont
		{L{\"u}schen}, \citenamefont {Fischer}, \citenamefont {Vosk}, \citenamefont
		{Altman}, \citenamefont {Schneider},\ and\ \citenamefont
		{Bloch}}]{Schreiber15}%
	\BibitemOpen
	\bibfield  {author} {\bibinfo {author} {\bibfnamefont {M.}~\bibnamefont
			{Schreiber}}, \bibinfo {author} {\bibfnamefont {S.~S.}\ \bibnamefont
			{Hodgman}}, \bibinfo {author} {\bibfnamefont {P.}~\bibnamefont {Bordia}},
		\bibinfo {author} {\bibfnamefont {H.~P.}\ \bibnamefont {L{\"u}schen}},
		\bibinfo {author} {\bibfnamefont {M.~H.}\ \bibnamefont {Fischer}}, \bibinfo
		{author} {\bibfnamefont {R.}~\bibnamefont {Vosk}}, \bibinfo {author}
		{\bibfnamefont {E.}~\bibnamefont {Altman}}, \bibinfo {author} {\bibfnamefont
			{U.}~\bibnamefont {Schneider}},\ and\ \bibinfo {author} {\bibfnamefont
			{I.}~\bibnamefont {Bloch}},\ }\bibfield  {title} {\bibinfo {title}
		{Observation of many-body localization of interacting fermions in a
			quasirandom optical lattice},\ }\href
	{https://doi.org/10.1126/science.aaa7432} {\bibfield  {journal} {\bibinfo
			{journal} {Science}\ }\textbf {\bibinfo {volume} {349}},\ \bibinfo {pages}
		{842} (\bibinfo {year} {2015})}\BibitemShut {NoStop}%
	\bibitem [{\citenamefont {Choi}\ \emph {et~al.}(2016)\citenamefont {Choi},
		\citenamefont {Hild}, \citenamefont {Zeiher}, \citenamefont {Schau{\ss}},
		\citenamefont {Rubio-Abadal}, \citenamefont {Yefsah}, \citenamefont
		{Khemani}, \citenamefont {Huse}, \citenamefont {Bloch},\ and\ \citenamefont
		{Gross}}]{Choi16}%
	\BibitemOpen
	\bibfield  {author} {\bibinfo {author} {\bibfnamefont {J.-y.}\ \bibnamefont
			{Choi}}, \bibinfo {author} {\bibfnamefont {S.}~\bibnamefont {Hild}}, \bibinfo
		{author} {\bibfnamefont {J.}~\bibnamefont {Zeiher}}, \bibinfo {author}
		{\bibfnamefont {P.}~\bibnamefont {Schau{\ss}}}, \bibinfo {author}
		{\bibfnamefont {A.}~\bibnamefont {Rubio-Abadal}}, \bibinfo {author}
		{\bibfnamefont {T.}~\bibnamefont {Yefsah}}, \bibinfo {author} {\bibfnamefont
			{V.}~\bibnamefont {Khemani}}, \bibinfo {author} {\bibfnamefont {D.~A.}\
			\bibnamefont {Huse}}, \bibinfo {author} {\bibfnamefont {I.}~\bibnamefont
			{Bloch}},\ and\ \bibinfo {author} {\bibfnamefont {C.}~\bibnamefont {Gross}},\
	}\bibfield  {title} {\bibinfo {title} {Exploring the many-body localization
			transition in two dimensions},\ }\href
	{https://doi.org/10.1126/science.aaf8834} {\bibfield  {journal} {\bibinfo
			{journal} {Science}\ }\textbf {\bibinfo {volume} {352}},\ \bibinfo {pages}
		{1547} (\bibinfo {year} {2016})}\BibitemShut {NoStop}%
	\bibitem [{\citenamefont {L\"uschen}\ \emph {et~al.}(2017)\citenamefont
		{L\"uschen}, \citenamefont {Bordia}, \citenamefont {Scherg}, \citenamefont
		{Alet}, \citenamefont {Altman}, \citenamefont {Schneider},\ and\
		\citenamefont {Bloch}}]{Luschen17}%
	\BibitemOpen
	\bibfield  {author} {\bibinfo {author} {\bibfnamefont {H.~P.}\ \bibnamefont
			{L\"uschen}}, \bibinfo {author} {\bibfnamefont {P.}~\bibnamefont {Bordia}},
		\bibinfo {author} {\bibfnamefont {S.}~\bibnamefont {Scherg}}, \bibinfo
		{author} {\bibfnamefont {F.}~\bibnamefont {Alet}}, \bibinfo {author}
		{\bibfnamefont {E.}~\bibnamefont {Altman}}, \bibinfo {author} {\bibfnamefont
			{U.}~\bibnamefont {Schneider}},\ and\ \bibinfo {author} {\bibfnamefont
			{I.}~\bibnamefont {Bloch}},\ }\bibfield  {title} {\bibinfo {title}
		{Observation of slow dynamics near the many-body localization transition in
			one-dimensional quasiperiodic systems},\ }\href
	{https://doi.org/10.1103/PhysRevLett.119.260401} {\bibfield  {journal}
		{\bibinfo  {journal} {Phys. Rev. Lett.}\ }\textbf {\bibinfo {volume} {119}},\
		\bibinfo {pages} {260401} (\bibinfo {year} {2017})}\BibitemShut {NoStop}%
	\bibitem [{\citenamefont {L\"uschen}\ \emph {et~al.}(2018)\citenamefont
		{L\"uschen}, \citenamefont {Scherg}, \citenamefont {Kohlert}, \citenamefont
		{Schreiber}, \citenamefont {Bordia}, \citenamefont {Li}, \citenamefont
		{Das~Sarma},\ and\ \citenamefont {Bloch}}]{Luschen18}%
	\BibitemOpen
	\bibfield  {author} {\bibinfo {author} {\bibfnamefont {H.~P.}\ \bibnamefont
			{L\"uschen}}, \bibinfo {author} {\bibfnamefont {S.}~\bibnamefont {Scherg}},
		\bibinfo {author} {\bibfnamefont {T.}~\bibnamefont {Kohlert}}, \bibinfo
		{author} {\bibfnamefont {M.}~\bibnamefont {Schreiber}}, \bibinfo {author}
		{\bibfnamefont {P.}~\bibnamefont {Bordia}}, \bibinfo {author} {\bibfnamefont
			{X.}~\bibnamefont {Li}}, \bibinfo {author} {\bibfnamefont {S.}~\bibnamefont
			{Das~Sarma}},\ and\ \bibinfo {author} {\bibfnamefont {I.}~\bibnamefont
			{Bloch}},\ }\bibfield  {title} {\bibinfo {title} {Single-particle mobility
			edge in a one-dimensional quasiperiodic optical lattice},\ }\href
	{https://doi.org/10.1103/PhysRevLett.120.160404} {\bibfield  {journal}
		{\bibinfo  {journal} {Phys. Rev. Lett.}\ }\textbf {\bibinfo {volume} {120}},\
		\bibinfo {pages} {160404} (\bibinfo {year} {2018})}\BibitemShut {NoStop}%
	\bibitem [{\citenamefont {Lukin}\ \emph {et~al.}(2019)\citenamefont {Lukin},
		\citenamefont {Rispoli}, \citenamefont {Schittko}, \citenamefont {Tai},
		\citenamefont {Kaufman}, \citenamefont {Choi}, \citenamefont {Khemani},
		\citenamefont {L{\'e}onard},\ and\ \citenamefont {Greiner}}]{Lukin19}%
	\BibitemOpen
	\bibfield  {author} {\bibinfo {author} {\bibfnamefont {A.}~\bibnamefont
			{Lukin}}, \bibinfo {author} {\bibfnamefont {M.}~\bibnamefont {Rispoli}},
		\bibinfo {author} {\bibfnamefont {R.}~\bibnamefont {Schittko}}, \bibinfo
		{author} {\bibfnamefont {M.~E.}\ \bibnamefont {Tai}}, \bibinfo {author}
		{\bibfnamefont {A.~M.}\ \bibnamefont {Kaufman}}, \bibinfo {author}
		{\bibfnamefont {S.}~\bibnamefont {Choi}}, \bibinfo {author} {\bibfnamefont
			{V.}~\bibnamefont {Khemani}}, \bibinfo {author} {\bibfnamefont
			{J.}~\bibnamefont {L{\'e}onard}},\ and\ \bibinfo {author} {\bibfnamefont
			{M.}~\bibnamefont {Greiner}},\ }\bibfield  {title} {\bibinfo {title} {Probing
			entanglement in a many-body{\textendash}localized system},\ }\href
	{https://doi.org/10.1126/science.aau0818} {\bibfield  {journal} {\bibinfo
			{journal} {Science}\ }\textbf {\bibinfo {volume} {364}},\ \bibinfo {pages}
		{256} (\bibinfo {year} {2019})}\BibitemShut {NoStop}%
	\bibitem [{\citenamefont {Rispoli}\ \emph {et~al.}(2019)\citenamefont
		{Rispoli}, \citenamefont {Lukin}, \citenamefont {Schittko}, \citenamefont
		{Kim}, \citenamefont {Tai}, \citenamefont {L{\'e}onard},\ and\ \citenamefont
		{Greiner}}]{Rispoli19}%
	\BibitemOpen
	\bibfield  {author} {\bibinfo {author} {\bibfnamefont {M.}~\bibnamefont
			{Rispoli}}, \bibinfo {author} {\bibfnamefont {A.}~\bibnamefont {Lukin}},
		\bibinfo {author} {\bibfnamefont {R.}~\bibnamefont {Schittko}}, \bibinfo
		{author} {\bibfnamefont {S.}~\bibnamefont {Kim}}, \bibinfo {author}
		{\bibfnamefont {M.~E.}\ \bibnamefont {Tai}}, \bibinfo {author} {\bibfnamefont
			{J.}~\bibnamefont {L{\'e}onard}},\ and\ \bibinfo {author} {\bibfnamefont
			{M.}~\bibnamefont {Greiner}},\ }\bibfield  {title} {\bibinfo {title} {Quantum
			critical behaviour at the many-body localization transition},\ }\href
	{https://doi.org/10.1038/s41586-019-1527-2} {\bibfield  {journal} {\bibinfo
			{journal} {Nature}\ }\textbf {\bibinfo {volume} {573}},\ \bibinfo {pages}
		{385} (\bibinfo {year} {2019})}\BibitemShut {NoStop}%
	\bibitem [{\citenamefont {Smith}\ \emph {et~al.}(2016)\citenamefont {Smith},
		\citenamefont {Lee}, \citenamefont {Richerme}, \citenamefont {Neyenhuis},
		\citenamefont {Hess}, \citenamefont {Hauke}, \citenamefont {Heyl},
		\citenamefont {Huse},\ and\ \citenamefont {Monroe}}]{Smith16}%
	\BibitemOpen
	\bibfield  {author} {\bibinfo {author} {\bibfnamefont {J.}~\bibnamefont
			{Smith}}, \bibinfo {author} {\bibfnamefont {A.}~\bibnamefont {Lee}}, \bibinfo
		{author} {\bibfnamefont {P.}~\bibnamefont {Richerme}}, \bibinfo {author}
		{\bibfnamefont {B.}~\bibnamefont {Neyenhuis}}, \bibinfo {author}
		{\bibfnamefont {P.~W.}\ \bibnamefont {Hess}}, \bibinfo {author}
		{\bibfnamefont {P.}~\bibnamefont {Hauke}}, \bibinfo {author} {\bibfnamefont
			{M.}~\bibnamefont {Heyl}}, \bibinfo {author} {\bibfnamefont {D.~A.}\
			\bibnamefont {Huse}},\ and\ \bibinfo {author} {\bibfnamefont
			{C.}~\bibnamefont {Monroe}},\ }\bibfield  {title} {\bibinfo {title}
		{Many-body localization in a quantum simulator with programmable random
			disorder},\ }\href {https://doi.org/10.1038/nphys3783} {\bibfield  {journal}
		{\bibinfo  {journal} {Nature Physics}\ }\textbf {\bibinfo {volume} {12}},\
		\bibinfo {pages} {907} (\bibinfo {year} {2016})}\BibitemShut {NoStop}%
	\bibitem [{\citenamefont {Roushan}\ \emph {et~al.}(2017)\citenamefont
		{Roushan}, \citenamefont {Neill}, \citenamefont {Tangpanitanon},
		\citenamefont {Bastidas}, \citenamefont {Megrant}, \citenamefont {Barends},
		\citenamefont {Chen}, \citenamefont {Chen}, \citenamefont {Chiaro},
		\citenamefont {Dunsworth}, \citenamefont {Fowler}, \citenamefont {Foxen},
		\citenamefont {Giustina}, \citenamefont {Jeffrey}, \citenamefont {Kelly},
		\citenamefont {Lucero}, \citenamefont {Mutus}, \citenamefont {Neeley},
		\citenamefont {Quintana}, \citenamefont {Sank}, \citenamefont {Vainsencher},
		\citenamefont {Wenner}, \citenamefont {White}, \citenamefont {Neven},
		\citenamefont {Angelakis},\ and\ \citenamefont {Martinis}}]{Roushan17}%
	\BibitemOpen
	\bibfield  {author} {\bibinfo {author} {\bibfnamefont {P.}~\bibnamefont
			{Roushan}}, \bibinfo {author} {\bibfnamefont {C.}~\bibnamefont {Neill}},
		\bibinfo {author} {\bibfnamefont {J.}~\bibnamefont {Tangpanitanon}}, \bibinfo
		{author} {\bibfnamefont {V.~M.}\ \bibnamefont {Bastidas}}, \bibinfo {author}
		{\bibfnamefont {A.}~\bibnamefont {Megrant}}, \bibinfo {author} {\bibfnamefont
			{R.}~\bibnamefont {Barends}}, \bibinfo {author} {\bibfnamefont
			{Y.}~\bibnamefont {Chen}}, \bibinfo {author} {\bibfnamefont {Z.}~\bibnamefont
			{Chen}}, \bibinfo {author} {\bibfnamefont {B.}~\bibnamefont {Chiaro}},
		\bibinfo {author} {\bibfnamefont {A.}~\bibnamefont {Dunsworth}}, \bibinfo
		{author} {\bibfnamefont {A.}~\bibnamefont {Fowler}}, \bibinfo {author}
		{\bibfnamefont {B.}~\bibnamefont {Foxen}}, \bibinfo {author} {\bibfnamefont
			{M.}~\bibnamefont {Giustina}}, \bibinfo {author} {\bibfnamefont
			{E.}~\bibnamefont {Jeffrey}}, \bibinfo {author} {\bibfnamefont
			{J.}~\bibnamefont {Kelly}}, \bibinfo {author} {\bibfnamefont
			{E.}~\bibnamefont {Lucero}}, \bibinfo {author} {\bibfnamefont
			{J.}~\bibnamefont {Mutus}}, \bibinfo {author} {\bibfnamefont
			{M.}~\bibnamefont {Neeley}}, \bibinfo {author} {\bibfnamefont
			{C.}~\bibnamefont {Quintana}}, \bibinfo {author} {\bibfnamefont
			{D.}~\bibnamefont {Sank}}, \bibinfo {author} {\bibfnamefont {A.}~\bibnamefont
			{Vainsencher}}, \bibinfo {author} {\bibfnamefont {J.}~\bibnamefont {Wenner}},
		\bibinfo {author} {\bibfnamefont {T.}~\bibnamefont {White}}, \bibinfo
		{author} {\bibfnamefont {H.}~\bibnamefont {Neven}}, \bibinfo {author}
		{\bibfnamefont {D.~G.}\ \bibnamefont {Angelakis}},\ and\ \bibinfo {author}
		{\bibfnamefont {J.}~\bibnamefont {Martinis}},\ }\bibfield  {title} {\bibinfo
		{title} {Spectroscopic signatures of localization with interacting photons in
			superconducting qubits},\ }\href {https://doi.org/10.1126/science.aao1401}
	{\bibfield  {journal} {\bibinfo  {journal} {Science}\ }\textbf {\bibinfo
			{volume} {358}},\ \bibinfo {pages} {1175} (\bibinfo {year}
		{2017})}\BibitemShut {NoStop}%
	\bibitem [{\citenamefont {Guo}\ \emph {et~al.}(2020)\citenamefont {Guo},
		\citenamefont {Tran}, \citenamefont {Childs}, \citenamefont {Gorshkov},\ and\
		\citenamefont {Gong}}]{Guo20}%
	\BibitemOpen
	\bibfield  {author} {\bibinfo {author} {\bibfnamefont {A.~Y.}\ \bibnamefont
			{Guo}}, \bibinfo {author} {\bibfnamefont {M.~C.}\ \bibnamefont {Tran}},
		\bibinfo {author} {\bibfnamefont {A.~M.}\ \bibnamefont {Childs}}, \bibinfo
		{author} {\bibfnamefont {A.~V.}\ \bibnamefont {Gorshkov}},\ and\ \bibinfo
		{author} {\bibfnamefont {Z.-X.}\ \bibnamefont {Gong}},\ }\bibfield  {title}
	{\bibinfo {title} {Signaling and scrambling with strongly long-range
			interactions},\ }\href {https://doi.org/10.1103/PhysRevA.102.010401}
	{\bibfield  {journal} {\bibinfo  {journal} {Phys. Rev. A}\ }\textbf {\bibinfo
			{volume} {102}},\ \bibinfo {pages} {010401} (\bibinfo {year}
		{2020})}\BibitemShut {NoStop}%
	\bibitem [{\citenamefont {Guo}\ \emph {et~al.}(2021)\citenamefont {Guo},
		\citenamefont {Cheng}, \citenamefont {Li}, \citenamefont {Xu}, \citenamefont
		{Zhang}, \citenamefont {Wang}, \citenamefont {Song}, \citenamefont {Liu},
		\citenamefont {Ren}, \citenamefont {Dong}, \citenamefont {Mondaini},\ and\
		\citenamefont {Wang}}]{Guo21}%
	\BibitemOpen
	\bibfield  {author} {\bibinfo {author} {\bibfnamefont {Q.}~\bibnamefont
			{Guo}}, \bibinfo {author} {\bibfnamefont {C.}~\bibnamefont {Cheng}}, \bibinfo
		{author} {\bibfnamefont {H.}~\bibnamefont {Li}}, \bibinfo {author}
		{\bibfnamefont {S.}~\bibnamefont {Xu}}, \bibinfo {author} {\bibfnamefont
			{P.}~\bibnamefont {Zhang}}, \bibinfo {author} {\bibfnamefont
			{Z.}~\bibnamefont {Wang}}, \bibinfo {author} {\bibfnamefont {C.}~\bibnamefont
			{Song}}, \bibinfo {author} {\bibfnamefont {W.}~\bibnamefont {Liu}}, \bibinfo
		{author} {\bibfnamefont {W.}~\bibnamefont {Ren}}, \bibinfo {author}
		{\bibfnamefont {H.}~\bibnamefont {Dong}}, \bibinfo {author} {\bibfnamefont
			{R.}~\bibnamefont {Mondaini}},\ and\ \bibinfo {author} {\bibfnamefont
			{H.}~\bibnamefont {Wang}},\ }\bibfield  {title} {\bibinfo {title} {Stark
			many-body localization on a superconducting quantum processor},\ }\href
	{https://doi.org/10.1103/PhysRevLett.127.240502} {\bibfield  {journal}
		{\bibinfo  {journal} {Phys. Rev. Lett.}\ }\textbf {\bibinfo {volume} {127}},\
		\bibinfo {pages} {240502} (\bibinfo {year} {2021})}\BibitemShut {NoStop}%
	\bibitem [{\citenamefont {Morong}\ \emph {et~al.}(2021)\citenamefont {Morong},
		\citenamefont {Liu}, \citenamefont {Becker}, \citenamefont {Collins},
		\citenamefont {Feng}, \citenamefont {Kyprianidis}, \citenamefont {Pagano},
		\citenamefont {You}, \citenamefont {Gorshkov},\ and\ \citenamefont
		{Monroe}}]{Morong21}%
	\BibitemOpen
	\bibfield  {author} {\bibinfo {author} {\bibfnamefont {W.}~\bibnamefont
			{Morong}}, \bibinfo {author} {\bibfnamefont {F.}~\bibnamefont {Liu}},
		\bibinfo {author} {\bibfnamefont {P.}~\bibnamefont {Becker}}, \bibinfo
		{author} {\bibfnamefont {K.~S.}\ \bibnamefont {Collins}}, \bibinfo {author}
		{\bibfnamefont {L.}~\bibnamefont {Feng}}, \bibinfo {author} {\bibfnamefont
			{A.}~\bibnamefont {Kyprianidis}}, \bibinfo {author} {\bibfnamefont
			{G.}~\bibnamefont {Pagano}}, \bibinfo {author} {\bibfnamefont
			{T.}~\bibnamefont {You}}, \bibinfo {author} {\bibfnamefont {A.~V.}\
			\bibnamefont {Gorshkov}},\ and\ \bibinfo {author} {\bibfnamefont
			{C.}~\bibnamefont {Monroe}},\ }\bibfield  {title} {\bibinfo {title}
		{{Observation of Stark many-body localization without disorder}},\ }\href
	{https://doi.org/10.1038/s41586-021-03988-0} {\bibfield  {journal} {\bibinfo
			{journal} {Nature}\ }\textbf {\bibinfo {volume} {599}},\ \bibinfo {pages}
		{393} (\bibinfo {year} {2021})}\BibitemShut {NoStop}%
	\bibitem [{\citenamefont {Nandkishore}\ and\ \citenamefont
		{Huse}(2015)}]{Nandkishore15}%
	\BibitemOpen
	\bibfield  {author} {\bibinfo {author} {\bibfnamefont {R.}~\bibnamefont
			{Nandkishore}}\ and\ \bibinfo {author} {\bibfnamefont {D.~A.}\ \bibnamefont
			{Huse}},\ }\bibfield  {title} {\bibinfo {title} {Mny-body-localization and
			thermalization in quantum statistical mechanics},\ }\href@noop {} {\bibfield
		{journal} {\bibinfo  {journal} {Ann. Rev. Cond. Mat. Phys.}\ }\textbf
		{\bibinfo {volume} {6}},\ \bibinfo {pages} {15} (\bibinfo {year}
		{2015})}\BibitemShut {NoStop}%
	\bibitem [{\citenamefont {Alet}\ and\ \citenamefont
		{Laflorencie}(2018)}]{Alet18}%
	\BibitemOpen
	\bibfield  {author} {\bibinfo {author} {\bibfnamefont {F.}~\bibnamefont
			{Alet}}\ and\ \bibinfo {author} {\bibfnamefont {N.}~\bibnamefont
			{Laflorencie}},\ }\bibfield  {title} {\bibinfo {title} {Many-body
			localization: An introduction and selected topics},\ }\href
	{https://doi.org/https://doi.org/10.1016/j.crhy.2018.03.003} {\bibfield
		{journal} {\bibinfo  {journal} {Comptes Rendus Physique}\ }\textbf {\bibinfo
			{volume} {19}},\ \bibinfo {pages} {498 } (\bibinfo {year}
		{2018})}\BibitemShut {NoStop}%
	\bibitem [{\citenamefont {Abanin}\ \emph {et~al.}(2019)\citenamefont {Abanin},
		\citenamefont {Altman}, \citenamefont {Bloch},\ and\ \citenamefont
		{Serbyn}}]{Abanin19}%
	\BibitemOpen
	\bibfield  {author} {\bibinfo {author} {\bibfnamefont {D.~A.}\ \bibnamefont
			{Abanin}}, \bibinfo {author} {\bibfnamefont {E.}~\bibnamefont {Altman}},
		\bibinfo {author} {\bibfnamefont {I.}~\bibnamefont {Bloch}},\ and\ \bibinfo
		{author} {\bibfnamefont {M.}~\bibnamefont {Serbyn}},\ }\bibfield  {title}
	{\bibinfo {title} {Colloquium: Many-body localization, thermalization, and
			entanglement},\ }\href {https://doi.org/10.1103/RevModPhys.91.021001}
	{\bibfield  {journal} {\bibinfo  {journal} {Rev. Mod. Phys.}\ }\textbf
		{\bibinfo {volume} {91}},\ \bibinfo {pages} {021001} (\bibinfo {year}
		{2019})}\BibitemShut {NoStop}%
	\bibitem [{\citenamefont {Huse}\ \emph {et~al.}(2014)\citenamefont {Huse},
		\citenamefont {Nandkishore},\ and\ \citenamefont {Oganesyan}}]{Huse14}%
	\BibitemOpen
	\bibfield  {author} {\bibinfo {author} {\bibfnamefont {D.~A.}\ \bibnamefont
			{Huse}}, \bibinfo {author} {\bibfnamefont {R.}~\bibnamefont {Nandkishore}},\
		and\ \bibinfo {author} {\bibfnamefont {V.}~\bibnamefont {Oganesyan}},\
	}\bibfield  {title} {\bibinfo {title} {Phenomenology of fully
			many-body-localized systems},\ }\href
	{https://doi.org/10.1103/PhysRevB.90.174202} {\bibfield  {journal} {\bibinfo
			{journal} {Phys. Rev. B}\ }\textbf {\bibinfo {volume} {90}},\ \bibinfo
		{pages} {174202} (\bibinfo {year} {2014})}\BibitemShut {NoStop}%
	\bibitem [{\citenamefont {Ros}\ \emph {et~al.}(2015)\citenamefont {Ros},
		\citenamefont {Mueller},\ and\ \citenamefont {Scardicchio}}]{Ros15}%
	\BibitemOpen
	\bibfield  {author} {\bibinfo {author} {\bibfnamefont {V.}~\bibnamefont
			{Ros}}, \bibinfo {author} {\bibfnamefont {M.}~\bibnamefont {Mueller}},\ and\
		\bibinfo {author} {\bibfnamefont {A.}~\bibnamefont {Scardicchio}},\
	}\bibfield  {title} {\bibinfo {title} {Integrals of motion in the many-body
			localized phase},\ }\href
	{https://doi.org/https://doi.org/10.1016/j.nuclphysb.2014.12.014} {\bibfield
		{journal} {\bibinfo  {journal} {Nuclear Physics B}\ }\textbf {\bibinfo
			{volume} {891}},\ \bibinfo {pages} {420 } (\bibinfo {year}
		{2015})}\BibitemShut {NoStop}%
	\bibitem [{\citenamefont {Serbyn}\ \emph
		{et~al.}(2013{\natexlab{a}})\citenamefont {Serbyn}, \citenamefont
		{Papi\'{c}},\ and\ \citenamefont {Abanin}}]{Serbyn13b}%
	\BibitemOpen
	\bibfield  {author} {\bibinfo {author} {\bibfnamefont {M.}~\bibnamefont
			{Serbyn}}, \bibinfo {author} {\bibfnamefont {Z.}~\bibnamefont {Papi\'{c}}},\
		and\ \bibinfo {author} {\bibfnamefont {D.~A.}\ \bibnamefont {Abanin}},\
	}\bibfield  {title} {\bibinfo {title} {Local conservation laws and the
			structure of the many-body localized states},\ }\href
	{https://doi.org/10.1103/PhysRevLett.111.127201} {\bibfield  {journal}
		{\bibinfo  {journal} {Phys. Rev. Lett.}\ }\textbf {\bibinfo {volume} {111}},\
		\bibinfo {pages} {127201} (\bibinfo {year} {2013}{\natexlab{a}})}\BibitemShut
	{NoStop}%
	\bibitem [{\citenamefont {Imbrie}(2016)}]{Imbrie16}%
	\BibitemOpen
	\bibfield  {author} {\bibinfo {author} {\bibfnamefont {J.~Z.}\ \bibnamefont
			{Imbrie}},\ }\bibfield  {title} {\bibinfo {title} {Diagonalization and
			many-body localization for a disordered quantum spin chain},\ }\href
	{https://doi.org/10.1103/PhysRevLett.117.027201} {\bibfield  {journal}
		{\bibinfo  {journal} {Phys. Rev. Lett.}\ }\textbf {\bibinfo {volume} {117}},\
		\bibinfo {pages} {027201} (\bibinfo {year} {2016})}\BibitemShut {NoStop}%
	\bibitem [{\citenamefont {Wahl}\ \emph {et~al.}(2017)\citenamefont {Wahl},
		\citenamefont {Pal},\ and\ \citenamefont {Simon}}]{Wahl17}%
	\BibitemOpen
	\bibfield  {author} {\bibinfo {author} {\bibfnamefont {T.~B.}\ \bibnamefont
			{Wahl}}, \bibinfo {author} {\bibfnamefont {A.}~\bibnamefont {Pal}},\ and\
		\bibinfo {author} {\bibfnamefont {S.~H.}\ \bibnamefont {Simon}},\ }\bibfield
	{title} {\bibinfo {title} {Efficient representation of fully many-body
			localized systems using tensor networks},\ }\href
	{https://doi.org/10.1103/PhysRevX.7.021018} {\bibfield  {journal} {\bibinfo
			{journal} {Phys. Rev. X}\ }\textbf {\bibinfo {volume} {7}},\ \bibinfo {pages}
		{021018} (\bibinfo {year} {2017})}\BibitemShut {NoStop}%
	\bibitem [{\citenamefont {Mierzejewski}\ \emph {et~al.}(2018)\citenamefont
		{Mierzejewski}, \citenamefont {Kozarzewski},\ and\ \citenamefont
		{Prelov\v{s}ek}}]{Mierzejewski18}%
	\BibitemOpen
	\bibfield  {author} {\bibinfo {author} {\bibfnamefont {M.}~\bibnamefont
			{Mierzejewski}}, \bibinfo {author} {\bibfnamefont {M.}~\bibnamefont
			{Kozarzewski}},\ and\ \bibinfo {author} {\bibfnamefont {P.}~\bibnamefont
			{Prelov\v{s}ek}},\ }\bibfield  {title} {\bibinfo {title} {Counting local
			integrals of motion in disordered spinless-fermion and {H}ubbard chains},\
	}\href {https://doi.org/10.1103/PhysRevB.97.064204} {\bibfield  {journal}
		{\bibinfo  {journal} {Phys. Rev. B}\ }\textbf {\bibinfo {volume} {97}},\
		\bibinfo {pages} {064204} (\bibinfo {year} {2018})}\BibitemShut {NoStop}%
	\bibitem [{\citenamefont {Thomson}\ and\ \citenamefont
		{Schir\'o}(2018)}]{Thomson18}%
	\BibitemOpen
	\bibfield  {author} {\bibinfo {author} {\bibfnamefont {S.~J.}\ \bibnamefont
			{Thomson}}\ and\ \bibinfo {author} {\bibfnamefont {M.}~\bibnamefont
			{Schir\'o}},\ }\bibfield  {title} {\bibinfo {title} {Time evolution of
			many-body localized systems with the flow equation approach},\ }\href
	{https://doi.org/10.1103/PhysRevB.97.060201} {\bibfield  {journal} {\bibinfo
			{journal} {Phys. Rev. B}\ }\textbf {\bibinfo {volume} {97}},\ \bibinfo
		{pages} {060201} (\bibinfo {year} {2018})}\BibitemShut {NoStop}%
	\bibitem [{\citenamefont {\v{Z}nidari\v{c}}\ \emph {et~al.}(2016)\citenamefont
		{\v{Z}nidari\v{c}}, \citenamefont {Scardicchio},\ and\ \citenamefont
		{Varma}}]{Znidaric16}%
	\BibitemOpen
	\bibfield  {author} {\bibinfo {author} {\bibfnamefont {M.}~\bibnamefont
			{\v{Z}nidari\v{c}}}, \bibinfo {author} {\bibfnamefont {A.}~\bibnamefont
			{Scardicchio}},\ and\ \bibinfo {author} {\bibfnamefont {V.~K.}\ \bibnamefont
			{Varma}},\ }\bibfield  {title} {\bibinfo {title} {Diffusive and subdiffusive
			spin transport in the ergodic phase of a many-body localizable system},\
	}\href {https://doi.org/10.1103/PhysRevLett.117.040601} {\bibfield  {journal}
		{\bibinfo  {journal} {Phys. Rev. Lett.}\ }\textbf {\bibinfo {volume} {117}},\
		\bibinfo {pages} {040601} (\bibinfo {year} {2016})}\BibitemShut {NoStop}%
	\bibitem [{\citenamefont {Bardarson}\ \emph {et~al.}(2012)\citenamefont
		{Bardarson}, \citenamefont {Pollmann},\ and\ \citenamefont
		{Moore}}]{Bardarson12}%
	\BibitemOpen
	\bibfield  {author} {\bibinfo {author} {\bibfnamefont {J.~H.}\ \bibnamefont
			{Bardarson}}, \bibinfo {author} {\bibfnamefont {F.}~\bibnamefont
			{Pollmann}},\ and\ \bibinfo {author} {\bibfnamefont {J.~E.}\ \bibnamefont
			{Moore}},\ }\bibfield  {title} {\bibinfo {title} {Unbounded growth of
			entanglement in models of many-body localization},\ }\href
	{https://doi.org/10.1103/PhysRevLett.109.017202} {\bibfield  {journal}
		{\bibinfo  {journal} {Phys. Rev. Lett.}\ }\textbf {\bibinfo {volume} {109}},\
		\bibinfo {pages} {017202} (\bibinfo {year} {2012})}\BibitemShut {NoStop}%
	\bibitem [{\citenamefont {Serbyn}\ \emph
		{et~al.}(2013{\natexlab{b}})\citenamefont {Serbyn}, \citenamefont
		{Papi{\'c}},\ and\ \citenamefont {Abanin}}]{serbyn2013universal}%
	\BibitemOpen
	\bibfield  {author} {\bibinfo {author} {\bibfnamefont {M.}~\bibnamefont
			{Serbyn}}, \bibinfo {author} {\bibfnamefont {Z.}~\bibnamefont {Papi{\'c}}},\
		and\ \bibinfo {author} {\bibfnamefont {D.~A.}\ \bibnamefont {Abanin}},\
	}\bibfield  {title} {\bibinfo {title} {Universal slow growth of entanglement
			in interacting strongly disordered systems},\ }\href@noop {} {\bibfield
		{journal} {\bibinfo  {journal} {Physical review letters}\ }\textbf {\bibinfo
			{volume} {110}},\ \bibinfo {pages} {260601} (\bibinfo {year}
		{2013}{\natexlab{b}})}\BibitemShut {NoStop}%
	\bibitem [{\citenamefont {Iemini}\ \emph {et~al.}(2016)\citenamefont {Iemini},
		\citenamefont {Russomanno}, \citenamefont {Rossini}, \citenamefont
		{Scardicchio},\ and\ \citenamefont {Fazio}}]{iemini2016signatures}%
	\BibitemOpen
	\bibfield  {author} {\bibinfo {author} {\bibfnamefont {F.}~\bibnamefont
			{Iemini}}, \bibinfo {author} {\bibfnamefont {A.}~\bibnamefont {Russomanno}},
		\bibinfo {author} {\bibfnamefont {D.}~\bibnamefont {Rossini}}, \bibinfo
		{author} {\bibfnamefont {A.}~\bibnamefont {Scardicchio}},\ and\ \bibinfo
		{author} {\bibfnamefont {R.}~\bibnamefont {Fazio}},\ }\bibfield  {title}
	{\bibinfo {title} {Signatures of many-body localization in the dynamics of
			two-site entanglement},\ }\href {https://doi.org/10.1103/PhysRevB.94.214206}
	{\bibfield  {journal} {\bibinfo  {journal} {Phys. Rev. B}\ }\textbf {\bibinfo
			{volume} {94}},\ \bibinfo {pages} {214206} (\bibinfo {year}
		{2016})}\BibitemShut {NoStop}%
	\bibitem [{\citenamefont {Doggen}\ \emph {et~al.}(2018)\citenamefont {Doggen},
		\citenamefont {Schindler}, \citenamefont {Tikhonov}, \citenamefont {Mirlin},
		\citenamefont {Neupert}, \citenamefont {Polyakov},\ and\ \citenamefont
		{Gornyi}}]{Doggen18}%
	\BibitemOpen
	\bibfield  {author} {\bibinfo {author} {\bibfnamefont {E.~V.~H.}\
			\bibnamefont {Doggen}}, \bibinfo {author} {\bibfnamefont {F.}~\bibnamefont
			{Schindler}}, \bibinfo {author} {\bibfnamefont {K.~S.}\ \bibnamefont
			{Tikhonov}}, \bibinfo {author} {\bibfnamefont {A.~D.}\ \bibnamefont
			{Mirlin}}, \bibinfo {author} {\bibfnamefont {T.}~\bibnamefont {Neupert}},
		\bibinfo {author} {\bibfnamefont {D.~G.}\ \bibnamefont {Polyakov}},\ and\
		\bibinfo {author} {\bibfnamefont {I.~V.}\ \bibnamefont {Gornyi}},\ }\bibfield
	{title} {\bibinfo {title} {Many-body localization and delocalization in
			large quantum chains},\ }\href {https://doi.org/10.1103/PhysRevB.98.174202}
	{\bibfield  {journal} {\bibinfo  {journal} {Phys. Rev. B}\ }\textbf {\bibinfo
			{volume} {98}},\ \bibinfo {pages} {174202} (\bibinfo {year}
		{2018})}\BibitemShut {NoStop}%
	\bibitem [{\citenamefont {Chanda}\ \emph
		{et~al.}(2020{\natexlab{a}})\citenamefont {Chanda}, \citenamefont {Sierant},\
		and\ \citenamefont {Zakrzewski}}]{Chanda19}%
	\BibitemOpen
	\bibfield  {author} {\bibinfo {author} {\bibfnamefont {T.}~\bibnamefont
			{Chanda}}, \bibinfo {author} {\bibfnamefont {P.}~\bibnamefont {Sierant}},\
		and\ \bibinfo {author} {\bibfnamefont {J.}~\bibnamefont {Zakrzewski}},\
	}\bibfield  {title} {\bibinfo {title} {Time dynamics with matrix product
			states: Many-body localization transition of large systems revisited},\
	}\href {https://doi.org/10.1103/PhysRevB.101.035148} {\bibfield  {journal}
		{\bibinfo  {journal} {Phys. Rev. B}\ }\textbf {\bibinfo {volume} {101}},\
		\bibinfo {pages} {035148} (\bibinfo {year} {2020}{\natexlab{a}})}\BibitemShut
	{NoStop}%
	\bibitem [{\citenamefont {Doggen}\ and\ \citenamefont
		{Mirlin}(2019)}]{Doggen19}%
	\BibitemOpen
	\bibfield  {author} {\bibinfo {author} {\bibfnamefont {E.~V.~H.}\
			\bibnamefont {Doggen}}\ and\ \bibinfo {author} {\bibfnamefont {A.~D.}\
			\bibnamefont {Mirlin}},\ }\bibfield  {title} {\bibinfo {title} {Many-body
			delocalization dynamics in long aubry-andr\'e quasiperiodic chains},\ }\href
	{https://doi.org/10.1103/PhysRevB.100.104203} {\bibfield  {journal} {\bibinfo
			{journal} {Phys. Rev. B}\ }\textbf {\bibinfo {volume} {100}},\ \bibinfo
		{pages} {104203} (\bibinfo {year} {2019})}\BibitemShut {NoStop}%
	\bibitem [{\citenamefont {Chanda}\ \emph
		{et~al.}(2020{\natexlab{b}})\citenamefont {Chanda}, \citenamefont {Sierant},\
		and\ \citenamefont {Zakrzewski}}]{Chanda20m}%
	\BibitemOpen
	\bibfield  {author} {\bibinfo {author} {\bibfnamefont {T.}~\bibnamefont
			{Chanda}}, \bibinfo {author} {\bibfnamefont {P.}~\bibnamefont {Sierant}},\
		and\ \bibinfo {author} {\bibfnamefont {J.}~\bibnamefont {Zakrzewski}},\
	}\bibfield  {title} {\bibinfo {title} {Many-body localization transition in
			large quantum spin chains: The mobility edge},\ }\href
	{https://doi.org/10.1103/PhysRevResearch.2.032045} {\bibfield  {journal}
		{\bibinfo  {journal} {Phys. Rev. Research}\ }\textbf {\bibinfo {volume}
			{2}},\ \bibinfo {pages} {032045} (\bibinfo {year}
		{2020}{\natexlab{b}})}\BibitemShut {NoStop}%
	\bibitem [{\citenamefont {Panda}\ \emph {et~al.}(2020)\citenamefont {Panda},
		\citenamefont {Scardicchio}, \citenamefont {Schulz}, \citenamefont {Taylor},\
		and\ \citenamefont {{\v{Z}}nidari{\v{c}}}}]{Panda20}%
	\BibitemOpen
	\bibfield  {author} {\bibinfo {author} {\bibfnamefont {R.~K.}\ \bibnamefont
			{Panda}}, \bibinfo {author} {\bibfnamefont {A.}~\bibnamefont {Scardicchio}},
		\bibinfo {author} {\bibfnamefont {M.}~\bibnamefont {Schulz}}, \bibinfo
		{author} {\bibfnamefont {S.~R.}\ \bibnamefont {Taylor}},\ and\ \bibinfo
		{author} {\bibfnamefont {M.}~\bibnamefont {{\v{Z}}nidari{\v{c}}}},\
	}\bibfield  {title} {\bibinfo {title} {Can we study the many-body
			localisation transition?},\ }\href
	{https://doi.org/10.1209/0295-5075/128/67003} {\bibfield  {journal} {\bibinfo
			{journal} {{EPL} (Europhysics Letters)}\ }\textbf {\bibinfo {volume}
			{128}},\ \bibinfo {pages} {67003} (\bibinfo {year} {2020})}\BibitemShut
	{NoStop}%
	\bibitem [{\citenamefont {Sierant}\ \emph
		{et~al.}(2020{\natexlab{a}})\citenamefont {Sierant}, \citenamefont
		{Delande},\ and\ \citenamefont {Zakrzewski}}]{Sierant20thouless}%
	\BibitemOpen
	\bibfield  {author} {\bibinfo {author} {\bibfnamefont {P.}~\bibnamefont
			{Sierant}}, \bibinfo {author} {\bibfnamefont {D.}~\bibnamefont {Delande}},\
		and\ \bibinfo {author} {\bibfnamefont {J.}~\bibnamefont {Zakrzewski}},\
	}\bibfield  {title} {\bibinfo {title} {Thouless time analysis of anderson and
			many-body localization transitions},\ }\href
	{https://doi.org/10.1103/PhysRevLett.124.186601} {\bibfield  {journal}
		{\bibinfo  {journal} {Phys. Rev. Lett.}\ }\textbf {\bibinfo {volume} {124}},\
		\bibinfo {pages} {186601} (\bibinfo {year} {2020}{\natexlab{a}})}\BibitemShut
	{NoStop}%
	\bibitem [{\citenamefont {Sierant}\ \emph
		{et~al.}(2020{\natexlab{b}})\citenamefont {Sierant}, \citenamefont
		{Lewenstein},\ and\ \citenamefont {Zakrzewski}}]{Sierant20polfed}%
	\BibitemOpen
	\bibfield  {author} {\bibinfo {author} {\bibfnamefont {P.}~\bibnamefont
			{Sierant}}, \bibinfo {author} {\bibfnamefont {M.}~\bibnamefont
			{Lewenstein}},\ and\ \bibinfo {author} {\bibfnamefont {J.}~\bibnamefont
			{Zakrzewski}},\ }\bibfield  {title} {\bibinfo {title} {Polynomially filtered
			exact diagonalization approach to many-body localization},\ }\href
	{https://doi.org/10.1103/PhysRevLett.125.156601} {\bibfield  {journal}
		{\bibinfo  {journal} {Phys. Rev. Lett.}\ }\textbf {\bibinfo {volume} {125}},\
		\bibinfo {pages} {156601} (\bibinfo {year} {2020}{\natexlab{b}})}\BibitemShut
	{NoStop}%
	\bibitem [{\citenamefont {Kiefer-Emmanouilidis}\ \emph
		{et~al.}(2020)\citenamefont {Kiefer-Emmanouilidis}, \citenamefont {Unanyan},
		\citenamefont {Fleischhauer},\ and\ \citenamefont {Sirker}}]{Kiefer20}%
	\BibitemOpen
	\bibfield  {author} {\bibinfo {author} {\bibfnamefont {M.}~\bibnamefont
			{Kiefer-Emmanouilidis}}, \bibinfo {author} {\bibfnamefont {R.}~\bibnamefont
			{Unanyan}}, \bibinfo {author} {\bibfnamefont {M.}~\bibnamefont
			{Fleischhauer}},\ and\ \bibinfo {author} {\bibfnamefont {J.}~\bibnamefont
			{Sirker}},\ }\bibfield  {title} {\bibinfo {title} {Evidence for unbounded
			growth of the number entropy in many-body localized phases},\ }\href
	{https://doi.org/10.1103/PhysRevLett.124.243601} {\bibfield  {journal}
		{\bibinfo  {journal} {Phys. Rev. Lett.}\ }\textbf {\bibinfo {volume} {124}},\
		\bibinfo {pages} {243601} (\bibinfo {year} {2020})}\BibitemShut {NoStop}%
	\bibitem [{\citenamefont {Kiefer-Emmanouilidis}\ \emph
		{et~al.}(2021)\citenamefont {Kiefer-Emmanouilidis}, \citenamefont {Unanyan},
		\citenamefont {Fleischhauer},\ and\ \citenamefont {Sirker}}]{Kiefer21}%
	\BibitemOpen
	\bibfield  {author} {\bibinfo {author} {\bibfnamefont {M.}~\bibnamefont
			{Kiefer-Emmanouilidis}}, \bibinfo {author} {\bibfnamefont {R.}~\bibnamefont
			{Unanyan}}, \bibinfo {author} {\bibfnamefont {M.}~\bibnamefont
			{Fleischhauer}},\ and\ \bibinfo {author} {\bibfnamefont {J.}~\bibnamefont
			{Sirker}},\ }\bibfield  {title} {\bibinfo {title} {Slow delocalization of
			particles in many-body localized phases},\ }\href
	{https://doi.org/10.1103/PhysRevB.103.024203} {\bibfield  {journal} {\bibinfo
			{journal} {Phys. Rev. B}\ }\textbf {\bibinfo {volume} {103}},\ \bibinfo
		{pages} {024203} (\bibinfo {year} {2021})}\BibitemShut {NoStop}%
	\bibitem [{\citenamefont {Abanin}\ \emph {et~al.}(2021)\citenamefont {Abanin},
		\citenamefont {Bardarson}, \citenamefont {De~Tomasi}, \citenamefont
		{Gopalakrishnan}, \citenamefont {Khemani}, \citenamefont {Parameswaran},
		\citenamefont {Pollmann}, \citenamefont {Potter}, \citenamefont {Serbyn},\
		and\ \citenamefont {Vasseur}}]{Abanin21}%
	\BibitemOpen
	\bibfield  {author} {\bibinfo {author} {\bibfnamefont {D.}~\bibnamefont
			{Abanin}}, \bibinfo {author} {\bibfnamefont {J.}~\bibnamefont {Bardarson}},
		\bibinfo {author} {\bibfnamefont {G.}~\bibnamefont {De~Tomasi}}, \bibinfo
		{author} {\bibfnamefont {S.}~\bibnamefont {Gopalakrishnan}}, \bibinfo
		{author} {\bibfnamefont {V.}~\bibnamefont {Khemani}}, \bibinfo {author}
		{\bibfnamefont {S.}~\bibnamefont {Parameswaran}}, \bibinfo {author}
		{\bibfnamefont {F.}~\bibnamefont {Pollmann}}, \bibinfo {author}
		{\bibfnamefont {A.}~\bibnamefont {Potter}}, \bibinfo {author} {\bibfnamefont
			{M.}~\bibnamefont {Serbyn}},\ and\ \bibinfo {author} {\bibfnamefont
			{R.}~\bibnamefont {Vasseur}},\ }\bibfield  {title} {\bibinfo {title}
		{Distinguishing localization from chaos: Challenges in finite-size systems},\
	}\href {https://doi.org/10.1016/j.aop.2021.168415} {\bibfield  {journal}
		{\bibinfo  {journal} {Annals of Physics}\ }\textbf {\bibinfo {volume}
			{427}},\ \bibinfo {pages} {168415} (\bibinfo {year} {2021})}\BibitemShut
	{NoStop}%
	\bibitem [{\citenamefont {Sels}\ and\ \citenamefont
		{Polkovnikov}(2021)}]{Sels20}%
	\BibitemOpen
	\bibfield  {author} {\bibinfo {author} {\bibfnamefont {D.}~\bibnamefont
			{Sels}}\ and\ \bibinfo {author} {\bibfnamefont {A.}~\bibnamefont
			{Polkovnikov}},\ }\bibfield  {title} {\bibinfo {title} {Dynamical obstruction
			to localization in a disordered spin chain},\ }\href
	{https://doi.org/10.1103/PhysRevE.104.054105} {\bibfield  {journal} {\bibinfo
			{journal} {Phys. Rev. E}\ }\textbf {\bibinfo {volume} {104}},\ \bibinfo
		{pages} {054105} (\bibinfo {year} {2021})}\BibitemShut {NoStop}%
	\bibitem [{\citenamefont {Krajewski}\ \emph {et~al.}(2022)\citenamefont
		{Krajewski}, \citenamefont {Vidmar}, \citenamefont
		{Bon\ifmmode~\check{c}\else \v{c}\fi{}a},\ and\ \citenamefont
		{Mierzejewski}}]{Krajewski22}%
	\BibitemOpen
	\bibfield  {author} {\bibinfo {author} {\bibfnamefont {B.}~\bibnamefont
			{Krajewski}}, \bibinfo {author} {\bibfnamefont {L.}~\bibnamefont {Vidmar}},
		\bibinfo {author} {\bibfnamefont {J.}~\bibnamefont
			{Bon\ifmmode~\check{c}\else \v{c}\fi{}a}},\ and\ \bibinfo {author}
		{\bibfnamefont {M.}~\bibnamefont {Mierzejewski}},\ }\bibfield  {title}
	{\bibinfo {title} {Restoring ergodicity in a strongly disordered interacting
			chain},\ }\href {https://doi.org/10.1103/PhysRevLett.129.260601} {\bibfield
		{journal} {\bibinfo  {journal} {Phys. Rev. Lett.}\ }\textbf {\bibinfo
			{volume} {129}},\ \bibinfo {pages} {260601} (\bibinfo {year}
		{2022})}\BibitemShut {NoStop}%
	\bibitem [{\citenamefont {{\v S}untajs}\ \emph
		{et~al.}(2020{\natexlab{a}})\citenamefont {{\v S}untajs}, \citenamefont
		{Bon{\v c}a}, \citenamefont {Prosen},\ and\ \citenamefont
		{Vidmar}}]{Suntajs20e}%
	\BibitemOpen
	\bibfield  {author} {\bibinfo {author} {\bibfnamefont {J.}~\bibnamefont {{\v
					S}untajs}}, \bibinfo {author} {\bibfnamefont {J.}~\bibnamefont {Bon{\v c}a}},
		\bibinfo {author} {\bibfnamefont {T.}~\bibnamefont {Prosen}},\ and\ \bibinfo
		{author} {\bibfnamefont {L.}~\bibnamefont {Vidmar}},\ }\bibfield  {title}
	{\bibinfo {title} {Quantum chaos challenges many-body localization},\ }\href
	{https://doi.org/10.1103/PhysRevE.102.062144} {\bibfield  {journal} {\bibinfo
			{journal} {Phys. Rev. E}\ }\textbf {\bibinfo {volume} {102}},\ \bibinfo
		{pages} {062144} (\bibinfo {year} {2020}{\natexlab{a}})}\BibitemShut
	{NoStop}%
	\bibitem [{\citenamefont {Sierant}\ and\ \citenamefont
		{Zakrzewski}(2022)}]{Sierant22challenges}%
	\BibitemOpen
	\bibfield  {author} {\bibinfo {author} {\bibfnamefont {P.}~\bibnamefont
			{Sierant}}\ and\ \bibinfo {author} {\bibfnamefont {J.}~\bibnamefont
			{Zakrzewski}},\ }\bibfield  {title} {\bibinfo {title} {Challenges to
			observation of many-body localization},\ }\href
	{https://doi.org/10.1103/PhysRevB.105.224203} {\bibfield  {journal} {\bibinfo
			{journal} {Phys. Rev. B}\ }\textbf {\bibinfo {volume} {105}},\ \bibinfo
		{pages} {224203} (\bibinfo {year} {2022})}\BibitemShut {NoStop}%
	\bibitem [{\citenamefont {Morningstar}\ \emph {et~al.}(2022)\citenamefont
		{Morningstar}, \citenamefont {Colmenarez}, \citenamefont {Khemani},
		\citenamefont {Luitz},\ and\ \citenamefont {Huse}}]{Morningstar22}%
	\BibitemOpen
	\bibfield  {author} {\bibinfo {author} {\bibfnamefont {A.}~\bibnamefont
			{Morningstar}}, \bibinfo {author} {\bibfnamefont {L.}~\bibnamefont
			{Colmenarez}}, \bibinfo {author} {\bibfnamefont {V.}~\bibnamefont {Khemani}},
		\bibinfo {author} {\bibfnamefont {D.~J.}\ \bibnamefont {Luitz}},\ and\
		\bibinfo {author} {\bibfnamefont {D.~A.}\ \bibnamefont {Huse}},\ }\bibfield
	{title} {\bibinfo {title} {Avalanches and many-body resonances in many-body
			localized systems},\ }\href {https://doi.org/10.1103/PhysRevB.105.174205}
	{\bibfield  {journal} {\bibinfo  {journal} {Phys. Rev. B}\ }\textbf {\bibinfo
			{volume} {105}},\ \bibinfo {pages} {174205} (\bibinfo {year}
		{2022})}\BibitemShut {NoStop}%
	\bibitem [{\citenamefont {Sierant}\ \emph {et~al.}(2021)\citenamefont
		{Sierant}, \citenamefont {Lazo}, \citenamefont {Dalmonte}, \citenamefont
		{Scardicchio},\ and\ \citenamefont {Zakrzewski}}]{Sierant21constrained}%
	\BibitemOpen
	\bibfield  {author} {\bibinfo {author} {\bibfnamefont {P.}~\bibnamefont
			{Sierant}}, \bibinfo {author} {\bibfnamefont {E.~G.}\ \bibnamefont {Lazo}},
		\bibinfo {author} {\bibfnamefont {M.}~\bibnamefont {Dalmonte}}, \bibinfo
		{author} {\bibfnamefont {A.}~\bibnamefont {Scardicchio}},\ and\ \bibinfo
		{author} {\bibfnamefont {J.}~\bibnamefont {Zakrzewski}},\ }\bibfield  {title}
	{\bibinfo {title} {Constraint-induced delocalization},\ }\href
	{https://doi.org/10.1103/PhysRevLett.127.126603} {\bibfield  {journal}
		{\bibinfo  {journal} {Phys. Rev. Lett.}\ }\textbf {\bibinfo {volume} {127}},\
		\bibinfo {pages} {126603} (\bibinfo {year} {2021})}\BibitemShut {NoStop}%
	\bibitem [{\citenamefont {{De Roeck}}\ \emph {et~al.}(2023)\citenamefont {{De
				Roeck}}, \citenamefont {Huveneers}, \citenamefont {Meeus},\ and\
		\citenamefont {Prośniak}}]{Deroeck23}%
	\BibitemOpen
	\bibfield  {author} {\bibinfo {author} {\bibfnamefont {W.}~\bibnamefont {{De
					Roeck}}}, \bibinfo {author} {\bibfnamefont {F.}~\bibnamefont {Huveneers}},
		\bibinfo {author} {\bibfnamefont {B.}~\bibnamefont {Meeus}},\ and\ \bibinfo
		{author} {\bibfnamefont {O.~A.}\ \bibnamefont {Prośniak}},\ }\bibfield
	{title} {\bibinfo {title} {Rigorous and simple results on very slow
			thermalization, or quasi-localization, of the disordered quantum chain},\
	}\href {https://doi.org/https://doi.org/10.1016/j.physa.2023.129245}
	{\bibfield  {journal} {\bibinfo  {journal} {Physica A: Statistical Mechanics
				and its Applications}\ }\textbf {\bibinfo {volume} {631}},\ \bibinfo {pages}
		{129245} (\bibinfo {year} {2023})}\BibitemShut {NoStop}%
	\bibitem [{\citenamefont {Sierant}\ \emph {et~al.}(2023)\citenamefont
		{Sierant}, \citenamefont {Lewenstein}, \citenamefont {Scardicchio},\ and\
		\citenamefont {Zakrzewski}}]{Sierant22f}%
	\BibitemOpen
	\bibfield  {author} {\bibinfo {author} {\bibfnamefont {P.}~\bibnamefont
			{Sierant}}, \bibinfo {author} {\bibfnamefont {M.}~\bibnamefont {Lewenstein}},
		\bibinfo {author} {\bibfnamefont {A.}~\bibnamefont {Scardicchio}},\ and\
		\bibinfo {author} {\bibfnamefont {J.}~\bibnamefont {Zakrzewski}},\ }\bibfield
	{title} {\bibinfo {title} {Stability of many-body localization in floquet
			systems},\ }\href {https://doi.org/10.1103/PhysRevB.107.115132} {\bibfield
		{journal} {\bibinfo  {journal} {Phys. Rev. B}\ }\textbf {\bibinfo {volume}
			{107}},\ \bibinfo {pages} {115132} (\bibinfo {year} {2023})}\BibitemShut
	{NoStop}%
	\bibitem [{\citenamefont {Krajewski}\ \emph {et~al.}(2023)\citenamefont
		{Krajewski}, \citenamefont {Vidmar}, \citenamefont
		{Bon\ifmmode~\check{c}\else \v{c}\fi{}a},\ and\ \citenamefont
		{Mierzejewski}}]{Krajewski23}%
	\BibitemOpen
	\bibfield  {author} {\bibinfo {author} {\bibfnamefont {B.}~\bibnamefont
			{Krajewski}}, \bibinfo {author} {\bibfnamefont {L.}~\bibnamefont {Vidmar}},
		\bibinfo {author} {\bibfnamefont {J.}~\bibnamefont
			{Bon\ifmmode~\check{c}\else \v{c}\fi{}a}},\ and\ \bibinfo {author}
		{\bibfnamefont {M.}~\bibnamefont {Mierzejewski}},\ }\bibfield  {title}
	{\bibinfo {title} {Strongly disordered anderson insulator chains with generic
			two-body interaction},\ }\href {https://doi.org/10.1103/PhysRevB.108.064203}
	{\bibfield  {journal} {\bibinfo  {journal} {Phys. Rev. B}\ }\textbf {\bibinfo
			{volume} {108}},\ \bibinfo {pages} {064203} (\bibinfo {year}
		{2023})}\BibitemShut {NoStop}%
	\bibitem [{\citenamefont {Vojta}(2010)}]{Vojta10}%
	\BibitemOpen
	\bibfield  {author} {\bibinfo {author} {\bibfnamefont {T.}~\bibnamefont
			{Vojta}},\ }\bibfield  {title} {\bibinfo {title} {Quantum {G}riffiths effects
			and smeared phase transitions in metals: Theory and experiment},\ }\href
	{https://doi.org/10.1007/s10909-010-0205-4} {\bibfield  {journal} {\bibinfo
			{journal} {J. Low Temp. Phys.}\ }\textbf {\bibinfo {volume} {161}},\ \bibinfo
		{pages} {299} (\bibinfo {year} {2010})}\BibitemShut {NoStop}%
	\bibitem [{\citenamefont {Gopalakrishnan}\ \emph {et~al.}(2016)\citenamefont
		{Gopalakrishnan}, \citenamefont {Agarwal}, \citenamefont {Demler},
		\citenamefont {Huse},\ and\ \citenamefont {Knap}}]{Gopalakrishnan16}%
	\BibitemOpen
	\bibfield  {author} {\bibinfo {author} {\bibfnamefont {S.}~\bibnamefont
			{Gopalakrishnan}}, \bibinfo {author} {\bibfnamefont {K.}~\bibnamefont
			{Agarwal}}, \bibinfo {author} {\bibfnamefont {E.~A.}\ \bibnamefont {Demler}},
		\bibinfo {author} {\bibfnamefont {D.~A.}\ \bibnamefont {Huse}},\ and\
		\bibinfo {author} {\bibfnamefont {M.}~\bibnamefont {Knap}},\ }\bibfield
	{title} {\bibinfo {title} {Griffiths effects and slow dynamics in nearly
			many-body localized systems},\ }\href
	{https://doi.org/10.1103/PhysRevB.93.134206} {\bibfield  {journal} {\bibinfo
			{journal} {Phys. Rev. B}\ }\textbf {\bibinfo {volume} {93}},\ \bibinfo
		{pages} {134206} (\bibinfo {year} {2016})}\BibitemShut {NoStop}%
	\bibitem [{\citenamefont {Agarwal}\ \emph {et~al.}(2015)\citenamefont
		{Agarwal}, \citenamefont {Gopalakrishnan}, \citenamefont {Knap},
		\citenamefont {M\"uller},\ and\ \citenamefont {Demler}}]{Agarwal15}%
	\BibitemOpen
	\bibfield  {author} {\bibinfo {author} {\bibfnamefont {K.}~\bibnamefont
			{Agarwal}}, \bibinfo {author} {\bibfnamefont {S.}~\bibnamefont
			{Gopalakrishnan}}, \bibinfo {author} {\bibfnamefont {M.}~\bibnamefont
			{Knap}}, \bibinfo {author} {\bibfnamefont {M.}~\bibnamefont {M\"uller}},\
		and\ \bibinfo {author} {\bibfnamefont {E.}~\bibnamefont {Demler}},\
	}\bibfield  {title} {\bibinfo {title} {Anomalous diffusion and griffiths
			effects near the many-body localization transition},\ }\href
	{https://doi.org/10.1103/PhysRevLett.114.160401} {\bibfield  {journal}
		{\bibinfo  {journal} {Phys. Rev. Lett.}\ }\textbf {\bibinfo {volume} {114}},\
		\bibinfo {pages} {160401} (\bibinfo {year} {2015})}\BibitemShut {NoStop}%
	\bibitem [{\citenamefont {Agarwal}\ \emph {et~al.}(2017)\citenamefont
		{Agarwal}, \citenamefont {Altman}, \citenamefont {Demler}, \citenamefont
		{Gopalakrishnan}, \citenamefont {Huse},\ and\ \citenamefont
		{Knap}}]{Agarwal17}%
	\BibitemOpen
	\bibfield  {author} {\bibinfo {author} {\bibfnamefont {K.}~\bibnamefont
			{Agarwal}}, \bibinfo {author} {\bibfnamefont {E.}~\bibnamefont {Altman}},
		\bibinfo {author} {\bibfnamefont {E.}~\bibnamefont {Demler}}, \bibinfo
		{author} {\bibfnamefont {S.}~\bibnamefont {Gopalakrishnan}}, \bibinfo
		{author} {\bibfnamefont {D.~A.}\ \bibnamefont {Huse}},\ and\ \bibinfo
		{author} {\bibfnamefont {M.}~\bibnamefont {Knap}},\ }\bibfield  {title}
	{\bibinfo {title} {Rare-region effects and dynamics near the many-body
			localization transition},\ }\href {https://doi.org/10.1002/andp.201600326}
	{\bibfield  {journal} {\bibinfo  {journal} {Annalen der Physik}\ }\textbf
		{\bibinfo {volume} {529}},\ \bibinfo {pages} {1600326} (\bibinfo {year}
		{2017})}\BibitemShut {NoStop}%
	\bibitem [{\citenamefont {Pancotti}\ \emph {et~al.}(2018)\citenamefont
		{Pancotti}, \citenamefont {Knap}, \citenamefont {Huse}, \citenamefont
		{Cirac},\ and\ \citenamefont {Ba\~nuls}}]{Pancotti18}%
	\BibitemOpen
	\bibfield  {author} {\bibinfo {author} {\bibfnamefont {N.}~\bibnamefont
			{Pancotti}}, \bibinfo {author} {\bibfnamefont {M.}~\bibnamefont {Knap}},
		\bibinfo {author} {\bibfnamefont {D.~A.}\ \bibnamefont {Huse}}, \bibinfo
		{author} {\bibfnamefont {J.~I.}\ \bibnamefont {Cirac}},\ and\ \bibinfo
		{author} {\bibfnamefont {M.~C.}\ \bibnamefont {Ba\~nuls}},\ }\bibfield
	{title} {\bibinfo {title} {Almost conserved operators in nearly many-body
			localized systems},\ }\href {https://doi.org/10.1103/PhysRevB.97.094206}
	{\bibfield  {journal} {\bibinfo  {journal} {Phys. Rev. B}\ }\textbf {\bibinfo
			{volume} {97}},\ \bibinfo {pages} {094206} (\bibinfo {year}
		{2018})}\BibitemShut {NoStop}%
	\bibitem [{\citenamefont {De~Roeck}\ and\ \citenamefont
		{Huveneers}(2017)}]{DeRoeck17}%
	\BibitemOpen
	\bibfield  {author} {\bibinfo {author} {\bibfnamefont {W.}~\bibnamefont
			{De~Roeck}}\ and\ \bibinfo {author} {\bibfnamefont {F.}~\bibnamefont
			{Huveneers}},\ }\bibfield  {title} {\bibinfo {title} {Stability and
			instability towards delocalization in many-body localization systems},\
	}\href {https://doi.org/10.1103/PhysRevB.95.155129} {\bibfield  {journal}
		{\bibinfo  {journal} {Phys. Rev. B}\ }\textbf {\bibinfo {volume} {95}},\
		\bibinfo {pages} {155129} (\bibinfo {year} {2017})}\BibitemShut {NoStop}%
	\bibitem [{\citenamefont {Luitz}\ \emph {et~al.}(2017)\citenamefont {Luitz},
		\citenamefont {Huveneers},\ and\ \citenamefont {De~Roeck}}]{Luitz17}%
	\BibitemOpen
	\bibfield  {author} {\bibinfo {author} {\bibfnamefont {D.~J.}\ \bibnamefont
			{Luitz}}, \bibinfo {author} {\bibfnamefont {F.~m.~c.}\ \bibnamefont
			{Huveneers}},\ and\ \bibinfo {author} {\bibfnamefont {W.}~\bibnamefont
			{De~Roeck}},\ }\bibfield  {title} {\bibinfo {title} {How a small quantum bath
			can thermalize long localized chains},\ }\href
	{https://doi.org/10.1103/PhysRevLett.119.150602} {\bibfield  {journal}
		{\bibinfo  {journal} {Phys. Rev. Lett.}\ }\textbf {\bibinfo {volume} {119}},\
		\bibinfo {pages} {150602} (\bibinfo {year} {2017})}\BibitemShut {NoStop}%
	\bibitem [{\citenamefont {\v{S}untajs}\ and\ \citenamefont
		{Vidmar}(2022)}]{Suntajs22sun}%
	\BibitemOpen
	\bibfield  {author} {\bibinfo {author} {\bibfnamefont {J.}~\bibnamefont
			{\v{S}untajs}}\ and\ \bibinfo {author} {\bibfnamefont {L.}~\bibnamefont
			{Vidmar}},\ }\bibfield  {title} {\bibinfo {title} {Ergodicity breaking
			transition in zero dimensions},\ }\href
	{https://doi.org/10.1103/PhysRevLett.129.060602} {\bibfield  {journal}
		{\bibinfo  {journal} {Phys. Rev. Lett.}\ }\textbf {\bibinfo {volume} {129}},\
		\bibinfo {pages} {060602} (\bibinfo {year} {2022})}\BibitemShut {NoStop}%
	\bibitem [{\citenamefont {Šuntajs}\ \emph {et~al.}(2023)\citenamefont
		{Šuntajs}, \citenamefont {Hopjan}, \citenamefont {Roeck},\ and\
		\citenamefont {Vidmar}}]{Suntajs23}%
	\BibitemOpen
	\bibfield  {author} {\bibinfo {author} {\bibfnamefont {J.}~\bibnamefont
			{Šuntajs}}, \bibinfo {author} {\bibfnamefont {M.}~\bibnamefont {Hopjan}},
		\bibinfo {author} {\bibfnamefont {W.~D.}\ \bibnamefont {Roeck}},\ and\
		\bibinfo {author} {\bibfnamefont {L.}~\bibnamefont {Vidmar}},\ }\href@noop {}
	{\bibinfo {title} {Similarity between a many-body quantum avalanche model and
			the ultrametric random matrix model}} (\bibinfo {year} {2023}),\ \Eprint
	{https://arxiv.org/abs/2308.07431} {arXiv:2308.07431 [cond-mat.stat-mech]}
	\BibitemShut {NoStop}%
	\bibitem [{\citenamefont {Pawlik}\ \emph {et~al.}(2023)\citenamefont {Pawlik},
		\citenamefont {Sierant}, \citenamefont {Vidmar},\ and\ \citenamefont
		{Zakrzewski}}]{pawlik2023manybody}%
	\BibitemOpen
	\bibfield  {author} {\bibinfo {author} {\bibfnamefont {K.}~\bibnamefont
			{Pawlik}}, \bibinfo {author} {\bibfnamefont {P.}~\bibnamefont {Sierant}},
		\bibinfo {author} {\bibfnamefont {L.}~\bibnamefont {Vidmar}},\ and\ \bibinfo
		{author} {\bibfnamefont {J.}~\bibnamefont {Zakrzewski}},\ }\href@noop {}
	{\bibinfo {title} {Many-body mobility edge in quantum sun models}} (\bibinfo
	{year} {2023}),\ \Eprint {https://arxiv.org/abs/2308.01073} {arXiv:2308.01073
		[cond-mat.dis-nn]} \BibitemShut {NoStop}%
	\bibitem [{\citenamefont {Thiery}\ \emph {et~al.}(2018)\citenamefont {Thiery},
		\citenamefont {Huveneers}, \citenamefont {M\"uller},\ and\ \citenamefont
		{De~Roeck}}]{Thiery18}%
	\BibitemOpen
	\bibfield  {author} {\bibinfo {author} {\bibfnamefont {T.}~\bibnamefont
			{Thiery}}, \bibinfo {author} {\bibfnamefont {F.~m.~c.}\ \bibnamefont
			{Huveneers}}, \bibinfo {author} {\bibfnamefont {M.}~\bibnamefont
			{M\"uller}},\ and\ \bibinfo {author} {\bibfnamefont {W.}~\bibnamefont
			{De~Roeck}},\ }\bibfield  {title} {\bibinfo {title} {Many-body delocalization
			as a quantum avalanche},\ }\href
	{https://doi.org/10.1103/PhysRevLett.121.140601} {\bibfield  {journal}
		{\bibinfo  {journal} {Phys. Rev. Lett.}\ }\textbf {\bibinfo {volume} {121}},\
		\bibinfo {pages} {140601} (\bibinfo {year} {2018})}\BibitemShut {NoStop}%
	\bibitem [{\citenamefont {Goremykina}\ \emph {et~al.}(2019)\citenamefont
		{Goremykina}, \citenamefont {Vasseur},\ and\ \citenamefont
		{Serbyn}}]{Goremykina19}%
	\BibitemOpen
	\bibfield  {author} {\bibinfo {author} {\bibfnamefont {A.}~\bibnamefont
			{Goremykina}}, \bibinfo {author} {\bibfnamefont {R.}~\bibnamefont
			{Vasseur}},\ and\ \bibinfo {author} {\bibfnamefont {M.}~\bibnamefont
			{Serbyn}},\ }\bibfield  {title} {\bibinfo {title} {Analytically solvable
			renormalization group for the many-body localization transition},\ }\href
	{https://doi.org/10.1103/PhysRevLett.122.040601} {\bibfield  {journal}
		{\bibinfo  {journal} {Phys. Rev. Lett.}\ }\textbf {\bibinfo {volume} {122}},\
		\bibinfo {pages} {040601} (\bibinfo {year} {2019})}\BibitemShut {NoStop}%
	\bibitem [{\citenamefont {Dumitrescu}\ \emph {et~al.}(2019)\citenamefont
		{Dumitrescu}, \citenamefont {Goremykina}, \citenamefont {Parameswaran},
		\citenamefont {Serbyn},\ and\ \citenamefont {Vasseur}}]{Dumitrescu19}%
	\BibitemOpen
	\bibfield  {author} {\bibinfo {author} {\bibfnamefont {P.~T.}\ \bibnamefont
			{Dumitrescu}}, \bibinfo {author} {\bibfnamefont {A.}~\bibnamefont
			{Goremykina}}, \bibinfo {author} {\bibfnamefont {S.~A.}\ \bibnamefont
			{Parameswaran}}, \bibinfo {author} {\bibfnamefont {M.}~\bibnamefont
			{Serbyn}},\ and\ \bibinfo {author} {\bibfnamefont {R.}~\bibnamefont
			{Vasseur}},\ }\bibfield  {title} {\bibinfo {title} {Kosterlitz-thouless
			scaling at many-body localization phase transitions},\ }\href
	{https://doi.org/10.1103/PhysRevB.99.094205} {\bibfield  {journal} {\bibinfo
			{journal} {Phys. Rev. B}\ }\textbf {\bibinfo {volume} {99}},\ \bibinfo
		{pages} {094205} (\bibinfo {year} {2019})}\BibitemShut {NoStop}%
	\bibitem [{\citenamefont {Morningstar}\ and\ \citenamefont
		{Huse}(2019)}]{Morningstar19}%
	\BibitemOpen
	\bibfield  {author} {\bibinfo {author} {\bibfnamefont {A.}~\bibnamefont
			{Morningstar}}\ and\ \bibinfo {author} {\bibfnamefont {D.~A.}\ \bibnamefont
			{Huse}},\ }\bibfield  {title} {\bibinfo {title} {Renormalization-group study
			of the many-body localization transition in one dimension},\ }\href
	{https://doi.org/10.1103/PhysRevB.99.224205} {\bibfield  {journal} {\bibinfo
			{journal} {Phys. Rev. B}\ }\textbf {\bibinfo {volume} {99}},\ \bibinfo
		{pages} {224205} (\bibinfo {year} {2019})}\BibitemShut {NoStop}%
	\bibitem [{\citenamefont {Potirniche}\ \emph {et~al.}(2019)\citenamefont
		{Potirniche}, \citenamefont {Banerjee},\ and\ \citenamefont
		{Altman}}]{Potirniche19}%
	\BibitemOpen
	\bibfield  {author} {\bibinfo {author} {\bibfnamefont {I.-D.}\ \bibnamefont
			{Potirniche}}, \bibinfo {author} {\bibfnamefont {S.}~\bibnamefont
			{Banerjee}},\ and\ \bibinfo {author} {\bibfnamefont {E.}~\bibnamefont
			{Altman}},\ }\bibfield  {title} {\bibinfo {title} {Exploration of the
			stability of many-body localization in $d>1$},\ }\href
	{https://doi.org/10.1103/PhysRevB.99.205149} {\bibfield  {journal} {\bibinfo
			{journal} {Phys. Rev. B}\ }\textbf {\bibinfo {volume} {99}},\ \bibinfo
		{pages} {205149} (\bibinfo {year} {2019})}\BibitemShut {NoStop}%
	\bibitem [{\citenamefont {Herviou}\ \emph {et~al.}(2019)\citenamefont
		{Herviou}, \citenamefont {Bera},\ and\ \citenamefont
		{Bardarson}}]{Herviou19}%
	\BibitemOpen
	\bibfield  {author} {\bibinfo {author} {\bibfnamefont {L.}~\bibnamefont
			{Herviou}}, \bibinfo {author} {\bibfnamefont {S.}~\bibnamefont {Bera}},\ and\
		\bibinfo {author} {\bibfnamefont {J.~H.}\ \bibnamefont {Bardarson}},\
	}\bibfield  {title} {\bibinfo {title} {Multiscale entanglement clusters at
			the many-body localization phase transition},\ }\href
	{https://doi.org/10.1103/PhysRevB.99.134205} {\bibfield  {journal} {\bibinfo
			{journal} {Phys. Rev. B}\ }\textbf {\bibinfo {volume} {99}},\ \bibinfo
		{pages} {134205} (\bibinfo {year} {2019})}\BibitemShut {NoStop}%
	\bibitem [{\citenamefont {Szo\l{}dra}\ \emph {et~al.}(2021)\citenamefont
		{Szo\l{}dra}, \citenamefont {Sierant}, \citenamefont {Kottmann},
		\citenamefont {Lewenstein},\ and\ \citenamefont {Zakrzewski}}]{Szoldra21}%
	\BibitemOpen
	\bibfield  {author} {\bibinfo {author} {\bibfnamefont {T.}~\bibnamefont
			{Szo\l{}dra}}, \bibinfo {author} {\bibfnamefont {P.}~\bibnamefont {Sierant}},
		\bibinfo {author} {\bibfnamefont {K.}~\bibnamefont {Kottmann}}, \bibinfo
		{author} {\bibfnamefont {M.}~\bibnamefont {Lewenstein}},\ and\ \bibinfo
		{author} {\bibfnamefont {J.}~\bibnamefont {Zakrzewski}},\ }\bibfield  {title}
	{\bibinfo {title} {Detecting ergodic bubbles at the crossover to many-body
			localization using neural networks},\ }\href
	{https://doi.org/10.1103/PhysRevB.104.L140202} {\bibfield  {journal}
		{\bibinfo  {journal} {Phys. Rev. B}\ }\textbf {\bibinfo {volume} {104}},\
		\bibinfo {pages} {L140202} (\bibinfo {year} {2021})}\BibitemShut {NoStop}%
	\bibitem [{\citenamefont {H\'emery}\ \emph {et~al.}(2022)\citenamefont
		{H\'emery}, \citenamefont {Pollmann},\ and\ \citenamefont
		{Smith}}]{Hemery22}%
	\BibitemOpen
	\bibfield  {author} {\bibinfo {author} {\bibfnamefont {K.}~\bibnamefont
			{H\'emery}}, \bibinfo {author} {\bibfnamefont {F.}~\bibnamefont {Pollmann}},\
		and\ \bibinfo {author} {\bibfnamefont {A.}~\bibnamefont {Smith}},\ }\bibfield
	{title} {\bibinfo {title} {Identifying correlation clusters in many-body
			localized systems},\ }\href {https://doi.org/10.1103/PhysRevB.105.064202}
	{\bibfield  {journal} {\bibinfo  {journal} {Phys. Rev. B}\ }\textbf {\bibinfo
			{volume} {105}},\ \bibinfo {pages} {064202} (\bibinfo {year}
		{2022})}\BibitemShut {NoStop}%
	\bibitem [{\citenamefont {Foo}\ \emph {et~al.}(2023)\citenamefont {Foo},
		\citenamefont {Swain}, \citenamefont {Sengupta}, \citenamefont {Lemari\'e},\
		and\ \citenamefont {Adam}}]{Foo23}%
	\BibitemOpen
	\bibfield  {author} {\bibinfo {author} {\bibfnamefont {D.~C.~W.}\
			\bibnamefont {Foo}}, \bibinfo {author} {\bibfnamefont {N.}~\bibnamefont
			{Swain}}, \bibinfo {author} {\bibfnamefont {P.}~\bibnamefont {Sengupta}},
		\bibinfo {author} {\bibfnamefont {G.}~\bibnamefont {Lemari\'e}},\ and\
		\bibinfo {author} {\bibfnamefont {S.}~\bibnamefont {Adam}},\ }\bibfield
	{title} {\bibinfo {title} {Stabilization mechanism for many-body localization
			in two dimensions},\ }\href
	{https://doi.org/10.1103/PhysRevResearch.5.L032011} {\bibfield  {journal}
		{\bibinfo  {journal} {Phys. Rev. Res.}\ }\textbf {\bibinfo {volume} {5}},\
		\bibinfo {pages} {L032011} (\bibinfo {year} {2023})}\BibitemShut {NoStop}%
	\bibitem [{\citenamefont {Colmenarez}\ \emph {et~al.}(2024)\citenamefont
		{Colmenarez}, \citenamefont {Luitz},\ and\ \citenamefont
		{De~Roeck}}]{Colmenarez24}%
	\BibitemOpen
	\bibfield  {author} {\bibinfo {author} {\bibfnamefont {L.}~\bibnamefont
			{Colmenarez}}, \bibinfo {author} {\bibfnamefont {D.~J.}\ \bibnamefont
			{Luitz}},\ and\ \bibinfo {author} {\bibfnamefont {W.}~\bibnamefont
			{De~Roeck}},\ }\bibfield  {title} {\bibinfo {title} {Ergodic inclusions in
			many-body localized systems},\ }\bibfield  {journal} {\bibinfo  {journal}
		{Physical Review B}\ }\textbf {\bibinfo {volume} {109}},\ \href
	{https://doi.org/10.1103/physrevb.109.l081117} {10.1103/physrevb.109.l081117}
	(\bibinfo {year} {2024})\BibitemShut {NoStop}%
	\bibitem [{\citenamefont {Sels}(2022)}]{Sels22bath}%
	\BibitemOpen
	\bibfield  {author} {\bibinfo {author} {\bibfnamefont {D.}~\bibnamefont
			{Sels}},\ }\bibfield  {title} {\bibinfo {title} {Bath-induced delocalization
			in interacting disordered spin chains},\ }\href
	{https://doi.org/10.1103/PhysRevB.106.L020202} {\bibfield  {journal}
		{\bibinfo  {journal} {Phys. Rev. B}\ }\textbf {\bibinfo {volume} {106}},\
		\bibinfo {pages} {L020202} (\bibinfo {year} {2022})}\BibitemShut {NoStop}%
	\bibitem [{\citenamefont {Ha}\ \emph {et~al.}(2023)\citenamefont {Ha},
		\citenamefont {Morningstar},\ and\ \citenamefont {Huse}}]{Ha23}%
	\BibitemOpen
	\bibfield  {author} {\bibinfo {author} {\bibfnamefont {H.}~\bibnamefont
			{Ha}}, \bibinfo {author} {\bibfnamefont {A.}~\bibnamefont {Morningstar}},\
		and\ \bibinfo {author} {\bibfnamefont {D.~A.}\ \bibnamefont {Huse}},\
	}\bibfield  {title} {\bibinfo {title} {Many-body resonances in the avalanche
			instability of many-body localization},\ }\href
	{https://doi.org/10.1103/PhysRevLett.130.250405} {\bibfield  {journal}
		{\bibinfo  {journal} {Phys. Rev. Lett.}\ }\textbf {\bibinfo {volume} {130}},\
		\bibinfo {pages} {250405} (\bibinfo {year} {2023})}\BibitemShut {NoStop}%
	\bibitem [{\citenamefont {Tu}\ \emph {et~al.}(2023{\natexlab{a}})\citenamefont
		{Tu}, \citenamefont {Vu},\ and\ \citenamefont {Das~Sarma}}]{Tu23avalanche}%
	\BibitemOpen
	\bibfield  {author} {\bibinfo {author} {\bibfnamefont {Y.-T.}\ \bibnamefont
			{Tu}}, \bibinfo {author} {\bibfnamefont {D.}~\bibnamefont {Vu}},\ and\
		\bibinfo {author} {\bibfnamefont {S.}~\bibnamefont {Das~Sarma}},\ }\bibfield
	{title} {\bibinfo {title} {Avalanche stability transition in interacting
			quasiperiodic systems},\ }\href {https://doi.org/10.1103/PhysRevB.107.014203}
	{\bibfield  {journal} {\bibinfo  {journal} {Phys. Rev. B}\ }\textbf {\bibinfo
			{volume} {107}},\ \bibinfo {pages} {014203} (\bibinfo {year}
		{2023}{\natexlab{a}})}\BibitemShut {NoStop}%
	\bibitem [{\citenamefont {L{\'e}onard}\ \emph {et~al.}(2023)\citenamefont
		{L{\'e}onard}, \citenamefont {Kim}, \citenamefont {Rispoli}, \citenamefont
		{Lukin}, \citenamefont {Schittko}, \citenamefont {Kwan}, \citenamefont
		{Demler}, \citenamefont {Sels},\ and\ \citenamefont {Greiner}}]{Leonard23}%
	\BibitemOpen
	\bibfield  {author} {\bibinfo {author} {\bibfnamefont {J.}~\bibnamefont
			{L{\'e}onard}}, \bibinfo {author} {\bibfnamefont {S.}~\bibnamefont {Kim}},
		\bibinfo {author} {\bibfnamefont {M.}~\bibnamefont {Rispoli}}, \bibinfo
		{author} {\bibfnamefont {A.}~\bibnamefont {Lukin}}, \bibinfo {author}
		{\bibfnamefont {R.}~\bibnamefont {Schittko}}, \bibinfo {author}
		{\bibfnamefont {J.}~\bibnamefont {Kwan}}, \bibinfo {author} {\bibfnamefont
			{E.}~\bibnamefont {Demler}}, \bibinfo {author} {\bibfnamefont
			{D.}~\bibnamefont {Sels}},\ and\ \bibinfo {author} {\bibfnamefont
			{M.}~\bibnamefont {Greiner}},\ }\bibfield  {title} {\bibinfo {title} {Probing
			the onset of quantum avalanches in a many-body localized system},\ }\href
	{https://doi.org/10.1038/s41567-022-01887-3} {\bibfield  {journal} {\bibinfo
			{journal} {Nature Physics}\ }\textbf {\bibinfo {volume} {19}},\ \bibinfo
		{pages} {481} (\bibinfo {year} {2023})}\BibitemShut {NoStop}%
	\bibitem [{\citenamefont {Peacock}\ and\ \citenamefont
		{Sels}(2023)}]{Peacock23}%
	\BibitemOpen
	\bibfield  {author} {\bibinfo {author} {\bibfnamefont {J.~C.}\ \bibnamefont
			{Peacock}}\ and\ \bibinfo {author} {\bibfnamefont {D.}~\bibnamefont {Sels}},\
	}\bibfield  {title} {\bibinfo {title} {Many-body delocalization from embedded
			thermal inclusion},\ }\href {https://doi.org/10.1103/PhysRevB.108.L020201}
	{\bibfield  {journal} {\bibinfo  {journal} {Phys. Rev. B}\ }\textbf {\bibinfo
			{volume} {108}},\ \bibinfo {pages} {L020201} (\bibinfo {year}
		{2023})}\BibitemShut {NoStop}%
	\bibitem [{\citenamefont {Tu}\ \emph {et~al.}(2023{\natexlab{b}})\citenamefont
		{Tu}, \citenamefont {Vu},\ and\ \citenamefont {Das~Sarma}}]{Tu23}%
	\BibitemOpen
	\bibfield  {author} {\bibinfo {author} {\bibfnamefont {Y.-T.}\ \bibnamefont
			{Tu}}, \bibinfo {author} {\bibfnamefont {D.}~\bibnamefont {Vu}},\ and\
		\bibinfo {author} {\bibfnamefont {S.}~\bibnamefont {Das~Sarma}},\ }\bibfield
	{title} {\bibinfo {title} {Localization spectrum of a bath-coupled
			generalized aubry-andr\'e model in the presence of interactions},\ }\href
	{https://doi.org/10.1103/PhysRevB.108.064313} {\bibfield  {journal} {\bibinfo
			{journal} {Phys. Rev. B}\ }\textbf {\bibinfo {volume} {108}},\ \bibinfo
		{pages} {064313} (\bibinfo {year} {2023}{\natexlab{b}})}\BibitemShut
	{NoStop}%
	\bibitem [{\citenamefont {Srednicki}(1999)}]{Srednicki99}%
	\BibitemOpen
	\bibfield  {author} {\bibinfo {author} {\bibfnamefont {M.}~\bibnamefont
			{Srednicki}},\ }\bibfield  {title} {\bibinfo {title} {The approach to thermal
			equilibrium in quantized chaotic systems},\ }\href
	{https://doi.org/10.1088/0305-4470/32/7/007} {\bibfield  {journal} {\bibinfo
			{journal} {Journal of Physics A: Mathematical and General}\ }\textbf
		{\bibinfo {volume} {32}},\ \bibinfo {pages} {1163} (\bibinfo {year}
		{1999})}\BibitemShut {NoStop}%
	\bibitem [{\citenamefont {Burke}\ and\ \citenamefont {Haque}(2023)}]{Burke23}%
	\BibitemOpen
	\bibfield  {author} {\bibinfo {author} {\bibfnamefont {P.~C.}\ \bibnamefont
			{Burke}}\ and\ \bibinfo {author} {\bibfnamefont {M.}~\bibnamefont {Haque}},\
	}\bibfield  {title} {\bibinfo {title} {Entropy and temperature in finite
			isolated quantum systems},\ }\href
	{https://doi.org/10.1103/PhysRevE.107.034125} {\bibfield  {journal} {\bibinfo
			{journal} {Phys. Rev. E}\ }\textbf {\bibinfo {volume} {107}},\ \bibinfo
		{pages} {034125} (\bibinfo {year} {2023})}\BibitemShut {NoStop}%
	\bibitem [{\citenamefont {Tannoudji}\ \emph {et~al.}(2002)\citenamefont
		{Tannoudji}, \citenamefont {Diu},\ and\ \citenamefont
		{Lalo{\"e}}}]{Tannoudji02quantum}%
	\BibitemOpen
	\bibfield  {author} {\bibinfo {author} {\bibfnamefont {C.~C.}\ \bibnamefont
			{Tannoudji}}, \bibinfo {author} {\bibfnamefont {B.}~\bibnamefont {Diu}},\
		and\ \bibinfo {author} {\bibfnamefont {F.}~\bibnamefont {Lalo{\"e}}},\
	}\href@noop {} {\emph {\bibinfo {title} {Quantum Mechanics: Vol. Two}}}\
	(\bibinfo  {publisher} {John Wiley and Sons},\ \bibinfo {year}
	{2002})\BibitemShut {NoStop}%
	\bibitem [{\citenamefont {Chandran}\ \emph {et~al.}(2015)\citenamefont
		{Chandran}, \citenamefont {Kim}, \citenamefont {Vidal},\ and\ \citenamefont
		{Abanin}}]{Chandran15}%
	\BibitemOpen
	\bibfield  {author} {\bibinfo {author} {\bibfnamefont {A.}~\bibnamefont
			{Chandran}}, \bibinfo {author} {\bibfnamefont {I.~H.}\ \bibnamefont {Kim}},
		\bibinfo {author} {\bibfnamefont {G.}~\bibnamefont {Vidal}},\ and\ \bibinfo
		{author} {\bibfnamefont {D.~A.}\ \bibnamefont {Abanin}},\ }\bibfield  {title}
	{\bibinfo {title} {Constructing local integrals of motion in the many-body
			localized phase},\ }\href {https://doi.org/10.1103/PhysRevB.91.085425}
	{\bibfield  {journal} {\bibinfo  {journal} {Phys. Rev. B}\ }\textbf {\bibinfo
			{volume} {91}},\ \bibinfo {pages} {085425} (\bibinfo {year}
		{2015})}\BibitemShut {NoStop}%
	\bibitem [{\citenamefont {Rademaker}\ and\ \citenamefont
		{Ortu\~no}(2016)}]{Rademaker16}%
	\BibitemOpen
	\bibfield  {author} {\bibinfo {author} {\bibfnamefont {L.}~\bibnamefont
			{Rademaker}}\ and\ \bibinfo {author} {\bibfnamefont {M.}~\bibnamefont
			{Ortu\~no}},\ }\bibfield  {title} {\bibinfo {title} {Explicit local integrals
			of motion for the many-body localized state},\ }\href
	{https://doi.org/10.1103/PhysRevLett.116.010404} {\bibfield  {journal}
		{\bibinfo  {journal} {Phys. Rev. Lett.}\ }\textbf {\bibinfo {volume} {116}},\
		\bibinfo {pages} {010404} (\bibinfo {year} {2016})}\BibitemShut {NoStop}%
	\bibitem [{\citenamefont {Ortu\~no}\ \emph {et~al.}(2019)\citenamefont
		{Ortu\~no}, \citenamefont {Somoza},\ and\ \citenamefont
		{Rademaker}}]{Ortuno19}%
	\BibitemOpen
	\bibfield  {author} {\bibinfo {author} {\bibfnamefont {M.}~\bibnamefont
			{Ortu\~no}}, \bibinfo {author} {\bibfnamefont {A.~M.}\ \bibnamefont
			{Somoza}},\ and\ \bibinfo {author} {\bibfnamefont {L.}~\bibnamefont
			{Rademaker}},\ }\bibfield  {title} {\bibinfo {title} {Construction of
			many-body eigenstates with displacement transformations in disordered
			systems},\ }\href {https://doi.org/10.1103/PhysRevB.100.085115} {\bibfield
		{journal} {\bibinfo  {journal} {Phys. Rev. B}\ }\textbf {\bibinfo {volume}
			{100}},\ \bibinfo {pages} {085115} (\bibinfo {year} {2019})}\BibitemShut
	{NoStop}%
	\bibitem [{\citenamefont {Pekker}\ \emph {et~al.}(2017)\citenamefont {Pekker},
		\citenamefont {Clark}, \citenamefont {Oganesyan},\ and\ \citenamefont
		{Refael}}]{Pekker17}%
	\BibitemOpen
	\bibfield  {author} {\bibinfo {author} {\bibfnamefont {D.}~\bibnamefont
			{Pekker}}, \bibinfo {author} {\bibfnamefont {B.~K.}\ \bibnamefont {Clark}},
		\bibinfo {author} {\bibfnamefont {V.}~\bibnamefont {Oganesyan}},\ and\
		\bibinfo {author} {\bibfnamefont {G.}~\bibnamefont {Refael}},\ }\bibfield
	{title} {\bibinfo {title} {Fixed points of wegner-wilson flows and many-body
			localization},\ }\href {https://doi.org/10.1103/PhysRevLett.119.075701}
	{\bibfield  {journal} {\bibinfo  {journal} {Phys. Rev. Lett.}\ }\textbf
		{\bibinfo {volume} {119}},\ \bibinfo {pages} {075701} (\bibinfo {year}
		{2017})}\BibitemShut {NoStop}%
	\bibitem [{\citenamefont {Kelly}\ \emph {et~al.}(2020)\citenamefont {Kelly},
		\citenamefont {Nandkishore},\ and\ \citenamefont {Marino}}]{Kelly20}%
	\BibitemOpen
	\bibfield  {author} {\bibinfo {author} {\bibfnamefont {S.~P.}\ \bibnamefont
			{Kelly}}, \bibinfo {author} {\bibfnamefont {R.}~\bibnamefont {Nandkishore}},\
		and\ \bibinfo {author} {\bibfnamefont {J.}~\bibnamefont {Marino}},\
	}\bibfield  {title} {\bibinfo {title} {Exploring many-body localization in
			quantum systems coupled to an environment via wegner-wilson flows},\ }\href
	{https://doi.org/https://doi.org/10.1016/j.nuclphysb.2019.114886} {\bibfield
		{journal} {\bibinfo  {journal} {Nuclear Physics B}\ }\textbf {\bibinfo
			{volume} {951}},\ \bibinfo {pages} {114886} (\bibinfo {year}
		{2020})}\BibitemShut {NoStop}%
	\bibitem [{\citenamefont {Thomson}\ \emph {et~al.}(2021)\citenamefont
		{Thomson}, \citenamefont {Magano},\ and\ \citenamefont
		{Schirò}}]{Thomson21flow}%
	\BibitemOpen
	\bibfield  {author} {\bibinfo {author} {\bibfnamefont {S.~J.}\ \bibnamefont
			{Thomson}}, \bibinfo {author} {\bibfnamefont {D.}~\bibnamefont {Magano}},\
		and\ \bibinfo {author} {\bibfnamefont {M.}~\bibnamefont {Schirò}},\
	}\bibfield  {title} {\bibinfo {title} {{Flow equations for disordered Floquet
				systems}},\ }\href {https://doi.org/10.21468/SciPostPhys.11.2.028} {\bibfield
		{journal} {\bibinfo  {journal} {SciPost Phys.}\ }\textbf {\bibinfo {volume}
			{11}},\ \bibinfo {pages} {028} (\bibinfo {year} {2021})}\BibitemShut
	{NoStop}%
	\bibitem [{\citenamefont {Thomson}\ and\ \citenamefont
		{Schirò}(2023)}]{Thomson23}%
	\BibitemOpen
	\bibfield  {author} {\bibinfo {author} {\bibfnamefont {S.~J.}\ \bibnamefont
			{Thomson}}\ and\ \bibinfo {author} {\bibfnamefont {M.}~\bibnamefont
			{Schirò}},\ }\bibfield  {title} {\bibinfo {title} {{Local integrals of
				motion in quasiperiodic many-body localized systems}},\ }\href
	{https://doi.org/10.21468/SciPostPhys.14.5.125} {\bibfield  {journal}
		{\bibinfo  {journal} {SciPost Phys.}\ }\textbf {\bibinfo {volume} {14}},\
		\bibinfo {pages} {125} (\bibinfo {year} {2023})}\BibitemShut {NoStop}%
	\bibitem [{\citenamefont {O'Brien}\ \emph {et~al.}(2016)\citenamefont
		{O'Brien}, \citenamefont {Abanin}, \citenamefont {Vidal},\ and\ \citenamefont
		{Papi\'{c}}}]{Brien16}%
	\BibitemOpen
	\bibfield  {author} {\bibinfo {author} {\bibfnamefont {T.~E.}\ \bibnamefont
			{O'Brien}}, \bibinfo {author} {\bibfnamefont {D.~A.}\ \bibnamefont {Abanin}},
		\bibinfo {author} {\bibfnamefont {G.}~\bibnamefont {Vidal}},\ and\ \bibinfo
		{author} {\bibfnamefont {Z.}~\bibnamefont {Papi\'{c}}},\ }\bibfield  {title}
	{\bibinfo {title} {Explicit construction of local conserved operators in
			disordered many-body systems},\ }\href
	{https://doi.org/10.1103/PhysRevB.94.144208} {\bibfield  {journal} {\bibinfo
			{journal} {Phys. Rev. B}\ }\textbf {\bibinfo {volume} {94}},\ \bibinfo
		{pages} {144208} (\bibinfo {year} {2016})}\BibitemShut {NoStop}%
	\bibitem [{\citenamefont {Peng}\ \emph {et~al.}(2019)\citenamefont {Peng},
		\citenamefont {Li}, \citenamefont {Yan}, \citenamefont {Wei},\ and\
		\citenamefont {Cappellaro}}]{Peng19}%
	\BibitemOpen
	\bibfield  {author} {\bibinfo {author} {\bibfnamefont {P.}~\bibnamefont
			{Peng}}, \bibinfo {author} {\bibfnamefont {Z.}~\bibnamefont {Li}}, \bibinfo
		{author} {\bibfnamefont {H.}~\bibnamefont {Yan}}, \bibinfo {author}
		{\bibfnamefont {K.~X.}\ \bibnamefont {Wei}},\ and\ \bibinfo {author}
		{\bibfnamefont {P.}~\bibnamefont {Cappellaro}},\ }\bibfield  {title}
	{\bibinfo {title} {Comparing many-body localization lengths via
			nonperturbative construction of local integrals of motion},\ }\href
	{https://doi.org/10.1103/PhysRevB.100.214203} {\bibfield  {journal} {\bibinfo
			{journal} {Phys. Rev. B}\ }\textbf {\bibinfo {volume} {100}},\ \bibinfo
		{pages} {214203} (\bibinfo {year} {2019})}\BibitemShut {NoStop}%
	\bibitem [{\citenamefont {Leipner-Johns}\ and\ \citenamefont
		{Wortis}(2019)}]{Johns19}%
	\BibitemOpen
	\bibfield  {author} {\bibinfo {author} {\bibfnamefont {B.}~\bibnamefont
			{Leipner-Johns}}\ and\ \bibinfo {author} {\bibfnamefont {R.}~\bibnamefont
			{Wortis}},\ }\bibfield  {title} {\bibinfo {title} {Charge- and spin-specific
			local integrals of motion in a disordered hubbard model},\ }\href
	{https://doi.org/10.1103/PhysRevB.100.125132} {\bibfield  {journal} {\bibinfo
			{journal} {Phys. Rev. B}\ }\textbf {\bibinfo {volume} {100}},\ \bibinfo
		{pages} {125132} (\bibinfo {year} {2019})}\BibitemShut {NoStop}%
	\bibitem [{\citenamefont {Adami}\ \emph {et~al.}(2022)\citenamefont {Adami},
		\citenamefont {Amini},\ and\ \citenamefont {Soltani}}]{Adami22}%
	\BibitemOpen
	\bibfield  {author} {\bibinfo {author} {\bibfnamefont {S.}~\bibnamefont
			{Adami}}, \bibinfo {author} {\bibfnamefont {M.}~\bibnamefont {Amini}},\ and\
		\bibinfo {author} {\bibfnamefont {M.}~\bibnamefont {Soltani}},\ }\bibfield
	{title} {\bibinfo {title} {Structural properties of local integrals of motion
			across the many-body localization transition via a fast and efficient method
			for their construction},\ }\href
	{https://doi.org/10.1103/PhysRevB.106.054202} {\bibfield  {journal} {\bibinfo
			{journal} {Phys. Rev. B}\ }\textbf {\bibinfo {volume} {106}},\ \bibinfo
		{pages} {054202} (\bibinfo {year} {2022})}\BibitemShut {NoStop}%
	\bibitem [{\citenamefont {Szo\l{}dra}\ \emph {et~al.}(2022)\citenamefont
		{Szo\l{}dra}, \citenamefont {Sierant}, \citenamefont {Lewenstein},\ and\
		\citenamefont {Zakrzewski}}]{Szoldra22}%
	\BibitemOpen
	\bibfield  {author} {\bibinfo {author} {\bibfnamefont {T.}~\bibnamefont
			{Szo\l{}dra}}, \bibinfo {author} {\bibfnamefont {P.}~\bibnamefont {Sierant}},
		\bibinfo {author} {\bibfnamefont {M.}~\bibnamefont {Lewenstein}},\ and\
		\bibinfo {author} {\bibfnamefont {J.}~\bibnamefont {Zakrzewski}},\ }\bibfield
	{title} {\bibinfo {title} {Unsupervised detection of decoupled subspaces:
			Many-body scars and beyond},\ }\href
	{https://doi.org/10.1103/PhysRevB.105.224205} {\bibfield  {journal} {\bibinfo
			{journal} {Phys. Rev. B}\ }\textbf {\bibinfo {volume} {105}},\ \bibinfo
		{pages} {224205} (\bibinfo {year} {2022})}\BibitemShut {NoStop}%
	\bibitem [{\citenamefont {Berkelbach}\ and\ \citenamefont
		{Reichman}(2010)}]{Berkelbach10}%
	\BibitemOpen
	\bibfield  {author} {\bibinfo {author} {\bibfnamefont {T.~C.}\ \bibnamefont
			{Berkelbach}}\ and\ \bibinfo {author} {\bibfnamefont {D.~R.}\ \bibnamefont
			{Reichman}},\ }\bibfield  {title} {\bibinfo {title} {Conductivity of
			disordered quantum lattice models at infinite temperature: Many-body
			localization},\ }\href {https://doi.org/10.1103/PhysRevB.81.224429}
	{\bibfield  {journal} {\bibinfo  {journal} {Phys. Rev. B}\ }\textbf {\bibinfo
			{volume} {81}},\ \bibinfo {pages} {224429} (\bibinfo {year}
		{2010})}\BibitemShut {NoStop}%
	\bibitem [{\citenamefont {Bera}\ \emph {et~al.}(2015)\citenamefont {Bera},
		\citenamefont {Schomerus}, \citenamefont {Heidrich-Meisner},\ and\
		\citenamefont {Bardarson}}]{Bera15}%
	\BibitemOpen
	\bibfield  {author} {\bibinfo {author} {\bibfnamefont {S.}~\bibnamefont
			{Bera}}, \bibinfo {author} {\bibfnamefont {H.}~\bibnamefont {Schomerus}},
		\bibinfo {author} {\bibfnamefont {F.}~\bibnamefont {Heidrich-Meisner}},\ and\
		\bibinfo {author} {\bibfnamefont {J.~H.}\ \bibnamefont {Bardarson}},\
	}\bibfield  {title} {\bibinfo {title} {Many-body localization characterized
			from a one-particle perspective},\ }\href
	{https://doi.org/10.1103/PhysRevLett.115.046603} {\bibfield  {journal}
		{\bibinfo  {journal} {Phys. Rev. Lett.}\ }\textbf {\bibinfo {volume} {115}},\
		\bibinfo {pages} {046603} (\bibinfo {year} {2015})}\BibitemShut {NoStop}%
	\bibitem [{\citenamefont {Enss}\ \emph {et~al.}(2017)\citenamefont {Enss},
		\citenamefont {Andraschko},\ and\ \citenamefont {Sirker}}]{Enss17}%
	\BibitemOpen
	\bibfield  {author} {\bibinfo {author} {\bibfnamefont {T.}~\bibnamefont
			{Enss}}, \bibinfo {author} {\bibfnamefont {F.}~\bibnamefont {Andraschko}},\
		and\ \bibinfo {author} {\bibfnamefont {J.}~\bibnamefont {Sirker}},\
	}\bibfield  {title} {\bibinfo {title} {Many-body localization in infinite
			chains},\ }\href {https://doi.org/10.1103/PhysRevB.95.045121} {\bibfield
		{journal} {\bibinfo  {journal} {Phys. Rev. B}\ }\textbf {\bibinfo {volume}
			{95}},\ \bibinfo {pages} {045121} (\bibinfo {year} {2017})}\BibitemShut
	{NoStop}%
	\bibitem [{\citenamefont {Serbyn}\ and\ \citenamefont
		{Moore}(2016)}]{Serbyn16}%
	\BibitemOpen
	\bibfield  {author} {\bibinfo {author} {\bibfnamefont {M.}~\bibnamefont
			{Serbyn}}\ and\ \bibinfo {author} {\bibfnamefont {J.~E.}\ \bibnamefont
			{Moore}},\ }\bibfield  {title} {\bibinfo {title} {Spectral statistics across
			the many-body localization transition},\ }\href
	{https://doi.org/10.1103/PhysRevB.93.041424} {\bibfield  {journal} {\bibinfo
			{journal} {Phys. Rev. B}\ }\textbf {\bibinfo {volume} {93}},\ \bibinfo
		{pages} {041424} (\bibinfo {year} {2016})}\BibitemShut {NoStop}%
	\bibitem [{\citenamefont {Bertrand}\ and\ \citenamefont
		{Garc\'{\i}a-Garc\'{\i}a}(2016)}]{Bertrand16}%
	\BibitemOpen
	\bibfield  {author} {\bibinfo {author} {\bibfnamefont {C.~L.}\ \bibnamefont
			{Bertrand}}\ and\ \bibinfo {author} {\bibfnamefont {A.~M.}\ \bibnamefont
			{Garc\'{\i}a-Garc\'{\i}a}},\ }\bibfield  {title} {\bibinfo {title} {Anomalous
			thouless energy and critical statistics on the metallic side of the many-body
			localization transition},\ }\href
	{https://doi.org/10.1103/PhysRevB.94.144201} {\bibfield  {journal} {\bibinfo
			{journal} {Phys. Rev. B}\ }\textbf {\bibinfo {volume} {94}},\ \bibinfo
		{pages} {144201} (\bibinfo {year} {2016})}\BibitemShut {NoStop}%
	\bibitem [{\citenamefont {Santos}\ \emph {et~al.}(2004)\citenamefont {Santos},
		\citenamefont {Rigolin},\ and\ \citenamefont {Escobar}}]{Santos04a}%
	\BibitemOpen
	\bibfield  {author} {\bibinfo {author} {\bibfnamefont {L.~F.}\ \bibnamefont
			{Santos}}, \bibinfo {author} {\bibfnamefont {G.}~\bibnamefont {Rigolin}},\
		and\ \bibinfo {author} {\bibfnamefont {C.~O.}\ \bibnamefont {Escobar}},\
	}\bibfield  {title} {\bibinfo {title} {Entanglement versus chaos in
			disordered spin chains},\ }\href {https://doi.org/10.1103/PhysRevA.69.042304}
	{\bibfield  {journal} {\bibinfo  {journal} {Phys. Rev. A}\ }\textbf {\bibinfo
			{volume} {69}},\ \bibinfo {pages} {042304} (\bibinfo {year}
		{2004})}\BibitemShut {NoStop}%
	\bibitem [{\citenamefont {Santos}(2004)}]{Santos04}%
	\BibitemOpen
	\bibfield  {author} {\bibinfo {author} {\bibfnamefont {L.~F.}\ \bibnamefont
			{Santos}},\ }\bibfield  {title} {\bibinfo {title} {Integrability of a
			disordered heisenberg spin-1/2 chain},\ }\href
	{https://doi.org/10.1088/0305-4470/37/17/004} {\bibfield  {journal} {\bibinfo
			{journal} {Journal of Physics A: Mathematical and General}\ }\textbf
		{\bibinfo {volume} {37}},\ \bibinfo {pages} {4723} (\bibinfo {year}
		{2004})}\BibitemShut {NoStop}%
	\bibitem [{\citenamefont {Colmenarez}\ \emph {et~al.}(2019)\citenamefont
		{Colmenarez}, \citenamefont {McClarty}, \citenamefont {Haque},\ and\
		\citenamefont {Luitz}}]{Colmenarez19}%
	\BibitemOpen
	\bibfield  {author} {\bibinfo {author} {\bibfnamefont {L.}~\bibnamefont
			{Colmenarez}}, \bibinfo {author} {\bibfnamefont {P.~A.}\ \bibnamefont
			{McClarty}}, \bibinfo {author} {\bibfnamefont {M.}~\bibnamefont {Haque}},\
		and\ \bibinfo {author} {\bibfnamefont {D.~J.}\ \bibnamefont {Luitz}},\
	}\bibfield  {title} {\bibinfo {title} {{Statistics of correlation functions
				in the random Heisenberg chain}},\ }\href
	{https://doi.org/10.21468/SciPostPhys.7.5.064} {\bibfield  {journal}
		{\bibinfo  {journal} {SciPost Phys.}\ }\textbf {\bibinfo {volume} {7}},\
		\bibinfo {pages} {064} (\bibinfo {year} {2019})}\BibitemShut {NoStop}%
	\bibitem [{\citenamefont {Sierant}\ and\ \citenamefont
		{Zakrzewski}(2019)}]{Sierant19b}%
	\BibitemOpen
	\bibfield  {author} {\bibinfo {author} {\bibfnamefont {P.}~\bibnamefont
			{Sierant}}\ and\ \bibinfo {author} {\bibfnamefont {J.}~\bibnamefont
			{Zakrzewski}},\ }\bibfield  {title} {\bibinfo {title} {Level statistics
			across the many-body localization transition},\ }\href
	{https://doi.org/10.1103/PhysRevB.99.104205} {\bibfield  {journal} {\bibinfo
			{journal} {Phys. Rev. B}\ }\textbf {\bibinfo {volume} {99}},\ \bibinfo
		{pages} {104205} (\bibinfo {year} {2019})}\BibitemShut {NoStop}%
	\bibitem [{\citenamefont {Sierant}\ and\ \citenamefont
		{Zakrzewski}(2020)}]{Sierant20model}%
	\BibitemOpen
	\bibfield  {author} {\bibinfo {author} {\bibfnamefont {P.}~\bibnamefont
			{Sierant}}\ and\ \bibinfo {author} {\bibfnamefont {J.}~\bibnamefont
			{Zakrzewski}},\ }\bibfield  {title} {\bibinfo {title} {Model of level
			statistics for disordered interacting quantum many-body systems},\ }\href
	{https://doi.org/10.1103/PhysRevB.101.104201} {\bibfield  {journal} {\bibinfo
			{journal} {Phys. Rev. B}\ }\textbf {\bibinfo {volume} {101}},\ \bibinfo
		{pages} {104201} (\bibinfo {year} {2020})}\BibitemShut {NoStop}%
	\bibitem [{\citenamefont {Schiulaz}\ \emph {et~al.}(2020)\citenamefont
		{Schiulaz}, \citenamefont {Torres-Herrera}, \citenamefont {P\'erez-Bernal},\
		and\ \citenamefont {Santos}}]{Schiulaz20}%
	\BibitemOpen
	\bibfield  {author} {\bibinfo {author} {\bibfnamefont {M.}~\bibnamefont
			{Schiulaz}}, \bibinfo {author} {\bibfnamefont {E.~J.}\ \bibnamefont
			{Torres-Herrera}}, \bibinfo {author} {\bibfnamefont {F.}~\bibnamefont
			{P\'erez-Bernal}},\ and\ \bibinfo {author} {\bibfnamefont {L.~F.}\
			\bibnamefont {Santos}},\ }\bibfield  {title} {\bibinfo {title}
		{Self-averaging in many-body quantum systems out of equilibrium: Chaotic
			systems},\ }\href {https://doi.org/10.1103/PhysRevB.101.174312} {\bibfield
		{journal} {\bibinfo  {journal} {Phys. Rev. B}\ }\textbf {\bibinfo {volume}
			{101}},\ \bibinfo {pages} {174312} (\bibinfo {year} {2020})}\BibitemShut
	{NoStop}%
	\bibitem [{\citenamefont {Torres-Herrera}\ \emph {et~al.}(2020)\citenamefont
		{Torres-Herrera}, \citenamefont {De~Tomasi}, \citenamefont {Schiulaz},
		\citenamefont {P\'erez-Bernal},\ and\ \citenamefont
		{Santos}}]{TorresHerrera20}%
	\BibitemOpen
	\bibfield  {author} {\bibinfo {author} {\bibfnamefont {E.~J.}\ \bibnamefont
			{Torres-Herrera}}, \bibinfo {author} {\bibfnamefont {G.}~\bibnamefont
			{De~Tomasi}}, \bibinfo {author} {\bibfnamefont {M.}~\bibnamefont {Schiulaz}},
		\bibinfo {author} {\bibfnamefont {F.}~\bibnamefont {P\'erez-Bernal}},\ and\
		\bibinfo {author} {\bibfnamefont {L.~F.}\ \bibnamefont {Santos}},\ }\bibfield
	{title} {\bibinfo {title} {Self-averaging in many-body quantum systems out
			of equilibrium: Approach to the localized phase},\ }\href
	{https://doi.org/10.1103/PhysRevB.102.094310} {\bibfield  {journal} {\bibinfo
			{journal} {Phys. Rev. B}\ }\textbf {\bibinfo {volume} {102}},\ \bibinfo
		{pages} {094310} (\bibinfo {year} {2020})}\BibitemShut {NoStop}%
	\bibitem [{\citenamefont {Szo\l{}dra}\ \emph {et~al.}(2023)\citenamefont
		{Szo\l{}dra}, \citenamefont {Sierant}, \citenamefont {Lewenstein},\ and\
		\citenamefont {Zakrzewski}}]{Szoldra23}%
	\BibitemOpen
	\bibfield  {author} {\bibinfo {author} {\bibfnamefont {T.}~\bibnamefont
			{Szo\l{}dra}}, \bibinfo {author} {\bibfnamefont {P.}~\bibnamefont {Sierant}},
		\bibinfo {author} {\bibfnamefont {M.}~\bibnamefont {Lewenstein}},\ and\
		\bibinfo {author} {\bibfnamefont {J.}~\bibnamefont {Zakrzewski}},\ }\bibfield
	{title} {\bibinfo {title} {Tracking locality in the time evolution of
			disordered systems},\ }\href {https://doi.org/10.1103/PhysRevB.107.054204}
	{\bibfield  {journal} {\bibinfo  {journal} {Phys. Rev. B}\ }\textbf {\bibinfo
			{volume} {107}},\ \bibinfo {pages} {054204} (\bibinfo {year}
		{2023})}\BibitemShut {NoStop}%
	\bibitem [{\citenamefont {Mac\'e}\ \emph {et~al.}(2019)\citenamefont {Mac\'e},
		\citenamefont {Alet},\ and\ \citenamefont {Laflorencie}}]{Mace18}%
	\BibitemOpen
	\bibfield  {author} {\bibinfo {author} {\bibfnamefont {N.}~\bibnamefont
			{Mac\'e}}, \bibinfo {author} {\bibfnamefont {F.}~\bibnamefont {Alet}},\ and\
		\bibinfo {author} {\bibfnamefont {N.}~\bibnamefont {Laflorencie}},\
	}\bibfield  {title} {\bibinfo {title} {Multifractal scalings across the
			many-body localization transition},\ }\href
	{https://doi.org/10.1103/PhysRevLett.123.180601} {\bibfield  {journal}
		{\bibinfo  {journal} {Phys. Rev. Lett.}\ }\textbf {\bibinfo {volume} {123}},\
		\bibinfo {pages} {180601} (\bibinfo {year} {2019})}\BibitemShut {NoStop}%
	\bibitem [{\citenamefont {Laflorencie}\ \emph {et~al.}(2020)\citenamefont
		{Laflorencie}, \citenamefont {Lemari\'e},\ and\ \citenamefont
		{Mac\'e}}]{Laflorencie20}%
	\BibitemOpen
	\bibfield  {author} {\bibinfo {author} {\bibfnamefont {N.}~\bibnamefont
			{Laflorencie}}, \bibinfo {author} {\bibfnamefont {G.}~\bibnamefont
			{Lemari\'e}},\ and\ \bibinfo {author} {\bibfnamefont {N.}~\bibnamefont
			{Mac\'e}},\ }\bibfield  {title} {\bibinfo {title} {Chain breaking and
			kosterlitz-thouless scaling at the many-body localization transition in the
			random-field heisenberg spin chain},\ }\href
	{https://doi.org/10.1103/PhysRevResearch.2.042033} {\bibfield  {journal}
		{\bibinfo  {journal} {Phys. Rev. Research}\ }\textbf {\bibinfo {volume}
			{2}},\ \bibinfo {pages} {042033} (\bibinfo {year} {2020})}\BibitemShut
	{NoStop}%
	\bibitem [{\citenamefont {{\v S}untajs}\ \emph
		{et~al.}(2020{\natexlab{b}})\citenamefont {{\v S}untajs}, \citenamefont
		{Bon{\v c}a}, \citenamefont {Prosen},\ and\ \citenamefont
		{Vidmar}}]{Suntajs20}%
	\BibitemOpen
	\bibfield  {author} {\bibinfo {author} {\bibfnamefont {J.}~\bibnamefont {{\v
					S}untajs}}, \bibinfo {author} {\bibfnamefont {J.}~\bibnamefont {Bon{\v c}a}},
		\bibinfo {author} {\bibfnamefont {T.}~\bibnamefont {Prosen}},\ and\ \bibinfo
		{author} {\bibfnamefont {L.}~\bibnamefont {Vidmar}},\ }\bibfield  {title}
	{\bibinfo {title} {Ergodicity breaking transition in finite disordered spin
			chains},\ }\href {https://doi.org/10.1103/PhysRevB.102.064207} {\bibfield
		{journal} {\bibinfo  {journal} {Phys. Rev. B}\ }\textbf {\bibinfo {volume}
			{102}},\ \bibinfo {pages} {064207} (\bibinfo {year}
		{2020}{\natexlab{b}})}\BibitemShut {NoStop}%
	\bibitem [{\citenamefont {Gray}\ \emph {et~al.}(2018)\citenamefont {Gray},
		\citenamefont {Bose},\ and\ \citenamefont {Bayat}}]{Gray18}%
	\BibitemOpen
	\bibfield  {author} {\bibinfo {author} {\bibfnamefont {J.}~\bibnamefont
			{Gray}}, \bibinfo {author} {\bibfnamefont {S.}~\bibnamefont {Bose}},\ and\
		\bibinfo {author} {\bibfnamefont {A.}~\bibnamefont {Bayat}},\ }\bibfield
	{title} {\bibinfo {title} {Many-body localization transition: Schmidt gap,
			entanglement length, and scaling},\ }\href
	{https://doi.org/10.1103/PhysRevB.97.201105} {\bibfield  {journal} {\bibinfo
			{journal} {Phys. Rev. B}\ }\textbf {\bibinfo {volume} {97}},\ \bibinfo
		{pages} {201105} (\bibinfo {year} {2018})}\BibitemShut {NoStop}%
	\bibitem [{\citenamefont {Devakul}\ and\ \citenamefont
		{Singh}(2015)}]{Devakul15}%
	\BibitemOpen
	\bibfield  {author} {\bibinfo {author} {\bibfnamefont {T.}~\bibnamefont
			{Devakul}}\ and\ \bibinfo {author} {\bibfnamefont {R.~R.~P.}\ \bibnamefont
			{Singh}},\ }\bibfield  {title} {\bibinfo {title} {Early breakdown of area-law
			entanglement at the many-body delocalization transition},\ }\href
	{https://doi.org/10.1103/PhysRevLett.115.187201} {\bibfield  {journal}
		{\bibinfo  {journal} {Phys. Rev. Lett.}\ }\textbf {\bibinfo {volume} {115}},\
		\bibinfo {pages} {187201} (\bibinfo {year} {2015})}\BibitemShut {NoStop}%
	\bibitem [{\citenamefont {Mac{\'e}}\ \emph {et~al.}(2019)\citenamefont
		{Mac{\'e}}, \citenamefont {Laflorencie},\ and\ \citenamefont
		{Alet}}]{Mace19}%
	\BibitemOpen
	\bibfield  {author} {\bibinfo {author} {\bibfnamefont {N.}~\bibnamefont
			{Mac{\'e}}}, \bibinfo {author} {\bibfnamefont {N.}~\bibnamefont
			{Laflorencie}},\ and\ \bibinfo {author} {\bibfnamefont {F.}~\bibnamefont
			{Alet}},\ }\bibfield  {title} {\bibinfo {title} {{Many-body localization in a
				quasiperiodic Fibonacci chain}},\ }\href
	{https://doi.org/10.21468/SciPostPhys.6.4.050} {\bibfield  {journal}
		{\bibinfo  {journal} {SciPost Phys.}\ }\textbf {\bibinfo {volume} {6}},\
		\bibinfo {pages} {50} (\bibinfo {year} {2019})}\BibitemShut {NoStop}%
	\bibitem [{\citenamefont {Tal‐Ezer}\ and\ \citenamefont
		{Kosloff}(1984)}]{TalEzer84}%
	\BibitemOpen
	\bibfield  {author} {\bibinfo {author} {\bibfnamefont {H.}~\bibnamefont
			{Tal‐Ezer}}\ and\ \bibinfo {author} {\bibfnamefont {R.}~\bibnamefont
			{Kosloff}},\ }\bibfield  {title} {\bibinfo {title} {An accurate and efficient
			scheme for propagating the time dependent schrödinger equation},\ }\href
	{https://doi.org/10.1063/1.448136} {\bibfield  {journal} {\bibinfo  {journal}
			{The Journal of Chemical Physics}\ }\textbf {\bibinfo {volume} {81}},\
		\bibinfo {pages} {3967} (\bibinfo {year} {1984})}\BibitemShut {NoStop}%
	\bibitem [{\citenamefont {Fehske}\ and\ \citenamefont
		{Schneider}(2008)}]{Fehske08}%
	\BibitemOpen
	\bibfield  {author} {\bibinfo {author} {\bibfnamefont {H.}~\bibnamefont
			{Fehske}}\ and\ \bibinfo {author} {\bibfnamefont {R.}~\bibnamefont
			{Schneider}},\ }\href {http://dx.doi.org/10.1007/978-3-540-74686-7} {\emph
		{\bibinfo {title} {Computational many-particle physics}}}\ (\bibinfo
	{publisher} {Springer, Germany},\ \bibinfo {year} {2008})\BibitemShut
	{NoStop}%
	\bibitem [{\citenamefont {Pulikkottil}\ \emph {et~al.}(2023)\citenamefont
		{Pulikkottil}, \citenamefont {Lakshminarayan}, \citenamefont {Srivastava},
		\citenamefont {Kieler}, \citenamefont {B\"acker},\ and\ \citenamefont
		{Tomsovic}}]{Pulikkottil23}%
	\BibitemOpen
	\bibfield  {author} {\bibinfo {author} {\bibfnamefont {J.~J.}\ \bibnamefont
			{Pulikkottil}}, \bibinfo {author} {\bibfnamefont {A.}~\bibnamefont
			{Lakshminarayan}}, \bibinfo {author} {\bibfnamefont {S.~C.~L.}\ \bibnamefont
			{Srivastava}}, \bibinfo {author} {\bibfnamefont {M.~F.~I.}\ \bibnamefont
			{Kieler}}, \bibinfo {author} {\bibfnamefont {A.}~\bibnamefont {B\"acker}},\
		and\ \bibinfo {author} {\bibfnamefont {S.}~\bibnamefont {Tomsovic}},\
	}\bibfield  {title} {\bibinfo {title} {Quantum coherence controls the nature
			of equilibration and thermalization in coupled chaotic systems},\ }\href
	{https://doi.org/10.1103/PhysRevE.107.024124} {\bibfield  {journal} {\bibinfo
			{journal} {Phys. Rev. E}\ }\textbf {\bibinfo {volume} {107}},\ \bibinfo
		{pages} {024124} (\bibinfo {year} {2023})}\BibitemShut {NoStop}%
	\bibitem [{\citenamefont {Crowley}\ and\ \citenamefont
		{Chandran}(2020)}]{Crowley20}%
	\BibitemOpen
	\bibfield  {author} {\bibinfo {author} {\bibfnamefont {P.~J.~D.}\
			\bibnamefont {Crowley}}\ and\ \bibinfo {author} {\bibfnamefont
			{A.}~\bibnamefont {Chandran}},\ }\bibfield  {title} {\bibinfo {title}
		{Avalanche induced coexisting localized and thermal regions in disordered
			chains},\ }\href {https://doi.org/10.1103/PhysRevResearch.2.033262}
	{\bibfield  {journal} {\bibinfo  {journal} {Phys. Rev. Res.}\ }\textbf
		{\bibinfo {volume} {2}},\ \bibinfo {pages} {033262} (\bibinfo {year}
		{2020})}\BibitemShut {NoStop}%
	\bibitem [{\citenamefont {Crowley}\ and\ \citenamefont
		{Chandran}(2022)}]{Crowley22}%
	\BibitemOpen
	\bibfield  {author} {\bibinfo {author} {\bibfnamefont {P.~J.~D.}\
			\bibnamefont {Crowley}}\ and\ \bibinfo {author} {\bibfnamefont
			{A.}~\bibnamefont {Chandran}},\ }\bibfield  {title} {\bibinfo {title} {{A
				constructive theory of the numerically accessible many-body localized to
				thermal crossover}},\ }\href {https://doi.org/10.21468/SciPostPhys.12.6.201}
	{\bibfield  {journal} {\bibinfo  {journal} {SciPost Phys.}\ }\textbf
		{\bibinfo {volume} {12}},\ \bibinfo {pages} {201} (\bibinfo {year}
		{2022})}\BibitemShut {NoStop}%
	\bibitem [{\citenamefont {Wigner}(1955)}]{Wigner55}%
	\BibitemOpen
	\bibfield  {author} {\bibinfo {author} {\bibfnamefont {E.~P.}\ \bibnamefont
			{Wigner}},\ }\bibfield  {title} {\bibinfo {title} {Characteristic vectors of
			bordered matrices with infinite dimensions},\ }\href
	{http://www.jstor.org/stable/1970079} {\bibfield  {journal} {\bibinfo
			{journal} {Annals of Mathematics}\ }\textbf {\bibinfo {volume} {62}},\
		\bibinfo {pages} {548} (\bibinfo {year} {1955})}\BibitemShut {NoStop}%
	\bibitem [{\citenamefont {Wigner}(1958)}]{Wigner58}%
	\BibitemOpen
	\bibfield  {author} {\bibinfo {author} {\bibfnamefont {E.~P.}\ \bibnamefont
			{Wigner}},\ }\bibfield  {title} {\bibinfo {title} {On the distribution of the
			roots of certain symmetric matrices},\ }\href
	{http://www.jstor.org/stable/1970008} {\bibfield  {journal} {\bibinfo
			{journal} {Annals of Mathematics}\ }\textbf {\bibinfo {volume} {67}},\
		\bibinfo {pages} {325} (\bibinfo {year} {1958})}\BibitemShut {NoStop}%
	\bibitem [{\citenamefont {Sekino}\ and\ \citenamefont
		{Susskind}(2008)}]{Sekino08}%
	\BibitemOpen
	\bibfield  {author} {\bibinfo {author} {\bibfnamefont {Y.}~\bibnamefont
			{Sekino}}\ and\ \bibinfo {author} {\bibfnamefont {L.}~\bibnamefont
			{Susskind}},\ }\bibfield  {title} {\bibinfo {title} {Fast scramblers},\
	}\href {https://doi.org/10.1088/1126-6708/2008/10/065} {\bibfield  {journal}
		{\bibinfo  {journal} {Journal of High Energy Physics}\ }\textbf {\bibinfo
			{volume} {2008}},\ \bibinfo {pages} {065} (\bibinfo {year}
		{2008})}\BibitemShut {NoStop}%
	\bibitem [{\citenamefont {Cotler}\ \emph {et~al.}(2017)\citenamefont {Cotler},
		\citenamefont {Hunter-Jones}, \citenamefont {Liu},\ and\ \citenamefont
		{Yoshida}}]{Cotler17}%
	\BibitemOpen
	\bibfield  {author} {\bibinfo {author} {\bibfnamefont {J.}~\bibnamefont
			{Cotler}}, \bibinfo {author} {\bibfnamefont {N.}~\bibnamefont
			{Hunter-Jones}}, \bibinfo {author} {\bibfnamefont {J.}~\bibnamefont {Liu}},\
		and\ \bibinfo {author} {\bibfnamefont {B.}~\bibnamefont {Yoshida}},\
	}\bibfield  {title} {\bibinfo {title} {Chaos, complexity, and random
			matrices},\ }\href {https://doi.org/10.1007/JHEP11(2017)048} {\bibfield
		{journal} {\bibinfo  {journal} {Journal of High Energy Physics}\ }\textbf
		{\bibinfo {volume} {2017}},\ \bibinfo {pages} {48} (\bibinfo {year}
		{2017})}\BibitemShut {NoStop}%
	\bibitem [{\citenamefont {Nandkishore}(2015)}]{Nandkishore15proximity}%
	\BibitemOpen
	\bibfield  {author} {\bibinfo {author} {\bibfnamefont {R.}~\bibnamefont
			{Nandkishore}},\ }\bibfield  {title} {\bibinfo {title} {Many-body
			localization proximity effect},\ }\href
	{https://doi.org/10.1103/PhysRevB.92.245141} {\bibfield  {journal} {\bibinfo
			{journal} {Phys. Rev. B}\ }\textbf {\bibinfo {volume} {92}},\ \bibinfo
		{pages} {245141} (\bibinfo {year} {2015})}\BibitemShut {NoStop}%
	\bibitem [{\citenamefont {Haake}(2010)}]{Haake}%
	\BibitemOpen
	\bibfield  {author} {\bibinfo {author} {\bibfnamefont {F.}~\bibnamefont
			{Haake}},\ }\href@noop {} {\emph {\bibinfo {title} {Quantum Signatures of
				Chaos}}}\ (\bibinfo  {publisher} {Springer, Berlin},\ \bibinfo {year}
	{2010})\BibitemShut {NoStop}%
	\bibitem [{\citenamefont {Abanin}\ \emph {et~al.}(2017)\citenamefont {Abanin},
		\citenamefont {De~Roeck}, \citenamefont {Ho},\ and\ \citenamefont
		{Huveneers}}]{Abanin17}%
	\BibitemOpen
	\bibfield  {author} {\bibinfo {author} {\bibfnamefont {D.}~\bibnamefont
			{Abanin}}, \bibinfo {author} {\bibfnamefont {W.}~\bibnamefont {De~Roeck}},
		\bibinfo {author} {\bibfnamefont {W.~W.}\ \bibnamefont {Ho}},\ and\ \bibinfo
		{author} {\bibfnamefont {F.}~\bibnamefont {Huveneers}},\ }\bibfield  {title}
	{\bibinfo {title} {A rigorous theory of many-body prethermalization for
			periodically driven and closed quantum systems},\ }\href
	{https://doi.org/10.1007/s00220-017-2930-x} {\bibfield  {journal} {\bibinfo
			{journal} {Communications in Mathematical Physics}\ }\textbf {\bibinfo
			{volume} {354}},\ \bibinfo {pages} {809} (\bibinfo {year}
		{2017})}\BibitemShut {NoStop}%
	\bibitem [{\citenamefont {Liu}\ \emph {et~al.}(2023)\citenamefont {Liu},
		\citenamefont {Zhang}, \citenamefont {Hsieh}, \citenamefont {Zhang},\ and\
		\citenamefont {Yao}}]{Shuo23}%
	\BibitemOpen
	\bibfield  {author} {\bibinfo {author} {\bibfnamefont {S.}~\bibnamefont
			{Liu}}, \bibinfo {author} {\bibfnamefont {S.-X.}\ \bibnamefont {Zhang}},
		\bibinfo {author} {\bibfnamefont {C.-Y.}\ \bibnamefont {Hsieh}}, \bibinfo
		{author} {\bibfnamefont {S.}~\bibnamefont {Zhang}},\ and\ \bibinfo {author}
		{\bibfnamefont {H.}~\bibnamefont {Yao}},\ }\bibfield  {title} {\bibinfo
		{title} {Probing many-body localization by excited-state variational quantum
			eigensolver},\ }\href {https://doi.org/10.1103/PhysRevB.107.024204}
	{\bibfield  {journal} {\bibinfo  {journal} {Phys. Rev. B}\ }\textbf {\bibinfo
			{volume} {107}},\ \bibinfo {pages} {024204} (\bibinfo {year}
		{2023})}\BibitemShut {NoStop}%
\end{thebibliography}

%

\appendix

\setcounter{equation}{0}
\setcounter{figure}{0}
\setcounter{table}{0}
\makeatletter
\renewcommand{\theequation}{A\arabic{equation}}
\renewcommand{\thefigure}{A\arabic{figure}}
\renewcommand{\thetable}{A\arabic{table}}
\renewcommand{\bibnumfmt}[1]{[A#1]}

\section{Using median over disorder realizations}
\label{app:median}

\begin{figure}
    \centering
    \includegraphics{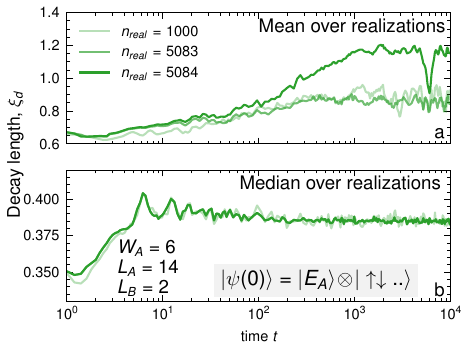}
    \caption{Growth of the decay length $\xi_d(t)$ calculated for non-interacting Anderson model ($\Delta_A=0$) with an interacting bath ($\Delta_B=1=\Delta_{AB}$) for different numbers of random realizations $n_{\text{real}}$. Including a single extra anomalous realization of disorder (low disorder close to the interface) affects the mean over realizations significantly (panel a), while the median stays unchanged (panel b).}
    \label{fig:mean_failure}
\end{figure}

Throughout the analysis of the data supporting the paper we have encountered issues with the convergence of the mean over realizations due to rare thermalization events. We have observed the largest difficulties with the convergence for the case of an Anderson insulator in subsystem $A$, $\Delta_A=0$, connected to an interacting bath, $\Delta_B=1=\Delta_{AB}$, see Eq.~(11) in the main text. Fig.~\ref{fig:mean_failure} shows the decay length $\xi_d(t)$ obtained by fitting an exponential decay to the the $\gtwo$ function calculated as a mean (Fig.~\ref{fig:mean_failure}a) or median (Fig.~\ref{fig:mean_failure}b) over random disorder realizations. We have found that even with more than $5000$ realizations, including  a single anomalous realization in the calculation leads to a drastic change in the mean value of $\gtwo$, while not affecting the median $\gtwo$. We have checked that this originates from a huge increase of the mean $\gtwo$ function due to thermalization caused by accidentally small values of disorder in part of subsystem $A$ close to the bath, i.e., an anomalous rare event. Thus, to stay consistent with the rest of the results presented in the main text, we use the median for all system configurations. It is important to note that for larger values of $\xi_d \gtrapprox 1$ we have observed no difference between the mean and the variance.

\section{Comment on the method of the decay length fitting}
\label{app:xi}
We define the correlation decay length $\xi_d$ according to Eq.~$(14)$ in the main text. To calculate $\xi_d$, we fit Eq.~$(14)$ to numerical data for $|g^{(2)}(i)|>10^{-8}$ in order not to include numerical noise. 

\begin{figure}
    \centering
    \includegraphics[width=.8\columnwidth]{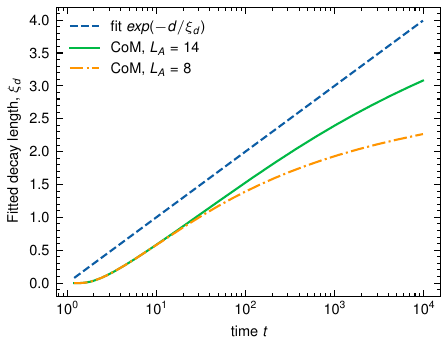}
\caption{Decay lengths obtained by fitting an exponential decay $\exp(-d/\xi_d)$ (solid line) (with $d$ - distance to interface) and calculating the center of mass, Eq.~\eqref{eq:xi_d_com} (dashed and dotted-dashed lines) assuming an artificial model of correlations with $\xi_d(t)=\log_{10} t$. Center-of-mass calculations, Eq.~\eqref{eq:xi_d_com}, would falsely suggest that the long-time growth of $\xi_d(t)$ is slower than $\xi_d\sim\log_{10}(t)$, and are visibly affected by finite size effects even for $\xi_d/L_A \sim 5$. Due to discreteness of the lattice, for $\xi_d\lessapprox 0.5$, the faster-than-logarithmic growth is erroneously implied by the CoM result.}
    \label{fig:fitexp_com}
\end{figure}

Another method, chosen in Ref.~\cite{Leonard23} for bosonic system, is to calculate the first moment,
\begin{equation}
    \xi_{d, \text{FM}} = -\sum_{i=1}^{L_A} (L_A-i) g^{(2)}(i).
\end{equation}
We believe this definition lacks an easily interpretable meaning in the spin system as it has a unit of magnetization multiplied by distance. Secondly, comparison between different system sizes is complicated as for a larger system size $L_A$ there are more terms in the sum above and the change in value of $\xi_{d, \text{FM}}$ could be related solely to the size change. Thus, for our purpose of measuring the distance of bath penetration into the MBL system, we propose an alternative to decay length obtained from exponential fit, i.e., the ``center-of-mass'' correlation length,
\begin{equation}
     \xi_{d, \text{CoM}} = \frac{\sum_{i=1}^{L_A} (L_A-i) |g^{(2)}(i)|}{\sum_{i=1}^{L_A}|g^{(2)}(i)|}.
     \label{eq:xi_d_com}
\end{equation}
Although the two definitions of $\xi_d$ and $\xi_{d, \text{CoM}}$ seem more or less equivalent, practical calculation of $\xi_{d, \text{CoM}}$ is more sensitive to i) discreteness of the lattice, and ii) finite size effects. To demonstrate that, we perform a simple numerical experiment in which we assume correlations with perfect artificial exponential decay with 
\begin{equation}
    \xi_d(t) = \log_{10} t
    \label{eq:xi_d_artificial}
\end{equation}
and use two methods to recover this value for finite system with $L_A=14$. Result is shown in Fig.~\ref{fig:fitexp_com}. The method of fitting an exponential decay gives a correct $\xi_d (t)$, and there are no finite size distortions. On the other hand, the center-of-mass method underestimates $\xi_d$ for the smallest values of $\xi_d \lessapprox 0.5$, while having an incorrect positive curvature instead of being a straight line. For systems deeply in the MBL phase, with a low $\xi_d$, this could be incorrectly interpreted as a speedup due to avalanche spreading, whereas the true decay length grows as in Eq.~\eqref{eq:xi_d_artificial}. Going further up to $\xi_d \lessapprox L_A/5$, the linear trend on the logarithmic scale is recovered and the center-of-mass result differs from the true value by an additive factor of $-0.5$. This is not problematic, as a constant shift does not change the conclusions about avalanches. However, at larger values, $\xi_d \gtrapprox L_A/5$, a clear negative curvature starts to appear due to finite size effects. This underestimation of the decay length could be interpreted as a lack of avalanches, in case the real decay length actually speeds up. As a consequence, in the main text we use exponential decay fits instead of the center-of-mass method. Even though the assumption of an exponential decay is needed here, the method does not falsely under- or overestimate the result, and change the conclusion about avalanches. 

Due to small values of the decay length $\xi_d$ in Fig.~3d in the main text, using the exponential fit revealed "jumps" of $\xi_d$ in time once new sites reached the threshold of $|g^{(2)}(i)|=10^{-8}$ and started to be included in the fit. In order to avoid this artifact, we calculated $\xi_{d, \text{CoM}}$, and translated the result back to $\xi_d$ by utilizing the empirical relation between $\xi_d$ and  $\xi_{d, \text{CoM}}$ found in Fig.~\ref{fig:fitexp_com}. In this way we are able to compare the result with the other results obtained solely by fitting an exponential decay. 

\section{Conditions for observing termination of avalanches}
\label{app:AL}

\begin{figure*}
    \centering
    \includegraphics[width=2\columnwidth]{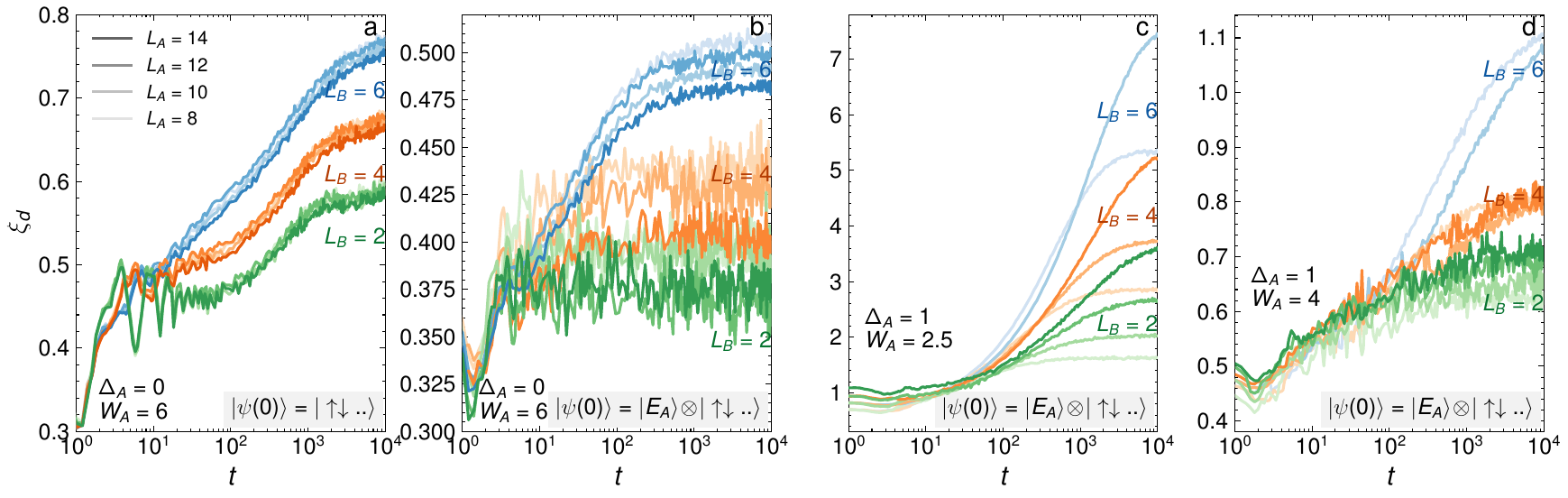}
    \caption{Correlation decay length $\xi_d(t)$ calculations in different regimes of the XXZ model, supplementing Fig.~3 in the main text. Panel a: Even the localized Anderson insulator ($\Delta_A=0$) shows a steady increase of the decay length with time for an initial N\'{e}el state. Panel b: Anderson insulator and an initial state being a product of an Anderson orbital in subsystem $A$ (we choose energy $E_A$ closest to $0$) and N\'{e}el state in the bath allows us to observe saturation in $\xi_d(t)$. Panel c: Starting from an special initial state still shows thermalization in the interacting system in its ETH regime, $W_A=2.5$. The same is observed in the critical regime with $W_A=4$ in panel d. 
    In all cases, $\Delta_B=1=\Delta_{AB}$.
    }
    \label{fig:more_xi_d_numerics}
\end{figure*}

In Fig.~3c in the main paper we have observed a rather unintuitive behavior: for $W_A=6$, $\xi_d(t)$ continues its growth until the latest times as $\xi_d(t)\propto\ln(t)$, in contrary to other measures from literature that suggest strong MBL in this regime of disorder strength and sizes. Thus, we speculated that this behavior is caused by either an initial state not being an eigenstate of $H_A$ assumed in theory from Sec.~II or by interactions. Results of these investigations are presented in Fig.~\ref{fig:more_xi_d_numerics}.

Figure~\ref{fig:more_xi_d_numerics}a with the Anderson insulator ($\Delta_A=0$) starting from an initial N\'{e}el state shows that the long-time dynamics does not disappear even in the case of no interactions. This is surprising as Anderson localization suppresses transport through the system. This suggests the explanation of the long-time growth of $\xi_d(t)$ for the MBL system in Fig.~3c of the main text - it is a feature of the dynamics and not an indication of the quantum avalanche. Indeed, setting the initial state in the Anderson insulator case to the product of an Anderson orbital (eigenstate with energy $E_A$ which we choose to be closest to $0$) and the N\'{e}el state in the bath, Fig.~\ref{fig:more_xi_d_numerics}b, we restore the saturation of $\xi_d(t)$ growth at large times.

Finally, we check whether the special choice of the initial state affects the behavior of the model in the ergodic and critical regime. In Fig.~\ref{fig:more_xi_d_numerics}c we demonstrate that for $W_A=2.5$ in the ergodic regime there is a speed up of $\xi_d(t)$ beyond $\xi_d(t) \propto \ln(t)$, a signature of a quantum avalanche. The same happens in the critical regime $W_A=4$ in Fig.~\ref{fig:more_xi_d_numerics} but the speed up appears later.

\section{GOE Hamiltonian - magnetization conservation and rescaling}
\label{app:goe}

In order to reproduce the finite spin chain bath by the GOE, the $H_{B, GOE}$ matrix is supposed to conserve the $Z$ component of the total magnetization in the whole system. The prescription we use for the GOE matrix construction is described as algorithm \ref{alg:cap}.  

\begin{figure}
    \centering
    \includegraphics[width=.9\columnwidth]{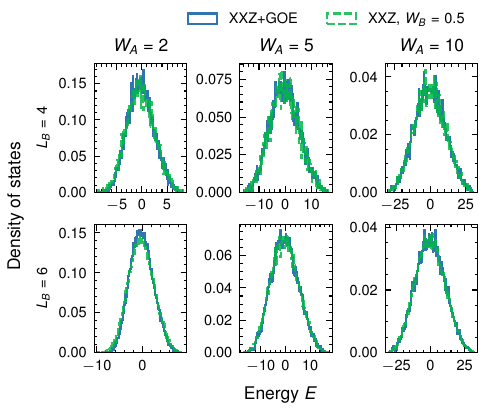}
\caption{When bath of size $L_B$ is modeled by a GOE matrix rescaled by a proper factor $\alpha$, the spectrum of the model remains approximately unchanged. Here, $L_A=10$. Single disorder realization.}
    \label{fig:dos_xxz_vs_goe_final}
\end{figure}

Furthermore, we rescale the energy spectrum of a GOE matrix from $-\sqrt{2} \leq E \leq \sqrt{2}$ to replicate the middle part of energy spectrum of the weakly disordered XXZ chain. We calculate the spectrum of the XXZ chain and find $\alpha = (p_{95} - p_{5}) / (2 \sqrt{2})$ where $p_{x}$ is the $x$-th percentile of the energy distribution. Values of $\alpha$ for different system sizes and disorder values are listed in Table~\ref{tab:alpha}. Result of the spectrum rescaling is presented in the main paper as Fig.~\ref{fig:dos_xxz_vs_goe_rescaling}. We have also verified in Fig.~\ref{fig:dos_xxz_vs_goe_final} that the total spectrum of the combined XXZ and GOE Hamiltonians, given by Eq.~\eqref{eq:H_GOE}, reproduces the original spectrum of the XXZ model with two regions of different disorder amplitudes, Eq.~\eqref{eq:H}.

\normalem 
\begin{algorithm*}
\DontPrintSemicolon
\caption{Embedding the GOE Hamiltonian in a $Z$-conserving basis}
\label{alg:cap}


 $basis \leftarrow $ Z-conserving basis of total dimension $N$, given by integers which contain spin configurations in binary format.

 $H_{XXZ} \leftarrow$ Standard XXZ Hamiltonian matrix of dimension $N\times N$ with hopping, interaction and disorder for the first $L_A$ spins, expressed in basis. It also includes hopping and interaction for the pair of spins $L_A$ (last in subsystem A) and $L_A+1$ (first in subsystem B).
 
 $\alpha \leftarrow$ GOE spectrum rescaling factor from Table~\ref{tab:alpha}.
 
 $A \leftarrow $ Matrix of $2^{L_B} \times 2^{L_B}$ random Gaussian variables of mean 0 and variance 1.
 
 $G \leftarrow 0.5*(A + A^T)$ \tcp*{GOE matrix of (excess) dimension $2^{L_B} \times 2^{L_B}$.}

 $maskB \leftarrow (1 \ll L_B) - 1$ \tcp*{Uses binary left shift $\ll$. Binary number that has 1 on the $B$ part of the spin configuration and 0 on A, eg. $000011$ for $L_A=4$, $L_B=2$.}

 $X \leftarrow $ Empty list of dimension $N$.

\For {i in range(N)} {\tcp*{To find normalization factor of the GOE matrix (number of non-zero elements) so that the final spectrum is the Wigner semicircle $-\sqrt{2}<E<\sqrt{2}$.}

     $X[i] \leftarrow basis[i] \& maskB$ \tcp*{Binary AND. Gives configuration on subsystem B.}
    
}
 $n \leftarrow$ Number of unique binary numbers in $X$.

\For {i in range(N)}
{

     $configAi \leftarrow basis[i] \gg L_B$
    
     $configBi \leftarrow basis[i] $\&$ maskB$

    \For {j in range(N)}
    {
         $configAj \leftarrow basis[j] \gg L_B$
        
         $configBj \leftarrow basis[j] \& maskB$
    
        \If{$configAi == configAj$} 
        {
             $H_{XXZ}[i, j] \leftarrow H_{XXZ}[i, j] + 0.5*\alpha*G[configBi, configBj] / \sqrt{n}$ \tcp*{Matrix elements of $G$ selected based on integers uniquely defining configuration in the $B$ part of the chain.}
            
             $H_{XXZ}[j, i] \leftarrow H_{XXZ}[i, j]$
         }   
     }   
}
\Return{$H_{XXZ}$}

\end{algorithm*}

\begin{table}
\centering
\begin{tabular}{lcc}
\toprule
$W$        & $L$  & scale $\alpha$ \\\midrule
0        & 2  & 0.354 \\
         & 4  & 0.837 \\
         & 6  & 1.113 \\
         & 8  & 1.353 \\
         & 10 & 1.574 \\
         & 12 & 1.720 \\\midrule
0.5      & 2  & 0.428 \\
         & 4  & 0.927 \\
         & 6  & 1.232 \\
         & 8  & 1.459 \\
         & 10 & 1.666 \\
         & 12 & 1.823 \\\midrule
1        & 2  & 0.600 \\
         & 4  & 1.115 \\
         & 5  & 1.468 \\
         & 8  & 1.684 \\
         & 10 & 1.915 \\
         & 12 & 2.091 \\\bottomrule
\end{tabular}
\caption{Factors used to rescale the GOE Hamiltonian. In the main paper, values for $W=0.5$ were used.}
\label{tab:alpha}
\end{table}

\section{Further numerical experiments}
\label{app:further}
In this section of Appendix, we perform further numerical studies of the interplay of the system and bath parameters and their effect on thermalization.

\subsection{Fixed bath, changing subsystem $A$ size}
\begin{figure}
    \centering
    \includegraphics[width=.9\columnwidth]{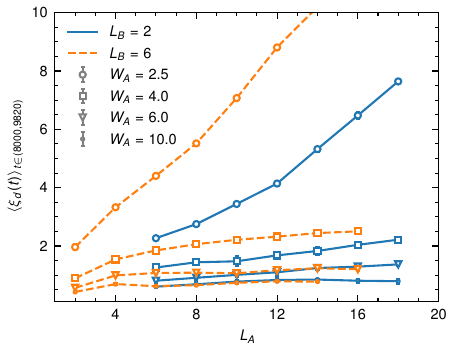}
    \caption{Fitted bath correlation decay length $\xi_d$ as a function of $A$ system size $L_A$, averaged over time window $t\in (8000,9820)$. Increasing the bath size from $L_B=2$ to $L_B=6$ enhances thermalization. 
    However, if the growth is not faster than linear with $L_A$, the thermalized part of chain $A$ will occupy at most a finite fraction of the chain A, and some localized sites far from the interface will not be affected by the bath. They may still be fully thermalized but at timescales exponential in system size $L_A$.}    
    \label{fig:xi_d_bath}
\end{figure}

In Fig.~\ref{fig:xi_d_bath} we check how the long-time value of $\xi_d(t)$ changes with $L_A$ for two bath sizes $L_B=2,6$. In particular, we observe that in the ETH phase ($W_A=2.5$) $\xi_d$ grows with $L_A$ faster than linearly. This means that for a larger system, a larger fraction of the system is correlated with the bath. Extrapolating this to the large system size limit, $L_A\gg 1$, almost the entire system is thermalized by a finite bath before reaching the Heisenberg time $t_H$. On the other hand, disorder values $W_A=4, 6, 10$ give a linear dependence of $\xi_d$ on $L_A$ and a constant fraction of the system stays localized for each setting, not being affected by the presence of the bath.

\subsection{Fixed bath and $L_A$, changing $W_A$}
\begin{figure}
    \centering
    \includegraphics[width=.8\columnwidth]{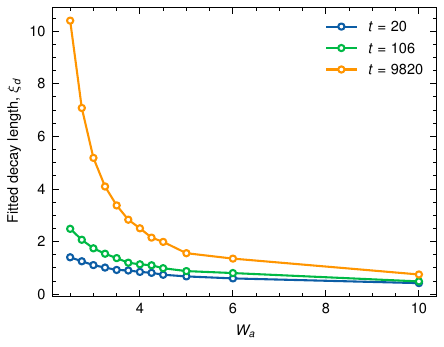}
    \caption{ Fitted bath correlation decay length $\xi_d$ as a function of disorder strength $W_A$. $L_A=14$, $L_B=6$.}
    \label{fig:xi_d_Wa}
\end{figure}

Restricting ourselves to a smaller system size of $L_A=14$, $L_B=6$, we check how the late-time decay length depends on the disorder strength $W_A$. The result is presented in Fig.~\ref{fig:xi_d_Wa}. This plot shows that the decay length decreases very fast between $W_A=2.5$ to $W_A\approx 5$, and then stays approximately constant around value $\xi_d\approx 0.5$ for higher $W_A$. It suggests that MBL prevents the spread of avalanches at strong disorder, yet no sharp transition can be identified due to finite sizes of the systems of interest.

\section{Comparison with other measures - entanglement entropy}
In order to get a view on thermalization process from another point, we calculate the von Neuman entanglement entropy of the system. In particular, we are interested in the number entropy part, which in the spinless fermions description of the XXZ chain measures the uncertainty in the number of particles in the subsystem. The formula reads
\begin{equation}
    \mathcal{S}_{num}(t)=-\sum_n p(n) \ln p(n),
\end{equation}
where $p(n)$ is the probability of total magnetization ${n=\sum_{i\in \text{cut}}Z_i}$ of a chosen subsystem of the whole system, and $\rho(n)$ is the block of the reduced density matrix in magnetization $n$ sector of the Hilbert space.

\begin{figure}
    \centering
    \includegraphics[width=.8\columnwidth]{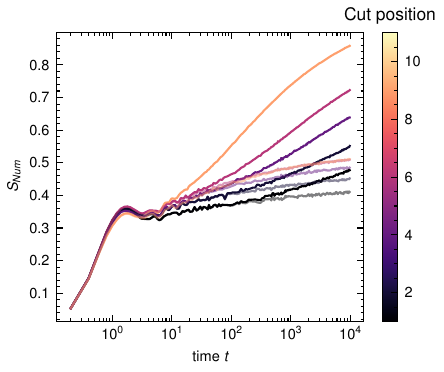}
    \caption{Number entropy $S_{Num}$ for $L_A=12$, $L_B=10$, $W_A=4$, and $W_B=0.5$ (solid lines) or $W_B=4$ (transparent lines). Depending on the position of the cut, a speedup beyond $\propto \ln(t)$ is observed in the case $W_B=0.5$ sooner or later, whereas the growth is slower than logarithmic when $W_B=4$. This suggests that due to the presence of the bath, a thermal avalanche is traveling through the system. Thus conclusions from the analysis of the number entropy and correlation decay length are fully compatible.}
    \label{fig:entropy}
\end{figure}

In Fig.~\ref{fig:entropy} we notice that in the case of no bath ($W_B=W_A=4$, $L_A=12$, $L_B=10$), the number entropy grows with time slower than $\ln t$. Bath with $W_B=0.5$ causes a speedup in the number entropy growth. While the first speedup for cut between sites 11 and 12 can be observed around $t=10$, the further the cut position is, the longer it takes for the speedup to be visible. Nevertheless, entanglement entropy for the furthest cut from the interface obtains a speedup beyond logarithmic growth when the bath is present. This can be interpreted as an avalanche reaching the boundary of the system. This conclusion stays in agreement with the result obtained for the same system by measuring simpler two-body correlation functions and their decay length, shown in Fig.~\ref{fig:xi_d_growth_4panels}b in the main paper.

\section{Universal scaling of time with disorder}
\label{app:scaling}
Here we present details on the determination of optimal constant $a$ in a universal time rescaling law $t\rightarrow t e^{-a W_A}$ in the discussion of Sec.~\ref{sec:comp}.

To find $a$ numerically, we fix the value of $a$, rescale time $t\rightarrow t e^{-a W_A}$, and perform a fourth degree polynomial fit to the imbalance from Fig.~\ref{fig:kicked_vs_notkicked} on the logarithmic time scale. The fit is performed either for a small interval of disorder values $W_A$, or for all data. Root-mean-squared error (RMSE) of this fit measures the quality of $a$ as a potential constant explaining the trends in the data that can be captured by a continuous curve. We repeat the fitting for a range of different values of $a$, and the optimal $a$ is the one with the minimal RMSE of the fit. 

\begin{figure*}
    \centering
    \includegraphics[width=2\columnwidth]{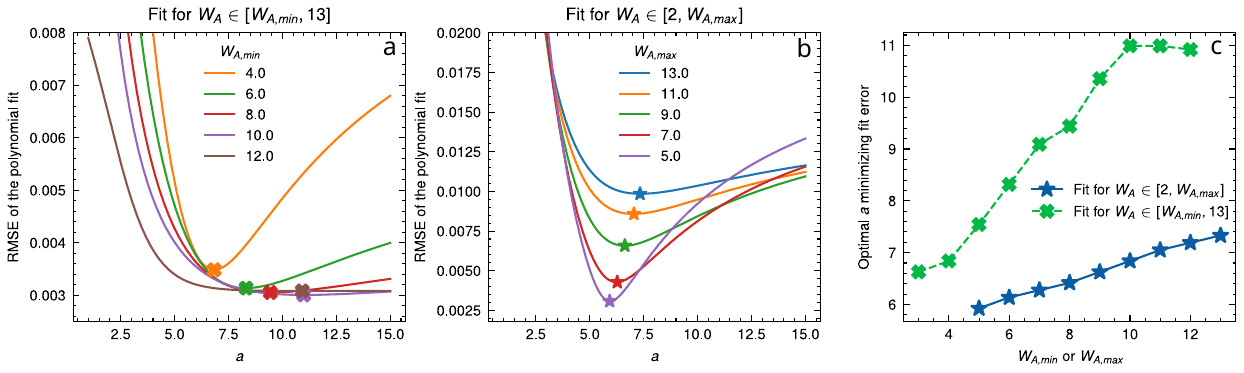}
\caption{Disordered XXZ model. \textbf{a, b }Root-mean-squared error (RMSE) of the 4-th degree polynomial fits to imbalance from Fig.~\ref{fig:kicked_vs_notkicked} with rescaled time axis according to $t\rightarrow t e^{-a W_A}$. Lowest values of the error, marked by crosses or stars, correspond to datapoints forming a universal curve due to rescaling time. Panel \textbf{a} shows fit errors when fitting the curve for the largest imbalances - interval of disorder amplitudes used for fitting is marked by the color, see legend. Minima are very shallow when fitting at large disorder values, signaling ambiguity of optimal $a$ in that case. Panel \textbf{b} shows the RMSE of the fit limited to the data for smallest disorder values, down to $W_A=2.0$. Panel \textbf{c}: Optimal $a$ that describes universality shifts from around $6.5$ to $11$ when fitting to largest disorders. It stays around $6-7$ when fitting to smallest disorder values, signaling the lack of a globally universal behavior for all disorder values. Fit was performed between times $t=90$ to $t=9820$ to not include initial oscillations of the imbalance, and for disorder values $W_A=2,3,4,...,13$. In all cases, $W_B=0.5$.}
    \label{fig:fit_a_XXZ}
\end{figure*}

Results for fitting $a$ to the data from our disordered XXZ model with a bath, decsribed by Eq.~\eqref{eq:H} in the main text, are presented in Fig.~\ref{fig:fit_a_XXZ}. When the fit is performed for the strongest disorder amplitudes up to $W_A=13$ (Fig.~\ref{fig:fit_a_XXZ}a), the optimal $a$ has values around $a\approx 6.5-11$, the RMSE minima shift significantly when changing the range of disorders used for fitting, and the minima are very shallow i.e. $a$ is not properly defined. Fitting to the data corresponding to the weakest disorder amplitudes, down to $W_A=2$, gives a different result: optimal $a$ stays around $6-7.5$ and the minima are deeper (see Fig.~\ref{fig:fit_a_XXZ}b). Optimal $a$ for all considered ranges of disorder are summarized in Fig.~\ref{fig:fit_a_XXZ}c: constant $a$ drifts between different ranges of disorder amplitudes used for fitting, and the two regimes of large and small disorder values give inconsistent $a$. This means that the model lacks a universal scaling of time $t\rightarrow t e^{-a W_A}$.

\begin{figure*}
    \centering
    \includegraphics[width=2\columnwidth]{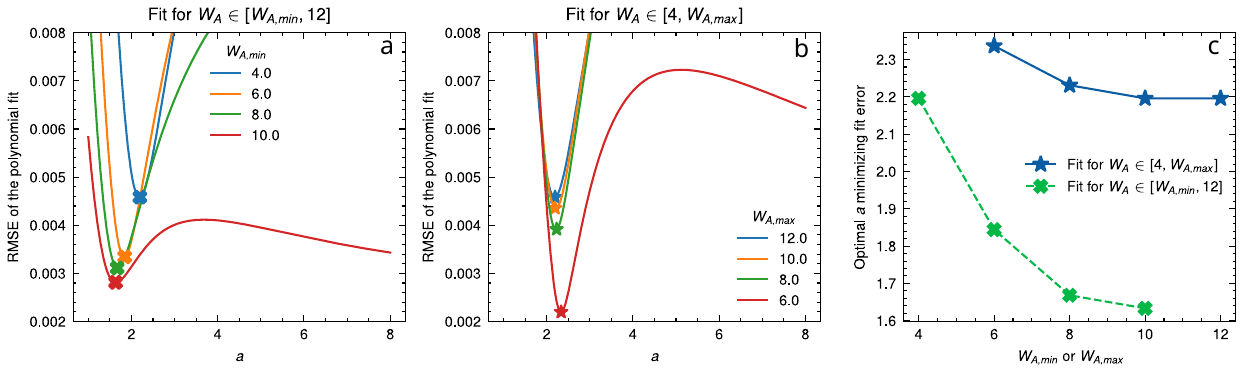}
\caption{Same as in Fig.~\ref{fig:fit_a_XXZ} but for the kicked XXZ model. Minima in panels \textbf{a}, \textbf{b} are well defined and do not change significantly depending on the fitting interval of disorders $W_A$. Panel \textbf{c} shows that an optimal $a$ minimizing the RMSE has an inverse trend when compared to the time-independent case in Fig.\ref{fig:fit_a_XXZ}. Nevertheless, its values stay consistently around $2$, while in the time-independent case they drift from $6.5$ to $11$ with the change of fitting disorder interval. Here, $W_A=4,6,8,10,12$, and $W_B=0.5$.}
    \label{fig:fit_a_kicked}
\end{figure*}

The same analysis, repeated for the kicked XXZ model of Peacock and Sels \cite{Peacock23}, confirms the presence of a single universal constant $a$ that drifts much less with the disorder interval considered. Figures~\ref{fig:fit_a_kicked}a (b) show RMSE of the fourth degree polynomial fit to the data which has large (small) disorder values. The minima are well defined in both cases, and the optimal $a$ stays around $2$ with an absolute error of around $0.3$. Thus, there exists a single $a$ that is universal irrespective of the disorder strength, in full agreement with the statements of Ref.~\cite{Peacock23}. It is interesting, though, that all trends observed in this case in Fig.~\ref{fig:fit_a_kicked}c, are reverse to trends for the time-independent model, Fig.~\ref{fig:fit_a_kicked}c. At present, it is not clear to us why such a change in trend is present.


\end{document}